\documentclass[iop,apj,appendixfloats]{emulateapj}
\usepackage{apjfonts}
\usepackage[latin1]{inputenc}
\usepackage[english]{babel}
\usepackage{amsmath}
\usepackage{amssymb}		
\usepackage{latexsym} 
\usepackage{epsfig}
\usepackage{subfigure}
\usepackage{natbib}
\usepackage{setspace}
\usepackage{graphicx}
\usepackage{longtable}
\usepackage{epstopdf}
\usepackage{verbatim}
\usepackage{color,soul}
\newcommand{\msun}{{M_{\odot}}}
\newcommand{\mstar}{{M_{\ast}}}
\newcommand{\ser}{S\'ersic }

\shorttitle{Connection Between Stellar Mass Distributions Within Galaxies and Quenching Since \textit{z} $=2$}
\shortauthors{Mosleh et al.}

\begin{document}

\title{Connection Between Stellar Mass Distributions Within Galaxies and Quenching Since \textit{z} $=2$}

\author{Moein Mosleh \altaffilmark{1,2}, Sandro Tacchella \altaffilmark{3}, Alvio Renzini \altaffilmark{4}, C. Marcella Carollo \altaffilmark{3}, Alireza Molaeinezhad \altaffilmark{1}, Masato Onodera \altaffilmark{5}, Habib G. Khosroshahi \altaffilmark{1}, Simon Lilly \altaffilmark{3}}

\altaffiltext{1}{School of Astronomy, Institute for Research in Fundamental Sciences (IPM), PO Box 19395-5531 Tehran, Iran}\email{Email: mosleh@ipm.ir}
\altaffiltext{2}{Physics Department and Biruni Observatory, Shiraz University, Shiraz 71454, Iran}
\altaffiltext{3}{Department of Physics, Institute for Astronomy, ETH Zurich, CH-8093 Zurich, Switzerland}
\altaffiltext{4}{INAF Osservatorio Astronomico di Padova, vicolo dell'Osservatorio 5, I-35122 Padova, Italy}
\altaffiltext{5}{Subaru Telescope, National Astronomical Observatory of Japan, 650 North A'ohoku Place, 96250 Hilo, HI, USA}

\slugcomment{Accepted for publication in ApJ}

\begin{abstract}

We study the history from $z\sim2$ to $z\sim0$ of the stellar mass assembly of quiescent and star-forming galaxies in a spatially resolved fashion. For this purpose we use multi-wavelength imaging data from the Hubble Space Telescope (HST) over the GOODS fields and the Sloan Digital Sky Survey (SDSS) for the local population. We present the radial stellar mass surface density profiles of galaxies with $\mstar>10^{10} \msun$, corrected for mass-to-light ratio ($\mstar/L$) variations, and derive the half-mass radius ($R_{m}$), central stellar mass surface density within 1 kpc ($\Sigma_{1}$) and surface density at $R_{m}$ ($\Sigma_{m}$) for star-forming and quiescent galaxies and study their evolution with redshift. At fixed stellar mass, the \textit{half-mass sizes} of quiescent galaxies increase from $z\sim2$ to $z\sim0$  by a factor of $\sim3-5$, whereas the half-mass sizes of star-forming galaxies increase only slightly, by a factor of $\sim2$. The central densities $\Sigma_{1}$ of quiescent galaxies decline slightly (by a factor of $\lesssim1.7$) from $z\sim2$ to $z\sim0$, while for star-forming galaxies $\Sigma_{1}$ increases with time, at fixed mass. We show that the central density $\Sigma_{1}$ has a tighter correlation with specific star-formation rate (sSFR) than $\Sigma_{m}$ and for all masses and redshifts galaxies with higher central density are more prone to be quenched. Reaching a high central density ($\Sigma_{1}\ga10^{10}~\msun~\mathrm{kpc}^2$) seems to be a prerequisite for the  cessation of star formation, though a causal link between high $\Sigma_{1}$ and quenching is difficult to prove and their correlation can have a different origin. \\ 

\end{abstract}

 
\section{Introduction}

Studying the stellar mass functions of quiescent and star-forming galaxies indicates that star-forming galaxies shut down their star-formation (``quenching'') with time and increase the number of quiescent galaxies \citep{Bundy2006, Penglilly2010, muzzin2013, moustakas2013, Ilbert2013, tomczak2014}. Consequently, the color bimodality of the galaxies exist in the local universe and persist up to the high redshifts \citep[e.g.,][]{strateva2001, blanton2003, Bell2004, williams2009a}. \cite{Penglilly2010} proposed distinct ``mass'' and ``environment'' processes or modes of quenching for central and satellite galaxies. However, as we focus on central galaxies in this paper, we do not look at environmental quenching mechanisms. To explain this bimodality, several quenching mechanisms are proposed to prevent cooling of gas in/into galaxies. These processes could e.g., expel the gas via stellar feedback or active galactic nuclei (AGN) feedback \citep[quasar-mode; e.g.,][]{spingel2005, Dimatteo2005, Dalla2008, ciotti2009, Vogelsberger2014} or prevent cooling of the gas in halo and keep the gas hot against further accretion to the galaxies, e.g., halo quenching \citep{brinboim2003,keres2005,dekel2006}, AGN radio-mode feedback \citep{croton2006} and gravitational heating \citep{Birnboim2011}. Morphological quenching \citep{martig2009} which stabilize the disk against gravitational collapse of gas and preventing formation of giant molecular clouds is another recently suggested mechanism.  

In addition, diverse morphologies of galaxies seen in the local universe have raised questions of how and when the galaxies assembled their stellar masses and structural components. Distinct properties of galaxies at high redshifts $(z\gtrsim1)$ compare to the local universe, indicates that galaxies of various types have gone through complex processes. Star-forming galaxies in the local universe could have consist of bulges (classical, pseudo- or peanut/boxy types), or bars in addition to the disk components. Processes that form these structural components of star-forming galaxies, are still matter of debate. Many processes including accretion of cold gas and formation/migration of giant gas clumps, major mergers or secular evolutionary mechanisms such as bar-instability are proposed for the formation of the galaxies central over densities or bulges \citep{kormendy2004, debattista2006, genzel2006, Carollo2007, genzel2008,  dekel2009, Elmegreen2008, forster2011, sales2012},  \citep[see also][for recent reviews.]{Bournaud2016, Brooks2016, Fisher2016, kormendy2016}. 

Quiescent galaxies, on the other hand, observed to have more concentrated light/mass profiles \citep[e.g.,][]{Driver2006, bell2008, bell2012} and have been grown in size since $z\sim2-3$, in particular, the average size of the population of massive quiescent galaxies has grown by a factor of $3-5$ since redshift of $z\approx2$ to the present \citep[e.g.,][]{daddi2005, trujillo2007, vandokkum2008, cimatti2008}, implying that these galaxies could have assembled their stellar masses via different mechanism at different epochs. One possible channel could be the formation of compact spheroids (or ``blue nuggets'') at high redshifts \citep{Barro2013, Barro2014a, vandokkum2015} and their evolution at later times via other processes such as major/minor mergers \citep[e.g.,][]{khochfar2006, bell2006, naab2009, bezanson2009, dokkum2010, oser2012, newman2012, hilz2013}. Considering the increasing number density of quiescent galaxies by factor of 10 since $z\sim2$ from studying the stellar mass functions, an alternative channel would be due to the progenitor bias, i.e, newly quenched galaxies increase the average size of the quiescent population \citep{poggianti2013a, carollo2013, Cassata2013, Belli2015}. An open question is how much quiescent galaxies grow individually in size (e.g., by merging) and how much of the average size-increase of the population can be explained by progenitor bias. Therefore, to understand the growth of quiescent galaxies, one has to study the growth of star-forming galaxies. 

The emergence of spheroidal-dominant morphology of quiescent galaxies -- while they migrate from the blue cloud to the red sequence -- encourages to investigate the correlation between structural parameters with the cessation of star-formation activity. Early studies by \citet{kauffmann2003} demonstrate that stellar mass surface density of galaxies are better correlated with the age of galaxies than stellar mass, in the local universe \citep[also][]{Kauffmann2006} using Sloan Digital Sky Survey (SDSS). The later works shows that galaxies with prominent bulges (higher \ser index $n$) are more prone to be quenched \citep{allen2006, Driver2006, schiminovich2007, bell2008, mendez2011}, implying that there might be a link between the star-formation activity and the stellar mass distribution within galaxies. \citet{franx2008} extended this study to high redshifts and demonstrate the better correlation of the specific star-formation rate with surface density and inferred velocity dispersion than stellar mass up to $z\sim3$ \citep[also][]{mosleh2011, wuyts2011, omand2014, bell2012}. 

\citet{wake2012} explored SDSS galaxies and showed that central velocity dispersion is the best indicator for predicting galaxy colors than stellar mass, surface density and \ser index. Moreover, \citet{cheung2012} used central stellar mass surface density ($\Sigma_{1}$), which is the stellar mass in the central 1 kpc region of galaxies and found that this parameter shows less scattered relation with galaxies color than surface density, stellar mass and \ser index at intermediate redshift $z\sim0.65$. Following that, \citet{fang2013} demonstrate this for local galaxies and stated the necessity of bulge formation for galaxies to be quenched, but not sufficient. In addition, they discuss that central velocity dispersion with 1 kpc is also correlated with $\Sigma_{1}$ and could be good predictor if measured robustly. \citet{woo2015} explored the formation of compact central density for central and satellite galaxies and their halo mass dependence, and argued on the fast quenching conditions related to central density compactness compare to the slow mechanism of halo for central galaxies. Nevertheless, results of these works suggest that determining distribution of stellar masses within galaxies, in particular central mass density, is important for better understanding of quenching mechanism. Though the good correlation between central mass density and quenching does not prove the causal relation. \citet{Lilly2016} have shown that a mass-dependent quenching mechanism acting on star-forming galaxies whose size follow the observed size-mass evolution leads automatically to a high central density at the onset of quenching, without however any causal connection between the central density and quenching.

In this paper, we study the stellar mass distributions of intermediate to massive galaxies from $z=2$ to $z\sim0$, by deriving point-spread function (PSF) corrected stellar mass profiles of these galaxies consistently at low and high redshifts, and further comparing the corresponding mass distribution parameters with star-formation rates of the galaxies. Many studies used the evolution of light profiles of galaxies to constrain the primary mechanism for assembling the observed structural properties of galaxies at a given stellar mass. The optical-near infrared light could be a good representative of stellar masses, though the existence of color-gradients at low and high redshifts \citep{franx1989, peletier1990, labarbera2005, dokkum2010, guo2011, gargiulo2011} could affect estimating the true complex underlying stellar population within galaxies. The origin of the color-gradient could be due to variation of the stellar population age, dust content and stellar metallicity. Consequently, as discussed by several authors \citep[e.g.,][]{szomoru2013, fang2013, Carollo2014, tacchella2015a, tacchella2015}, the mass-weighted profiles of galaxies are more robust than the light-weighted profiles, while comparing different galaxy populations. Hence, the constrains on the assembly history of various galaxies can be provided by studying their mass growth on the central and outer regions, at different epochs. Therefore, tracing and constraining the plausible evolutionary models of galaxies could benefits from studying the evolution of the stellar mass distributions within them. 

Several approaches are employed for deriving mass surface density of galaxies at low and high-$z$. In recent works by \citet{dokkum2010} and \citet{patel2013a}, the light profiles are converted to the mass profiles, without assuming mass-to-light ratio ($\mstar/L$) gradients. Some authors \citep[e.g.,][]{Zibetti2009, szomoru2013, fang2013, tacchella2015} overcome this by exploiting the empirical relation between rest-frame color and $\mstar/L$, however, as pointed by these authors, the effects of age, metallicty and dust are not exactly the same and hence introduce scatter around the color-$\mstar/L$ relation. To infer the variation of the stellar masses and stellar populations, \citet{wuyts2012} and \citet{hemmati2014} fitted stellar population models to the resolved 2-dimensional images of the galaxies. Thanks to the high-resolution, sensitive instruments on board of the \textit{Hubble Space Telescope} (HST) such as Advanced Camera for Survey (ACS) and Wide Field Camera 3 (WFC3), this could be feasible out to high-redshifts ($z\lesssim2-3$).  Although, the lack of high-resolution middle infra-red images could slightly limited this method, the multi-wavelength observations will allow for fitting the stellar population models to a large sample of galaxies at these redshifts. This approach is extended by \citet{morishita2015}, by converting 2-dimensional images to 1-dimension images and median-stacking sample of galaxies at each redshift bin, prior to the spectral energy distribution (SED) modeling. The derived stellar surface density are then de-convolved to reduce the effect of PSF.  Recently, \citet{Barro2015b} used observed 1-D profiles of galaxies to derive their mass-profiles. 

In this study, we exploit the SED fitting technique on one dimensional ``PSF-corrected'' light profiles of individual galaxies. Instead of deriving mass density profile from the stacked images \citep[e.g.,][]{dokkum2010, morishita2015}, we derived PSF-corrected mass profiles of individual galaxies from all their available HST filters, taking into account the  gradients of $\mstar/L$.

The layout of this paper is as following. In Section 2, we introduce the sample and data used in this work. The methods for deriving the stellar mass density are described in Section 3. We present the comparison of galaxies mass profiles and explore the mass profile parameters on star-formation activity in Section 4 \& 5 and discussed the results in Section 6. We summarize our work on Section 7. The cosmological parameters adopted in this work are $\Omega_{m}$ = 0.3, $\Omega_{\Lambda}$ = 0.7 and $H_{0} = 70$ $km$ $s^{-1}$ $Mpc^{-1}$.\\

\begin{figure*}
\includegraphics[width=0.9\textwidth]{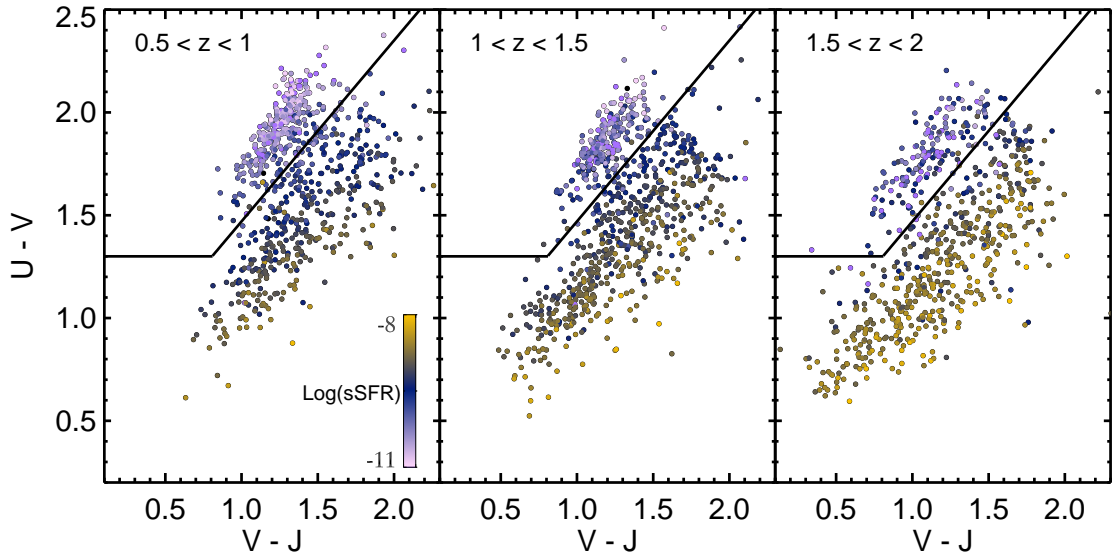}
\caption{Quiescent and star-forming galaxies are defined in different redshift bins based on their rest-frame $U-V$ and $V-J$ colors, known as UVJ method \citep{williams2009a}. Sources are color-coded according to their specific star formation rate (sSFR) from the 3D-HST catalog \citep{whitaker2014}. At all redshifts quiescent galaxies occupy separate region on the UVJ diagram.} 
\label{fig1}
\end{figure*}

\begin{figure}
\includegraphics[width=0.4\textwidth]{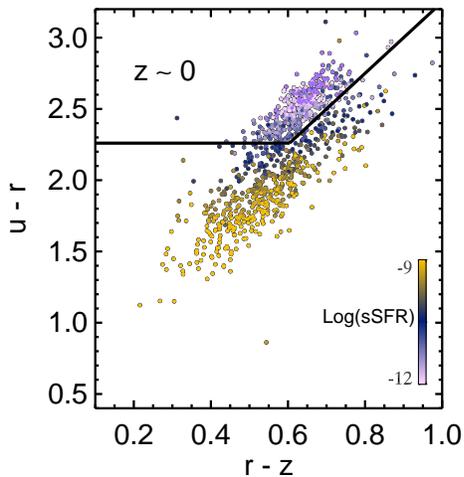}
\caption{Quiescent and star-forming galaxies in the local Universe are based on $urz$ color-color selection \citep{holden2012}, analogues to high-$z$ UVJ. Sources are color-coded according to their specific star formation rate (sSFR) from the MPA-JHU DR7 catalog \citep{brinchmann2004}.} 
\label{fig2}
\end{figure}

\section{DATA \& SAMPLE}

For our analysis, we have used the catalog and imaging of the 3D-HST Treasury Program \citep{brammer2012, skelton2014} on the Great Observatories Origins Deep Survey \citep[GOODS:][]{giavalisco2004}, GOODS-Northern and GOODS-Southern fields (GOODS-North and GOODS-South) of the Cosmic Assembly Near-IR Deep Extragalactic Legacy Survey \citep[CANDELS;][]{grogin2011, koekemoer2011}. The catalogs provided photometry, redshift (spectroscopic and photomteric), the stellar mass and color of galaxies, in addition to the HST imaging of these fields, which are all available at the 3D-HST website\footnote[1]{\url{http://3dhst.research.yale.edu/Home.html}}. We use sources with \texttt{use\_phot}$=1$ (photometric quality flag) in the 3D-HST catalogs to select a reliable sample. In order to reduce the systematic effects in our methodology (described in the next section) for deriving consistent light profiles at different filters, we used PSF-matched images at different filters (smoothed to the $H_{160}$ image resolution).   

In the \citet{skelton2014}, the photometric redshifts are determind using $EAZY$ code \citep{brammer2008} for objects without spectroscopic redshift and the stellar masses are measured using FAST code \citep{kriek2009}.  

The total area of GOODS-North and GOODS south fields are about 340 arcmin$^2$. These two GOODS fields, have observations on seven HST filters in common, $F435W$, $F606W$, $F775W$, $F850W$, $F125W$, $F140W$, $F160W$ ($B_{435}$,$V_{606}$, $i_{775}$, $z_{850}$, $J_{125}$, $JH_{140}$, $H_{160}$, hereafter), providing sufficient wavelength coverage for deriving stellar population modeling from SED fitting, in comparison to other CANDELS fields, which have less wavelengths range (five filters). 

We have divided our sample into quiescent and star-forming galaxies using UVJ color criteria \citep{williams2009a}, based on the $U-V$ and $V-J$ rest-frame colors. The EAZY code is used to derive $U-V$ and $V-J$ colors.  Our criteria for selecting quiescent galaxies at $ z < 2.0$, is as following:

\begin{equation}
 (U - V) > 0.88 \times (V-J) + 0.59 \quad [0.5 < z < 2.0]
\end{equation} 
 \begin{equation}
 (U - V) > 0.88 \times (V-J) + 0.69 \quad [0.0 < z < 0.5]
\end{equation}
\begin{equation}
 (U - V) > 1.3 \quad [0.0 < z < 2.0] 
\end{equation}

which is similar as the criteria defined by \citet{whitaker2011}. Any sources falling outside of these criteria are defined as star-forming galaxies.  The combination of these equations allows separation of un-obscured and dusty star-forming galaxies from the quiescent ones while preventing the contamination. In Figure \ref{fig1}, we show the color criteria (sold lines) used in this study to separate star-forming and quiescent galaxies at three different redshift intervals between $0.5 \leq z \leq 2.0$. The sources are color-coded according to their total specific star formation rate from 3D-HST catalog using star-formation rate (using combined UV+IR SFR) derived by \cite{whitaker2014}  \citep[see also][]{momcheva2015}.  As can be seen, the clumps of quiescent galaxies clearly falling in a separate regions of the UVJ diagrams. Our samples are then selected to have stellar masses $\geq 10^{10} \msun$. As shown by \citet{skelton2014} and \citet{vanderwel2014}, both star forming and quiescent galaxies are complete out to the redshift of 2 \citep[see also][]{morishita2015}.

We have also used a sample of local galaxies using SDSS Data Release 7 (DR7) images \citep{abazajian2009}. We randomly select 1000 galaxies at $0.06<z<0.08$, stellar mass matched with the total mass distributions of SDSS. The stellar masses are taken from Max-Planck-Institute for Astrophysics (MPA)-Johns Hopkins University (JHU) SDSS DR7 catalog  \citep{kauffmann2003,salim2007} and the star formation rate from \citep{brinchmann2004} \footnote[2]{\url{http://www.mpa-garching.mpg.de/SDSS/DR7/}}. In order to divide the local sample into star-forming and quiescent we use analogues method to the UVJ method, using $u-r$ versus $r-z$ color-color criteria, as defined by \citet{holden2012}:

\begin{equation}
(u -r)_{0} > 2.26
\end{equation}
\begin{equation}
(u -r)_{0} > 0.76 + 2.5 (r-z)_{0}
\end{equation}

As shown in Figure \ref{fig2}, quiescent galaxies occupy separate region from star-forming ones. Similar to Figure \ref{fig1}, galaxies are color-coded according to their specific SFR. We use the New York University Value-Added Galaxy Catalog \citep[NYU-VAGC;][]{Blanton2005} for local galaxies color and absolute magnitudes. Note that using similar color-color selection criteria to separate quiescent and star-forming galaxies at low and high redshifts reduces biases may caused by different methods of star-formation rate measurements. \\


\begin{figure}
\includegraphics[width=0.45\textwidth]{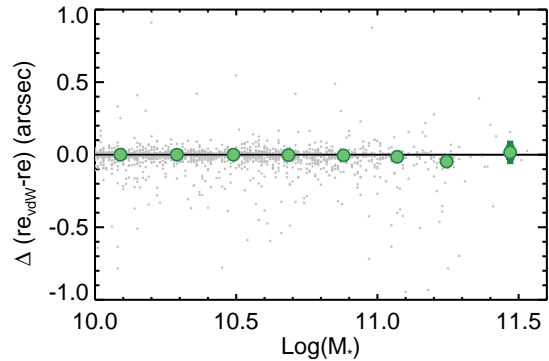}
\caption{The difference between $H_{160}$-band $r_{e}$ of \citet{vanderwel2014} and this work as a function of stellar masses. The green points represent the median differences at each mass bin and 1-$\sigma$ error bars. There is no systematic error in our size measurements. } 
\label{fig3}
\end{figure}

\begin{figure*}
\centering
\includegraphics[width=1.95 in]{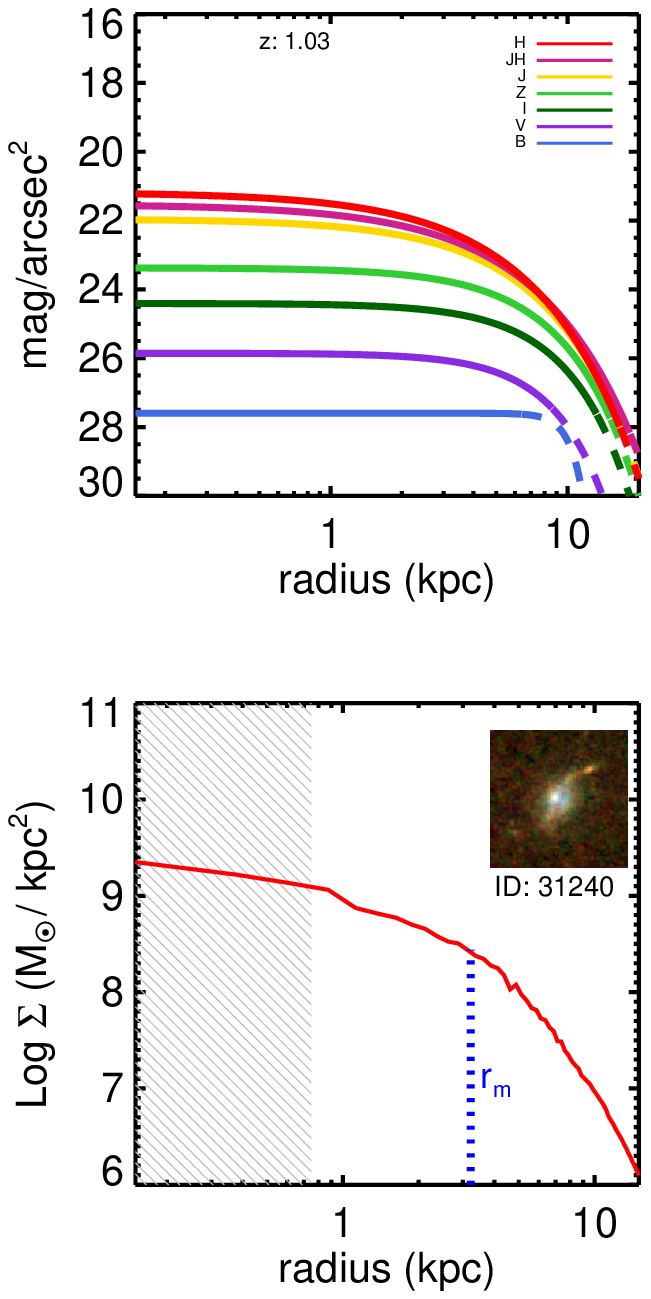}
\hspace{1. mm}
\includegraphics[width=1.95 in]{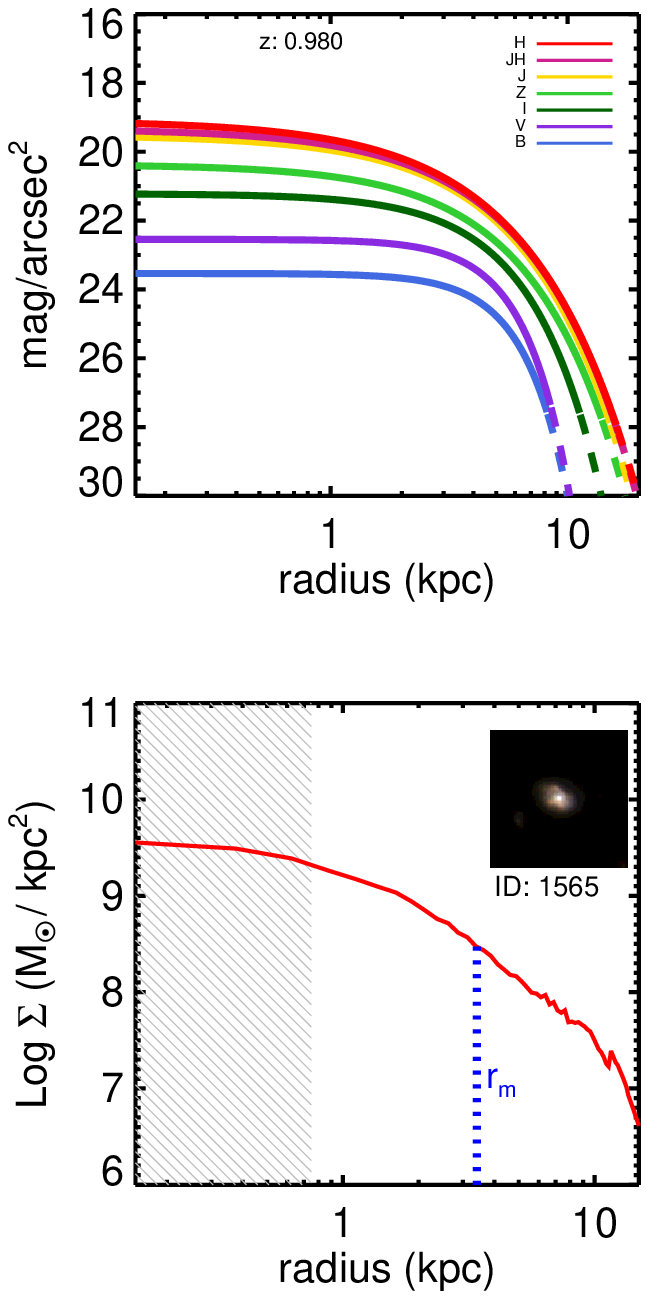}
\hspace{1. mm}
\includegraphics[width=1.95 in]{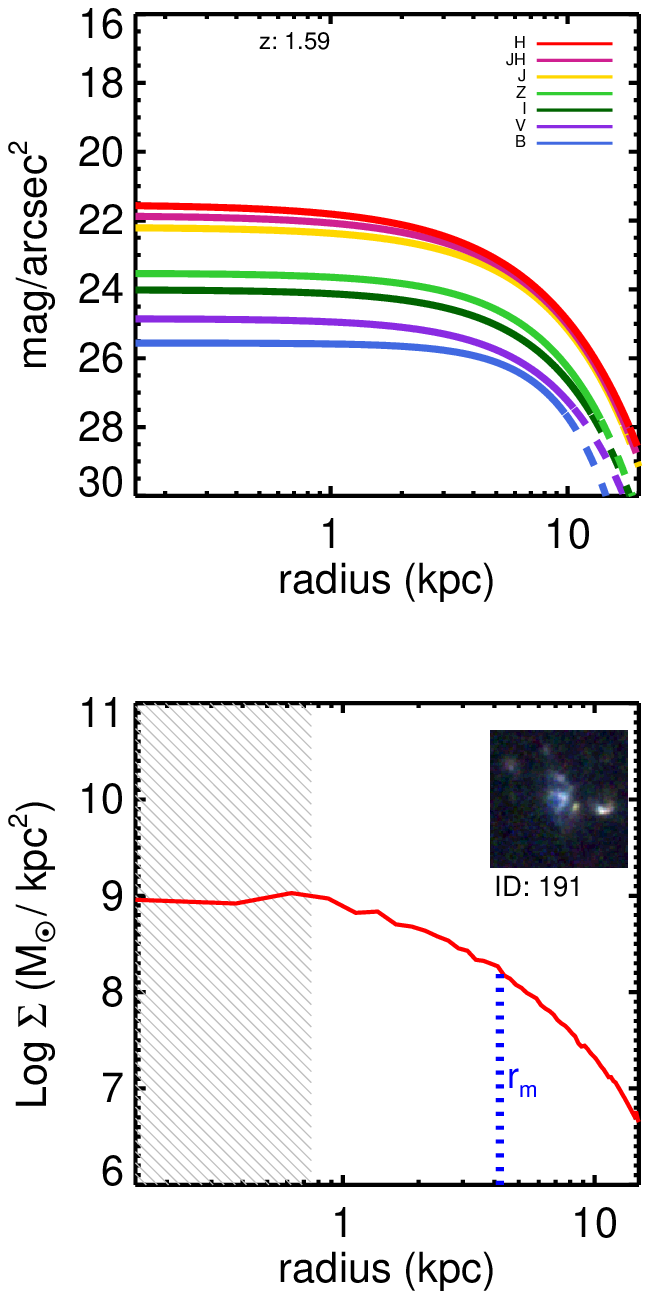}
\hspace{1. mm}
\includegraphics[width=1.95 in]{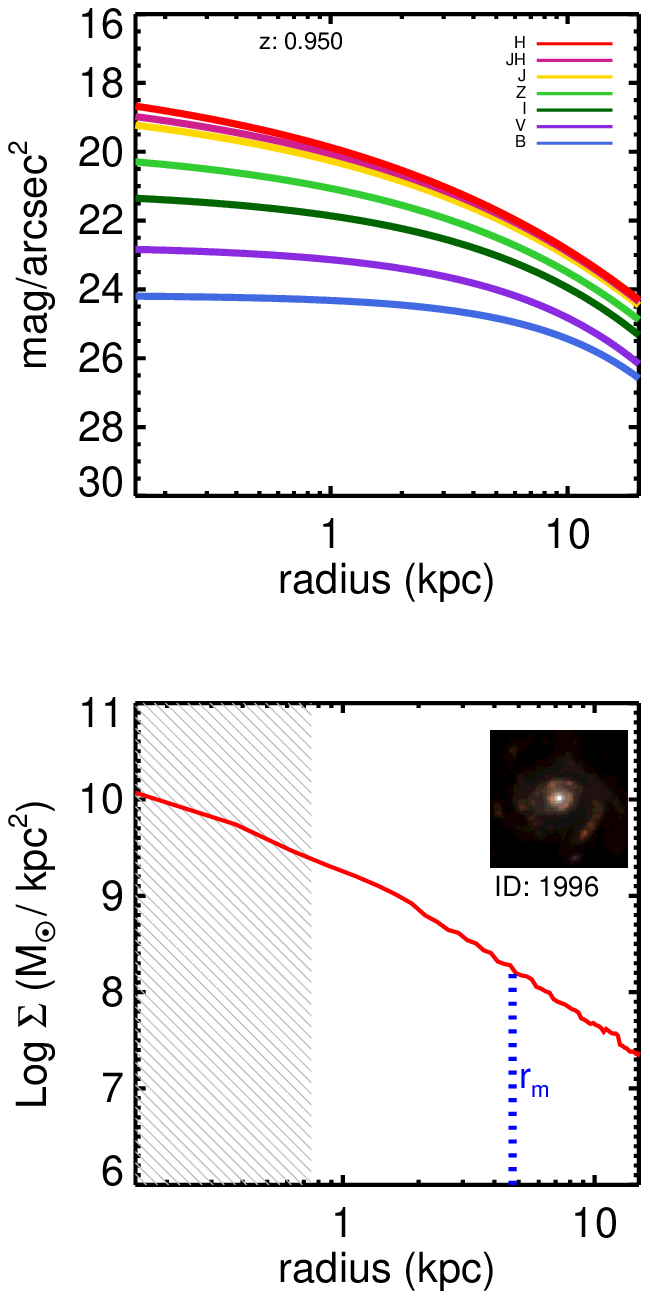}
\hspace{1. mm}
\includegraphics[width=1.95 in]{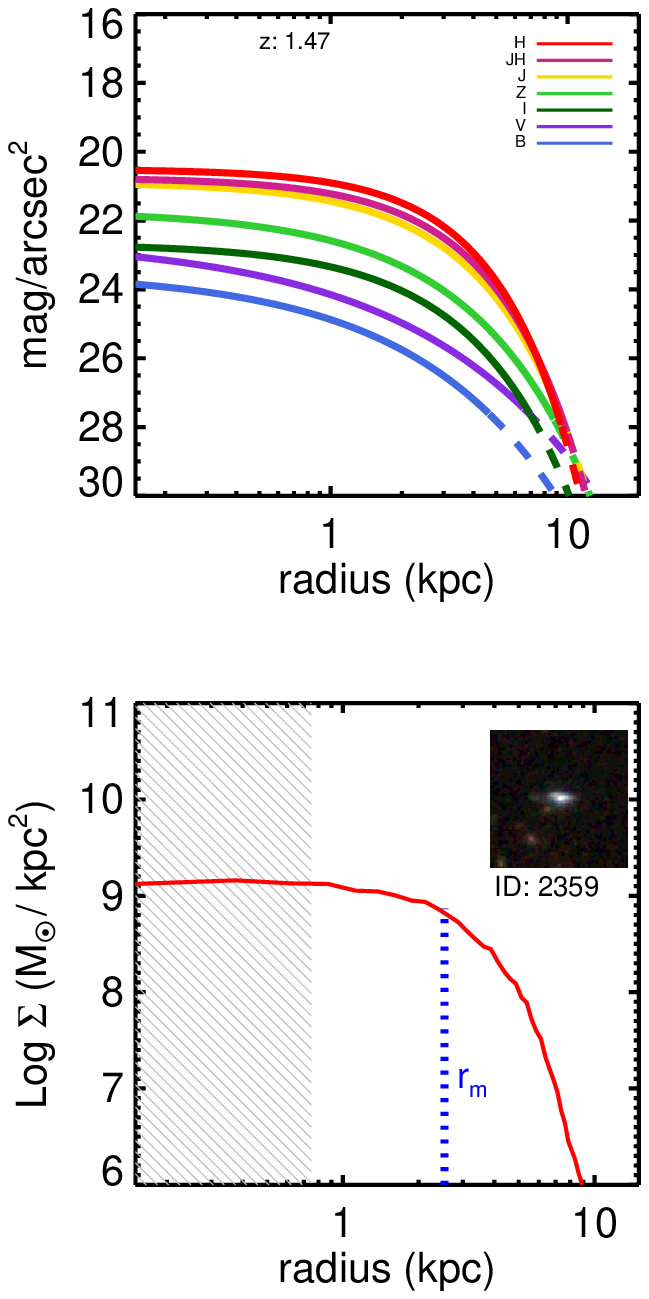}
\hspace{1. mm}
\includegraphics[width=1.95 in]{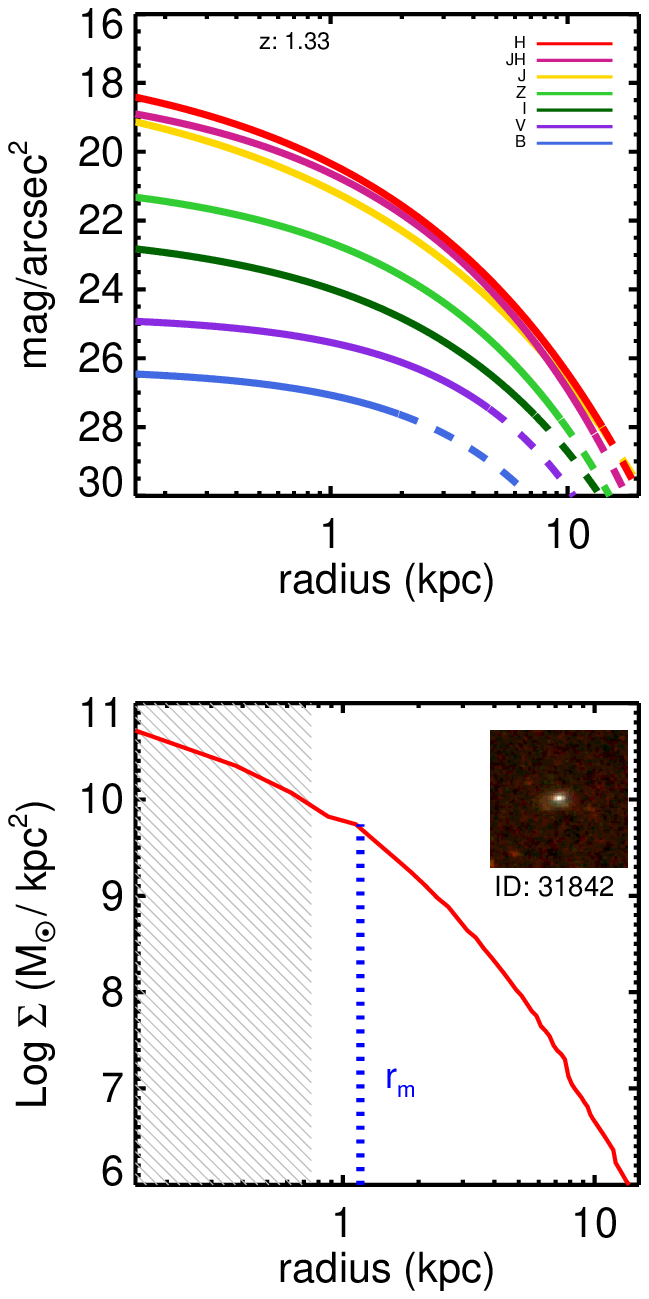}
\caption{The surface brightness profiles of few galaxies in our sample and their stellar mass surface densities are shown. The dotted- blue lines represent the half-mass radii and the gray hatch regions, show the HWHM of PSFs. The dashed-lines represent where the surface brightness profiles reaches the surface brightness limits of their images. }

\label{fig4}
\end{figure*}

\section{Methodology}

\subsection{Deriving Light Profiles}
In this section, we describe the method for deriving the galaxies surface brightness profiles in all available filters. We first derive the de-convolved profiles of galaxies at different filters, are first derived by fitting PSF convolved single \ser \citep{sersic1963, sersic1968} models to the observed two dimensional surface brightness images of galaxies. We use GALFIT v3 \citep{peng2010} to perform two-dimensional fitting while using $H_{160}$-band PSFs (taken from 3D-HST). The procedure is similar to the \cite{mosleh2012, mosleh2013}. We extract $48''\times48''$ cutout images around each galaxy prior to the fit, and masked neighboring objects during the fitting, using mask map provided by 3D-HST. We take initial guesses for position, ellipticity, effective radius $r_e$, position angle, from the original catalog and initially assumed \ser index of 2. We compare sizes of $H_{160}$-band images of our galaxies for derived from our method to the ones measured by \citet{vanderwel2014, vanderwel2012}, in Figure \ref{fig3}, which shows very good consistency between the results and depict the robustness of our size measurements. 

The procedure for deriving the surface brightness profiles at shorter wavelengths (i.e., $B_{435}$,$V_{606}$, $i_{775}$, $z_{850}$, $J_{125}$, $JH_{140}$), is similar to the $H_{160}$-band. However, to avoid systematic errors in deriving color and stellar-mass profiles it is essential that the light profiles in different filters originate from a common center. Hence, following \cite{Labarbera2009} and \cite{gargiulo2011, gargiulo2012}, the position, ellipticity and position angle of the best-fit \ser model for $H_{160}$-band images were adopted as priors while fitting two-dimensional single \ser models to galaxy images at shorter wavelengths. In case, fitting did not converged at these shorter wavelengths, (or for cases with large uncertainties of the results, i.e., converging to the boundaries), we fixed the \ser index and $r_e$ to the best-fit of its closest longer filter. We also note that the light profiles are all circularized to remove the effects of ellipticity.

Any deviation from a single \ser model (due to the variation of light distributions between different bands or sub-structural components) is a matter of concern for deriving the true color and $\mstar/L$ profiles. The residual-corrected method \citep{szomoru2010} could have help to mitigate the problem. However, this method is sensitive to the noise/background and works best for massive objects. In this study, we use a sample of galaxies in a wide range of stellar masses and, for the consistency, in our analysis we only use single \ser model profiles. In the Appendix, we show that using residual-corrected method should not change the overall results of this study. 

The total number of objects used in this study at high-$z$ is 2391. About $\sim 7\%$ of sources from original catalog which have less $\leq 3$ bands detection  (HST filters; in order to have sufficient wavelength coverage for deriving their mass profiles) and those with large uncertainties in their derived light profile parameters (including objects close to very bright sources) are excluded from our analysis. \\

The procedure for deriving light profiles of the local galaxies ($z\sim0$) is similar. We use the  $u, g, r, i, z$ images of SDSS DR7 galaxies. We use the $r$-band images as a reference (due to having high S/N) and derived the light profiles. We have used synthesized PSF images generated by the SDSS photo pipelines\footnote[3]{\url{http://www.sdss.org/dr7/products/images/read\_psf.html}} at the position of each object and for different filters, to derive the bets-fit \ser profiles using GALFIT \citep[see][for more details]{mosleh2013}. In the following analysis, we use the circularized 1D profiles.

\subsection{From Light Profile to Stellar-Mass Profile}

The best-model light profiles at different filters derived for each galaxy, can be used to measure stellar mass, color and mass-to-light ratio profiles. The stellar mass profile of galaxies were measured by finding the best spectral energy distribution (SED) model at each radius. We used the Fitting and Assessment of Synthesis Template (FAST) code \citep{kriek2009} and follow the \citet{skelton2014}, for finding the best SED models, i.e., using the \citet{BC2003} stellar population evolution models with exponential declining star formation history, assuming \cite{chabrier2003} initial mass function (IMF) and solar metallicity, \citet{calzetti2000} dust attenuation law and allowing the $A_{V}$ to vary between 0 and 4. We fixed the redshift of galaxies to the ones provided by the 3D-HST catalog.       

In details, we divided the light profiles into small bins of $0.25$ kpc out to 20 kpc and SED fitting each bin, separately. We assigned the flux errors at each bin, which determined using empty aperture technique around each object, in a similar way described by \citet{skelton2014}, appropriate to the bin sizes. However, at larger radii, the model fluxes are very low, hence it is crucial to increase the (model) signal-to-noise ($S/N_{model}$) for robust SED fitting at each profile location. We used an adaptive binning method \citep[similar to ][]{Cappellari2003} and created bins with sufficient $S/N_{model}$ ($\geqslant 10$) in the $J-H$ color.

However, at larger radii, where model fluxes are very low, we used adaptive binning method to create meshes of 10 model signal-to-noise ($S/N_{model}$) in the $J-H$ color, in order to increase the $S/N_{model}$ for robust SED fitting. The stellar masses are then measured by SED-fitting at each radius, and hence creating the stellar mass profiles of galaxies. Using $J-H$ color which straddles between, age, metallicity and dust sensitive 4000\AA \ and Balmer break feature and provides better constrain on SED modelling. We note our tests shows that the results are robust against using other colors such as $z-H$ color instead of $J-H$ or even using just $H$-band filter. 

Examples of the surface brightness profiles and stellar mass surface densities of six galaxies are shown in Figure \ref{fig4}. For each galaxy, the solid color lines in the upper panels, are best fit surface brightness profiles at different filters and the dashed lines are where the surface brightness profiles of galaxies reaches to the average surface brightness limit of their images. The bottom panel depicts their stellar mass profiles. 

For deriving the mass-profiles for $z\sim0$ galaxies, we use iSEDfit code \citep{moustakas2013} and applying similar \citet{chabrier2003} IMF to high-$z$ sample. We follow \citet{kaufmann2003a} to constrain star-formation history of local galaxies, i.e.,  allowing random star burst in addition to the exponential declining star formation history. Similar to high-$z$ sample, we use $g-r$ color for adaptive binning light profiles prior to the SED fitting and deriving their mass profiles. We note that using similar star-formation history and the same code for high redshift object will not alter the general results of this paper, as described in Appendixes. \\ 

\begin{figure}
\includegraphics[width=0.45\textwidth]{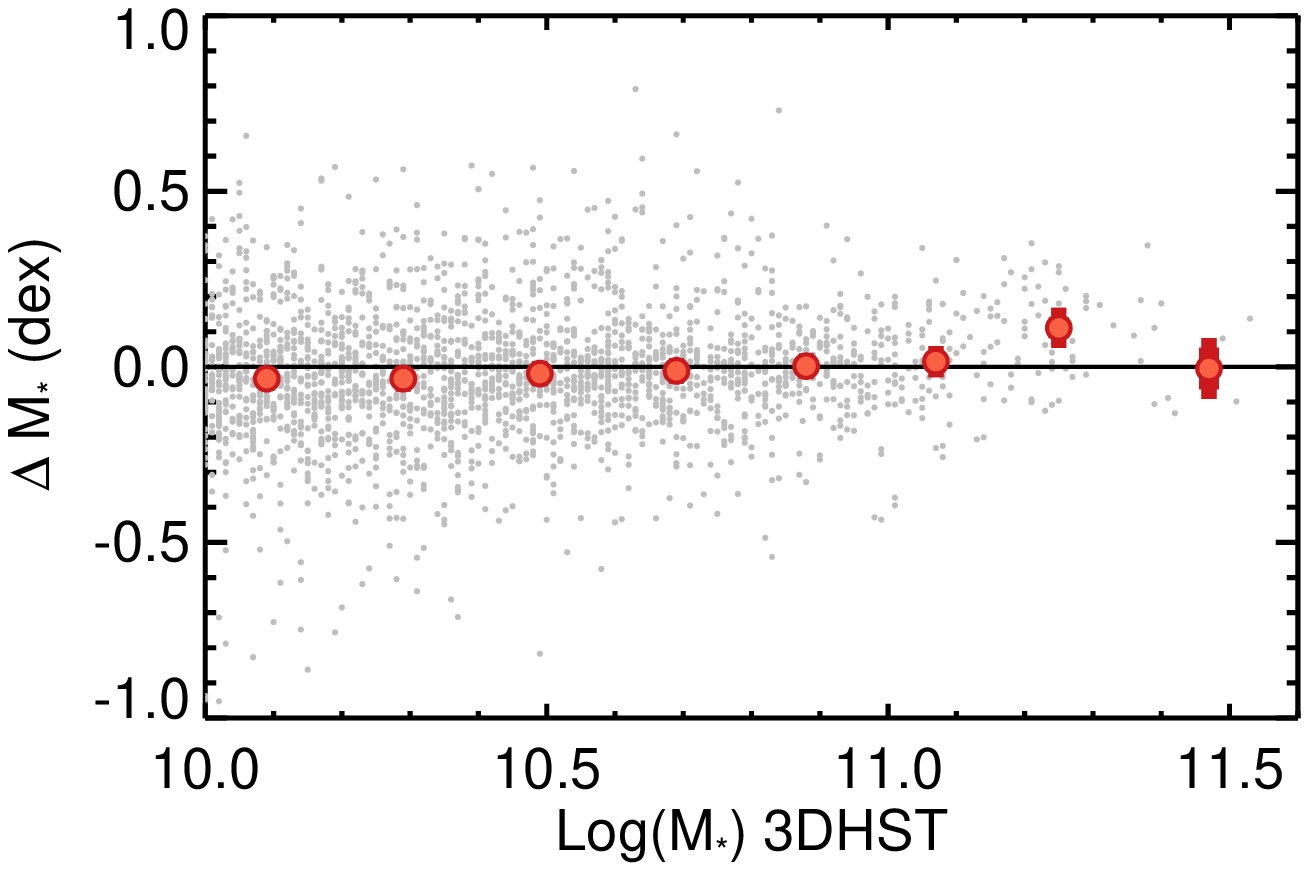}
\hspace{5. mm}
\includegraphics[width=0.45\textwidth]{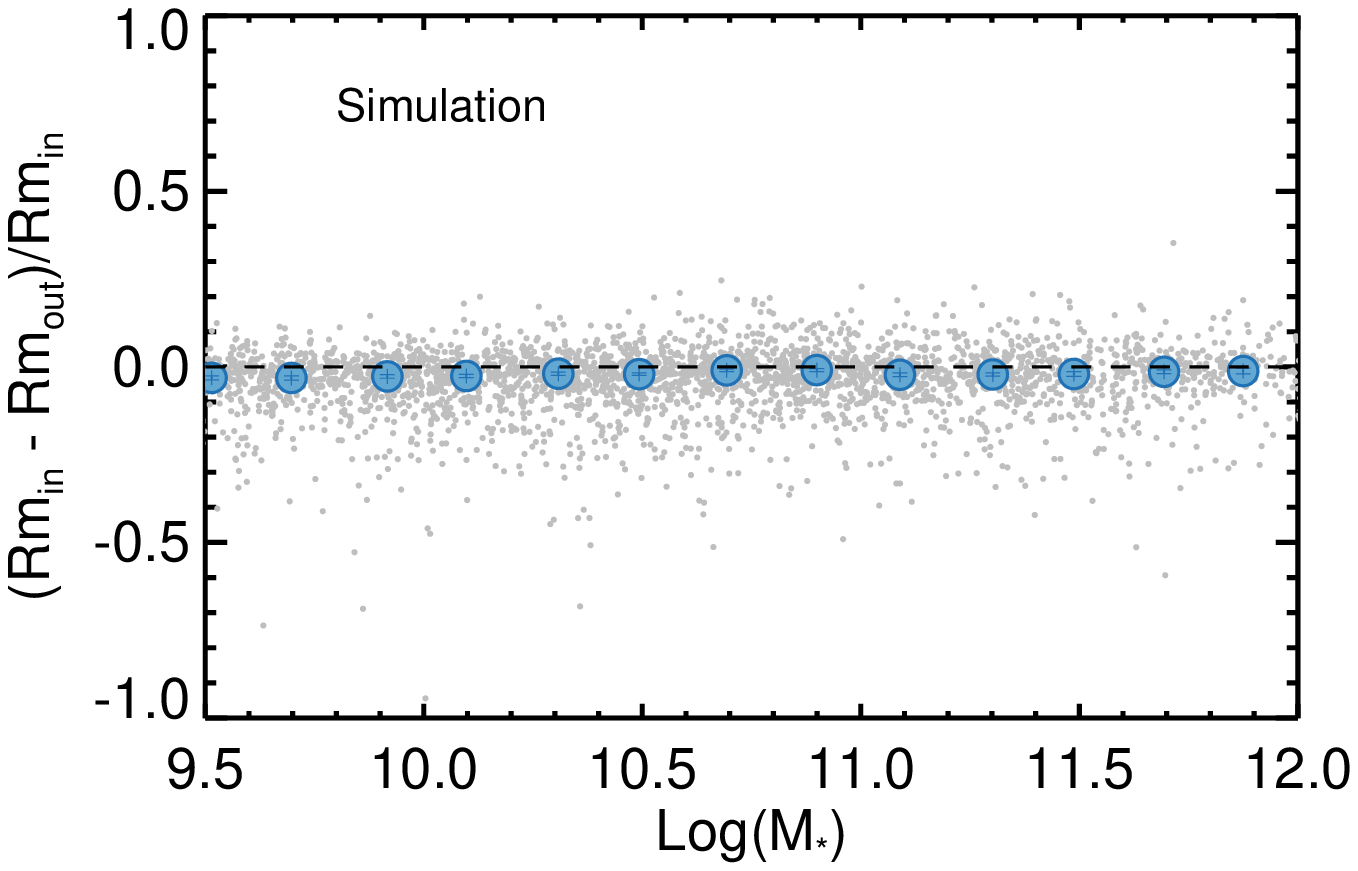}

\caption{\textit{Top Panel}: The total stellar masses derived from resolved profiles compared to the total stellar masses measured provided by the 3D-HST catalog. \textit{Bottom Panel}: Simulations to show the robustness of half-mass size measurements (see text for details). In general, there is no systematic in recovering total masses and the half-mass radii of our galaxies using the method described in this paper.} 
\label{fig5}
\end{figure}
  
\begin{figure*}
\centering
\includegraphics[width=0.48\textwidth]{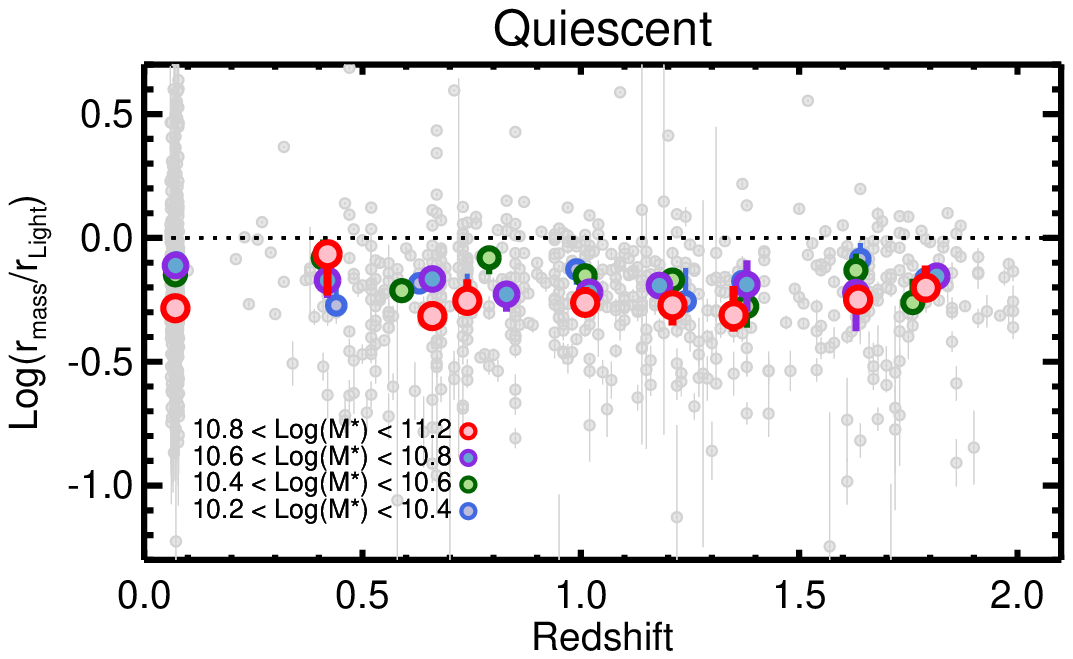}
\hspace{5. mm}
\includegraphics[width=0.48\textwidth]{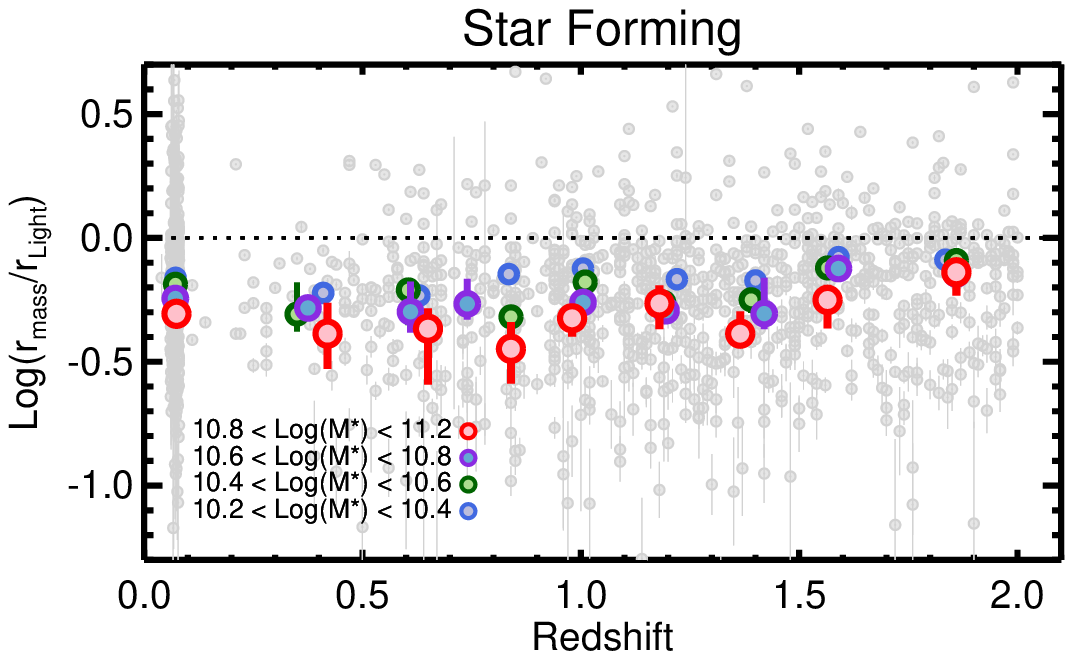}
\caption{The ratio of half-mass sizes to half-light sizes versus redshift for all quiescent (left-panel) and star-forming galaxies (right-panel) in our sample. At all our studied redshifts, the half-mass sizes are smaller than half-light sizes. This is also the case for star-forming and quiescent galaxies. The color points show the median of the size ratios at different redshift bins color-coded according to their stellar mass bins with their related standard errors. The ratio slightly depends on the stellar masses, i.e.,  for massive galaxies the half-mass to half-light size ration is smaller.}

\label{fig6}
\end{figure*}

\subsection{Half-Mass Radius}

The next step is to measure the half-mass radii from the mass profiles for each object. For that, we first interpolated between the radial mass bins and then extrapolated the mass density profiles out to $100$ kpc by fitting a single \ser model in the outer parts of mass profiles, and then integrating the mass profiles out to this radius to find the half-mass radii. Defining the radius at which the extrapolation should be started beyond that point, is not an straightforward task. In order to secure that the results were not strongly biased toward our guess of this connection point and the extrapolation process in general, we have measured the half mass radii for each galaxy by fitting about 1000 \ser profiles starting at different radii and finding the most frequent output results. In order to check the reliability of this method, we applied it on a set of 3000 two-component \ser models of density profiles, while random noise was added to the profiles to resemble the measured mass profiles of this study. The result of our simulation shows that our method is robust and half-mass radii can be recovered without any systematic over our studied stellar mass range (bottom panel of Figure \ref{fig5}). The blue points in Figure \ref{fig5} are the median of the relative error of the half-mass sizes with the error on the median and average scatter of $10\%$. Note that our test using simulated profiles shows that fitting a single \ser model to entire profiles cause large scatter on derived half-mass sizes, especially for objects with $R_{m} < 1$ kpc.

The total stellar masses derived by this method are compared with the total masses from the 3D-HST catalog in the top panel of Figure \ref{fig5}. There is no systematic differences between total stellar masses derived for sources using their mass density profiles and the ones from the original catalog. Note that we only used seven HST filters (at most) for each galaxy for SED fitting comparing to so many available complementary filters used in the 3D-HST catalog, including rest-frame near-infrared data from other instruments/telescopes. The differences in stellar mass estimates possibly arises from outshining effects or dusty regions \citep{Zibetti2009, maraston2010, wuyts2012, reddy2012, Sorba2015}.

Note that for the rest of the paper, we have corrected the stellar mass profile (and consequently their related parameters) to the total masses derived from integrated star-formation history using \cite{BC2003} which includes lived stellar masses, remnant and mass returned to interstellar medium (ISM), simply, the stellar mass is the mass of gas turned into stars. The advantage of this stellar mass definition is that quiescent galaxies do not suffer from mass loss.

\subsection{Half-Mass versus Half-Light Radius}
We have compared the half-mass radii and half-light radii of our star-forming and quiescent galaxies at different redshifts and in different stellar mass bins in Figure 6. The effect of morphological $K$-correction has been minimized by using the half-light sizes of object in the closest WFC3/IR or ACS band to the rest-frame $\sim$5000\AA \, i.e., using $H_{160}$ for sources at $1.5<z<2.0$, $J_{125}$ for objects at $1.0<z<1.5$ and $z_{850}$ for galaxies at $0.5<z<1.0$. As shown in Figure \ref{fig6}, at all studied redshifts half-mass sizes are always smaller than half-light radii for both star-forming and quiescent ones (right and left panels, respectively). The color symbols represent median ratio values for different stellar mass bins. The quiescent galaxies are on average have half-mass radii about $30-45\%$ smaller than their half-light radii over the entire studied redshifts, slightly depends on the stellar masses. Massive quiescent ones ($10^{10.8}-10^{11.2}\msun$) are $\sim45\%$ smaller size ratio compare to the  lower mass bins with $\sim30\%$ size ratios. Using galaxies with $>10^{10.7}\msun$ \citet{szomoru2013} also found smaller half-mass sizes compare to the half-light sizes but with average of $\sim25\%$ size ratios. 

The ratio of sizes for star-forming galaxies, on the other hand, (right panel of Figure \ref{fig6}) shows little dependence on redshift. Above $z>1.5$ the ratio is smaller that at lower redshifts, presumably light and stellar masses following more similar distributions, however, at lower redshifts,  half-mass sizes are about $30-50\%$ smaller than their half-light sizes, again depending on stellar masses, with the most massive ones shows larger differences, indicating the stellar mass profiles are more concentrated  compare to their light profiles. \\

\section{Evolution of Mass Distributions}

As discussed earlier the stellar mass distributions of galaxies at different epoch could help to study their stellar mass assembly. Therefore, in this Section, we first compare the stellar mass distributions of different types (quiescent and star forming) of galaxies at different redshifts, by means of their mass profiles and then we study how the mass distribution varies for each type by the changes of half-mass radii and surface densities at effective radii and central compactness of our samples. We have split the sample of galaxies into four stellar mass bins for each type between $10^{10.2} - 10^{11.2} \msun$ and compare them at different redshifts. \\

\begin{figure*}
\centering
\includegraphics[width=\textwidth]{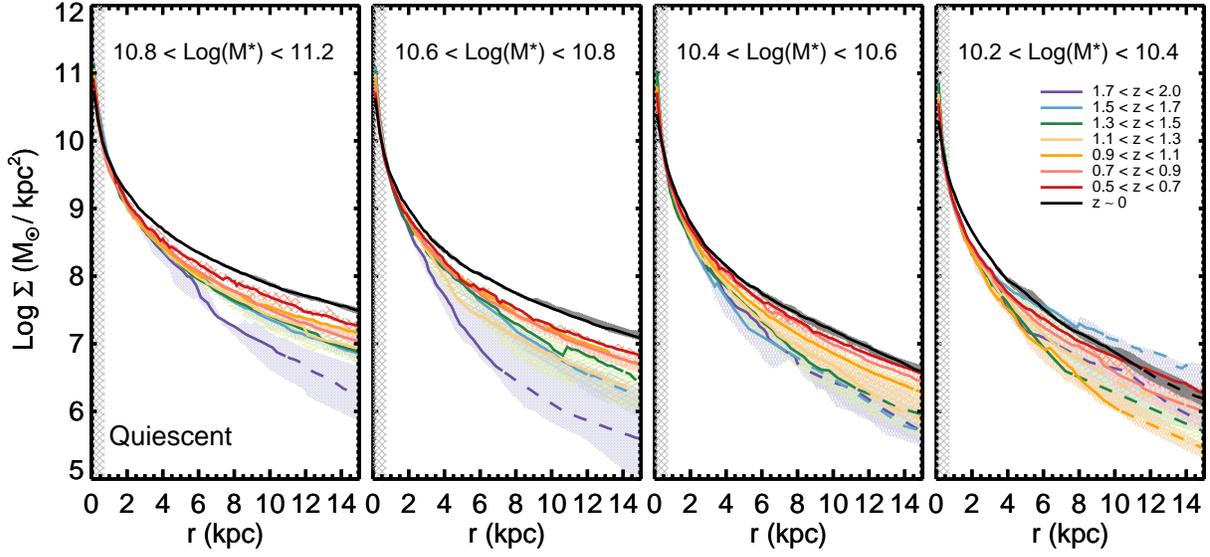}
\caption{The comparison of the stellar mass density profiles of quiescent galaxies from $z\sim2$ to $z\sim0$. Quiescent galaxies are split into four stellar mass bins, decreasing from left to right panel. The shaded regions depict the errors affecting the medians. At all mass bins, the most significant changes of their mass profiles can be seen in the outer regions of these galaxies, i.e., quiescent galaxies at low-$z$ have extended stellar mass profiles comparing to their counterparts at high-$z$. We caution that the time evolution of individual galaxies cannot easily be seen from this figure, as it compares galaxies at fixed mass bins and individual galaxies could have different evolution history.}

\label{fig7}
\end{figure*}

\begin{figure*}
\centering
\includegraphics[width=\textwidth]{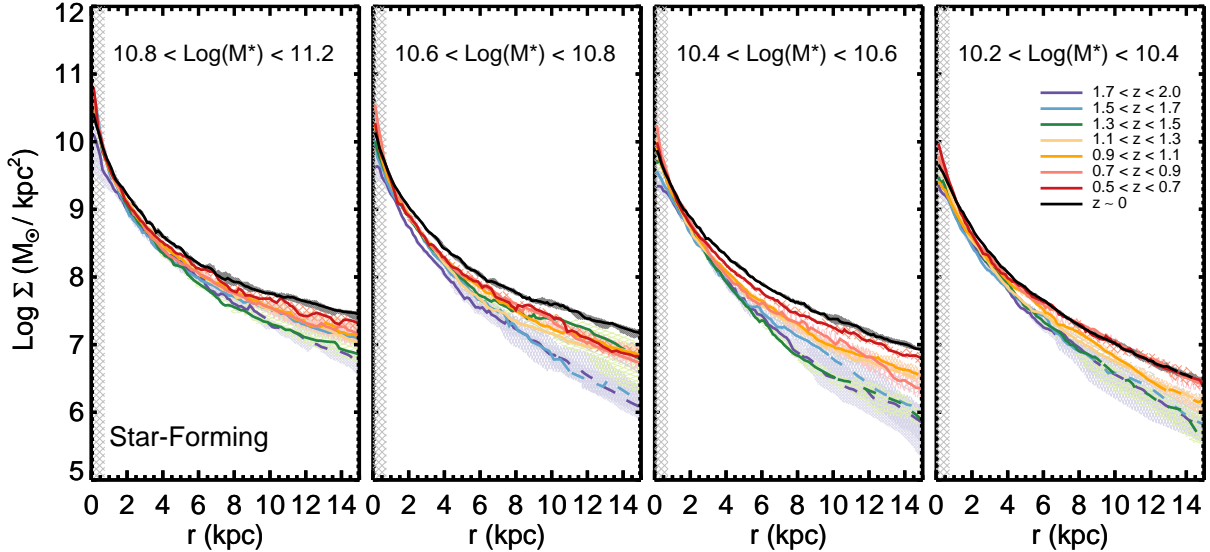}
\caption{The radial density profiles of star forming galaxies at different epochs (color-lines) and at different mass bins (different panels).  The comparison of density profiles at different epochs reveals that star forming galaxies at later times have assembled stellar masses at both inner and outer regions comparing to their high-$z$ counterparts, though the central density varies little with redshift while the density in the outer regions grows substantially. These differences is more prominent for objects within $\log(\mstar/\msun)$ $=10.4-10.8$. The massive star forming galaxies ($10.8-11.2 \msun$) have little differences (mostly in their outer regions), up to $z\sim1.5-2$ indicating that stellar mass profiles of these objects grow self-similarly at all epochs. }

\label{fig8}
\end{figure*}

\subsection{Stellar Mass Density Profiles}

The comparison of mass densities of galaxies at different epochs reveals how the distributions of stellar mass varies at fixed mass. The median radial distribution of stellar mass densities of quiescent galaxies at different redshifts are shown in Figure \ref{fig7}, for different mass bins (decreasing from left to right panels). 

Starting from the highest stellar mass bin, i.e., quiescent galaxies within $10.8 < \log(\mstar/\msun)< 11.2 $, in the left panel, the significant changes with redshift can be seen in the outer parts of these galaxies. The density profiles of these galaxies are higher in the outer regions of these galaxies at lower reshifts compared to their high-$z$ counterparts, meaning that these galaxies have more stellar masses at their larger radii. 
 
Clearly, it can also be seen from middle panels of Figure \ref{fig7} that intermediate massive quiescent galaxies with stellar masses $10^{10.6}< \mstar <  10^{10.8} \msun$  and $10^{10.4}< \mstar <  10^{10.6} \msun$ have similar differences in their density profiles in the outer regions at different epochs. In general, at later time, quiescent galaxies at massive and intermediate mass ranges have more assembled stellar masses at larger radii, compared to their counterparts at high-$z$. Accretion of mass via (minor) mergers or arrival of the newly quenched systems into red sequence, both are plausible scenarios for the differences \citep{naab2009, carollo2013}. This can be examined by studying the evolution of central and effective surface densities that we discuss in the following sections. It is also needed to study the evolution of the mass profiles of the star-forming galaxies. 

Note that for the lowest mass bin of quiescent objects within $10^{10.2} < \mstar <  10^{10.4} \msun$ (right panel of Figure \ref{fig7}), there is also a hint of having similar behavior, though their specific changes are small compared to the massive ones, and hence the way they changes looks slightly noisy. We further test this by comparing their half-mass radii in the next section.    

For the star-forming galaxies, the changes of density profiles have somewhat different behavior. Figure \ref{fig8} presents the comparison of median radial density profiles at different epochs and at fixes mass bins. The most massive star-forming galaxies with stellar masses  $10.8 < \log(\mstar/\msun)< 11.2 $, have similar radial distributions of stellar masses at different redshift bins (left panel of Figure \ref{fig8}) with subtle changes at larger radii at later times. This shows that at any epoch, the star-forming galaxies have very similar profiles, i.e., they grow self-similar at all epochs as predicted in simulations \citep{tacchella2016a}. 

However, star-forming galaxies with intermediate stellar masses $10^{10.2} < \mstar < 10^{10.8}\msun$, have different mass densities at inner and outer regions comparing to their counterparts at higher redshifts. These galaxies have higher stellar mass densities within $1$ kpc at later times, indicating the existence of more prominent central densities (can be expressed as bulges) at lower redshifts. Meanwhile the stellar mass densities in the outer regions also changes. Therefore, comparing low-$z$ and high-$z$ star-forming galaxies at intermediate masses, the mass densities varies at all radii, meaning that these objects have assembled stellar masses in both central and outer regions with time, with a different growth rate at the center and the outskirts. Hence it argues against the scenario of the formation of bulges prior to the disk at $z<2$.  

However, as we discuss below, central densities varies less at later time comparing to the mass profiles at larger radii, therefore, suggesting slightly different time scales for their changes. Perhaps, different mechanisms with different time scales could play separate roles for the evolution of mass profiles of star-forming galaxies, e.g., accretion of gas into disk (or maybe bulge) and migration of stars into the central regions via mechanisms such as disk instabilities. This needs to be addressed in more details. Nevertheless, studying the mass profiles delineates the existence of different mechanisms of stellar mass assembly for star-forming galaxies compared to quiescent ones which is reflected into their mass distributions.\\

\begin{figure*}
\centering
\includegraphics[width=0.48\textwidth]{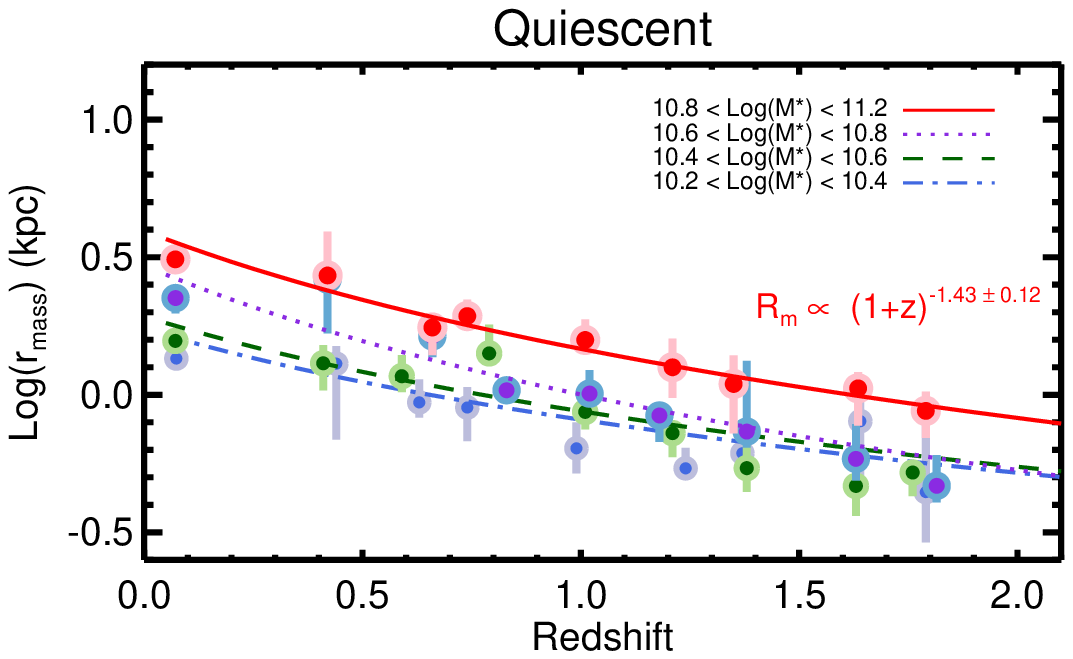}
\hspace{5. mm}
\includegraphics[width=0.48\textwidth]{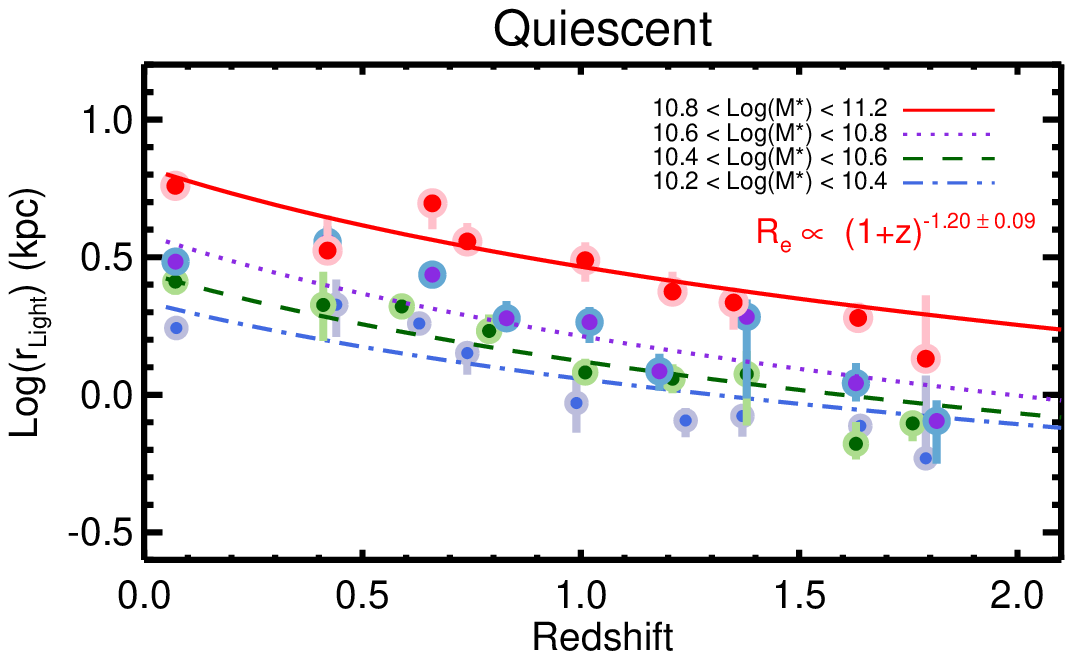}
\caption{Left Panel: The evolution of half-mass radii from $z\sim2$ to $z\sim0$ for quiescent galaxies, for different mass bins. Right Panel: The half-light size evolution for the same galaxies in the left panels. The symbols are the median sizes at each redshift bin and the error bars show the standard errors. The lines and are the best fit to the original data points assuming $(1+z)^{\alpha}$. The mass-weighted sizes of quiescent galaxies increases by a factor of $\sim4$ since $z\sim2$. The rate of half-mass size evolution is slightly faster than half-light sizes. Similar to Figure \ref{fig7} \& \ref{fig8}, this is not reflecting the size evolution of individual galaxies.}

\label{fig9}
\end{figure*}

\begin{figure*}
\centering
\includegraphics[width=0.48\textwidth]{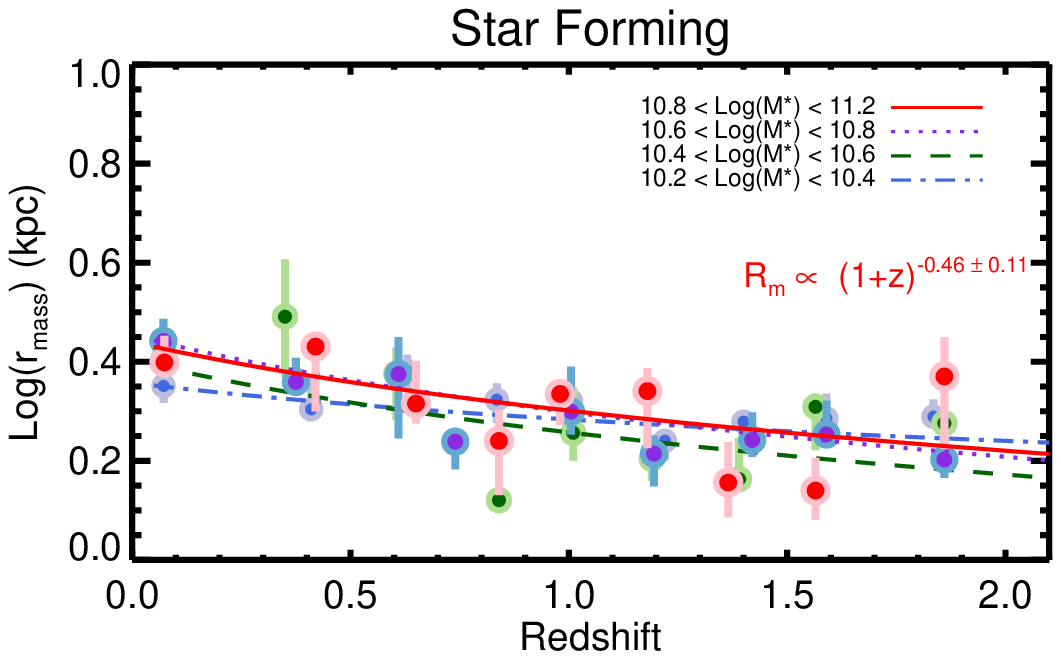}
\hspace{5. mm}
\includegraphics[width=0.48\textwidth]{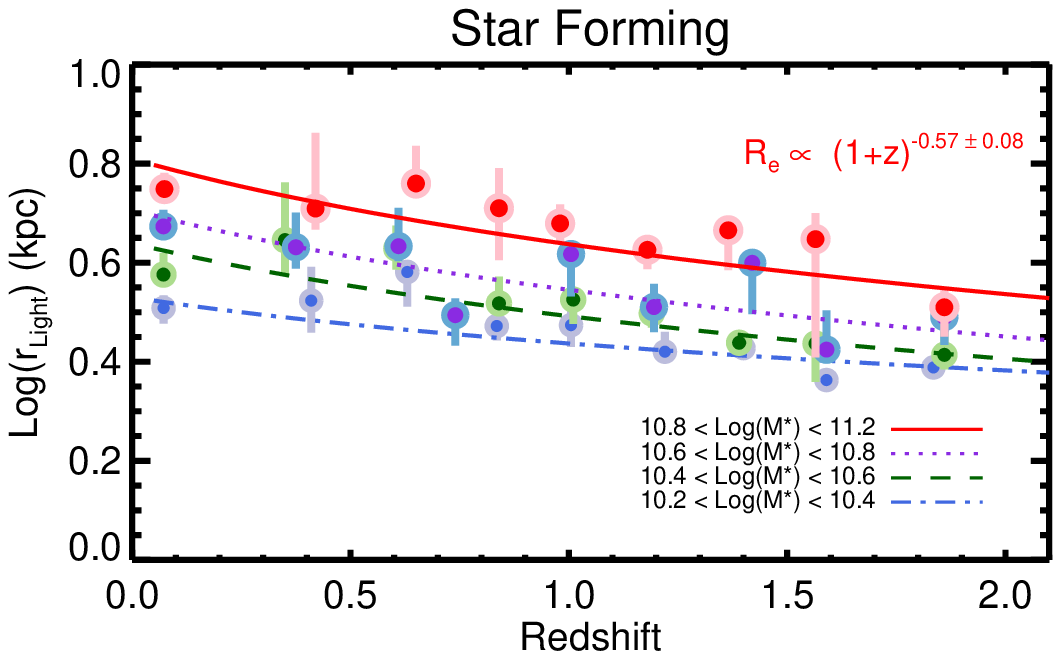}
\caption{Similar to Figure \ref{fig9} but for star-forming galaxies. The mass-weighted sizes of star-forming galaxies increase slowly with cosmic time and depend only weakly on stellar mass. This indicates that the central regions and the outskirts have to grow concurrently: the stellar mass profiles have to grow in the central region as well as in the outskirts.}

\label{fig10}
\end{figure*}

\subsection{Half-Mass Size Evolution} 

The evolution of the half-mass radii ($r_{m}$) of galaxies is an effective way of illustrating their structural evolution. In left panel of Figure \ref{fig9}, the evolution of median half-mass radii of quiescent galaxies between $0 < z < 2$ is illustrated for different mass bins (different colored symbols). As expected, the half-mass radii of the quiescent galaxies increases with time in a similar manner to their half-light radii (right panel of Figure \ref{fig9}). This is the case for galaxies at different mass bins. 

We have parameterized the rate of changes as a function of $(1+z)^{\alpha}$.  The change of mass-weighted sizes for massive ones is as $(1+z)^{-1.43\pm0.12}$. For highest mass bin (red circles), the half-mass sizes increase by a factor of $\sim 4$ from $z\sim 2 $ to $z \sim 0$, consistent with results from the simulation by \citet{naab2009}. The half-mass size evolution is marginally faster than the evolution of their light-weighted sizes with $\alpha = -1.20\pm0.09$.

The pace of evolution is also similar for quiescent galaxies at lower mass bins,  $\alpha = -1.56\pm0.13$, $-1.15\pm0.13$, $-1.10\pm0.17$, for mass bins of $10.6< \log(\mstar/\msun) <10.8$, $10.4<\log(\mstar/\msun)<10.6$ and  $10.2 < \log(\mstar/\msun) <10.4$, respectively. This shows that at fixed masses, quiescent galaxies at low redshifts have extended mass density profiles compared to their counterparts at higher redshifts. 

The rate of half-mass size evolution for star-forming galaxies is not as fast as quiescent ones. The left-panel of Figure \ref{fig10}, shows that the mass-weighted sizes of these galaxies evolve slowly for our studied mass ranges since $z\sim2$. The rate of evolution for massive star-forming ones is $\alpha = -0.46\pm0.11$ slightly slower than their half-light radii evolution with $\alpha = -0.57\pm0.08$.  The rate of half-mass size evolution is very similar for intermediate mass ranges of $10.6 < \log(\mstar/\msun) < 10.8$ and $10.4 < \log(\mstar/\msun) <10.6$, with  $\alpha = -0.51\pm0.11$  and $-0.48\pm0.08$, respectively. However, this is much slower for the lowest mass bin of $10.2< \log(\mstar/\msun) <10.4$ with $\alpha = -0.25\pm0.08$, consistent with very subtle to no evolution. The slow evolution of the half-mass sizes, in combination with the results presented in Section 4.1 and Figure \ref{fig8}, leads us to propose two evolutionary phase of $r_{m}$ in star-forming galaxies, (1) self-similar changes of mass-profiles at high-$z$  (at least $z\gtrsim 1$) \citep[e.g.,][]{tacchella2016a} and (2) increasing the disk sizes once galaxies central regions are saturated at later times ($z\lesssim 1$) \citep[see e.g.,][]{Nelson2012}. In detail, these trends may actually be mass dependent.  It worth noting that a flattening of the M/L gradient is expected, in particular for massive star-forming galaxies, if the effects at short rest-frame wavelengths of dust and clumpy structures are not fully taken into account. Hence, the single-\ser models and the methodology described in Section 3.1 could have introduced a bias affecting the rate of mass-weighted size evolution of the star-forming galaxies. However, we argue that our results do not change significantly when accounting for residuals from poor best fits (see Appendix for more details).

We note that at fixed redshift, the half-light radii of massive star-forming galaxies are larger than their less massive ones. However, the striking similarities between half-mass radii of these galaxies at different masses at fixed reshift bins, could either be a representative of the similarities between the shapes of their mass profiles (having exponential shapes with just different normalizations, perhaps at high-$z$) or the faster increase of the central mass densities of massive ones compared to less massive galaxies, therefore reducing the half-mass sizes for massive ones.

Overall, the changes in stellar mass-weighted radii of the star-forming galaxies is very gradual, compared to the quiescent ones at fixed mass. This has been pointed by \citet{lilly1998, simard1999,ravindranath2004, barden2005, vanderwel2014}, that mass-size relation of late-type galaxies has very little evolution since redshift of $\sim 1$ \citep[see also][]{dutton2011}. Concurrent formation of the bulge and growth of the disk could contribute to this gradual change of half-mass sizes. Based on \cite{tacchella2015}, the specific star formation rate (sSFR) of galaxies within $10^{10}$ and $10^{11}\msun$ are on average flat, which would cause very little mass-weighted size evolution. We also note that the observed rate of (light-weighted) size growth of star-forming galaxies is slightly faster: $\alpha=-0.57\pm0.08$ (but consistent within the errors in this study, see Figure \ref{fig10}). However, the exponent varies between different studies; \cite{vanderwel2014} found observed growth of $(1+z)^{-0.75}$ for the light-weighted sizes of star-forming galaxies \citep[see also][]{straatman2015}. \cite{mosleh2012} show $\alpha=-1.20$,  \cite{newman2012} observed $\alpha=-1$ and \cite{barden2005} observed $\alpha=-0.2$. This discrepancy could first be originated from different sample selection at different redshift ranges, e.g., \cite{barden2005} studied disk-like sample (morphological selection) at $z<1$ while \cite{mosleh2012} used UV-bright galaxies (color-based selection) at $z\sim 1-7$.  In addition, different mass-normalization of the $r_{e}$ (i.e., different value of $\beta$ in $r{_e} \propto M^{\beta}$) between different studies, might introduce some uncertainties. In this work, we selected star-forming galaxies based on the color-color selection method and used $z\sim0$ sample as a reference point for measuring the rate of (half-mass and half-light) size evolution. However, a larger sample of galaxies, in particular between $z\sim0$ and $z\sim0.5$ with similar resolution are required for a better estimation of the half-mass size growth rate of star-forming galaxies.  We should also mention that the accuracy of the M/L gradient is particularly important for measuring the rate of half-mass size evolution. To first order, correcting the light profiles using residual-corrected method shows that this evolution is real (Figure \ref{figA3} in the Appendix, illustrates this for each stellar mass bin).\\ 
 
\begin{figure*}
\centering
\includegraphics[width=0.48\textwidth]{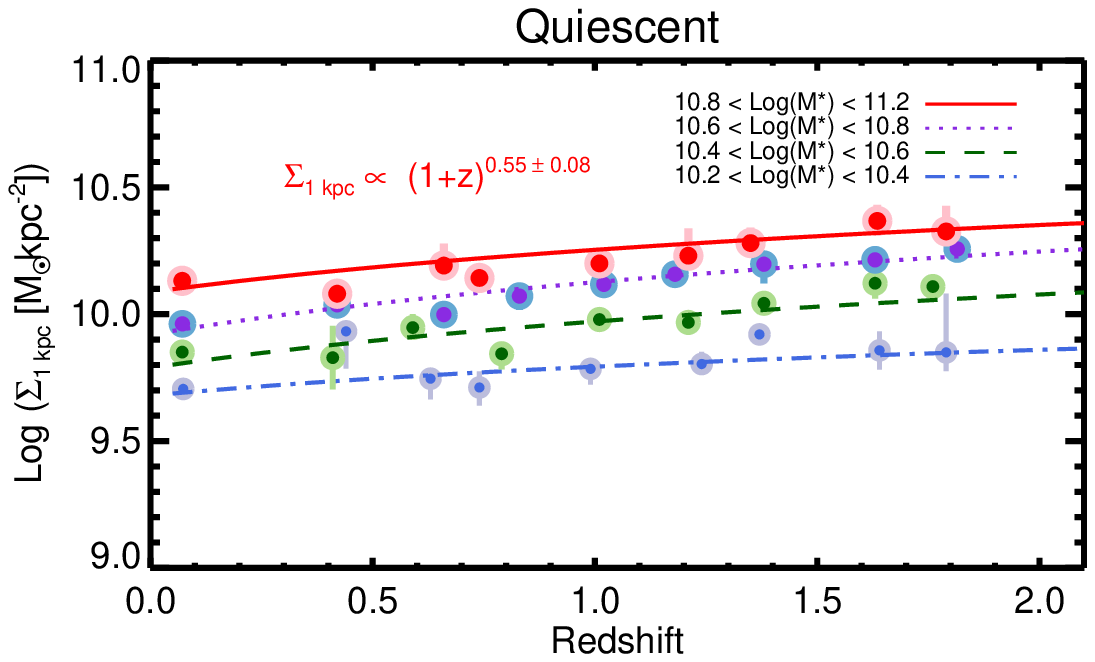}
\hspace{5. mm}
\includegraphics[width=0.48\textwidth]{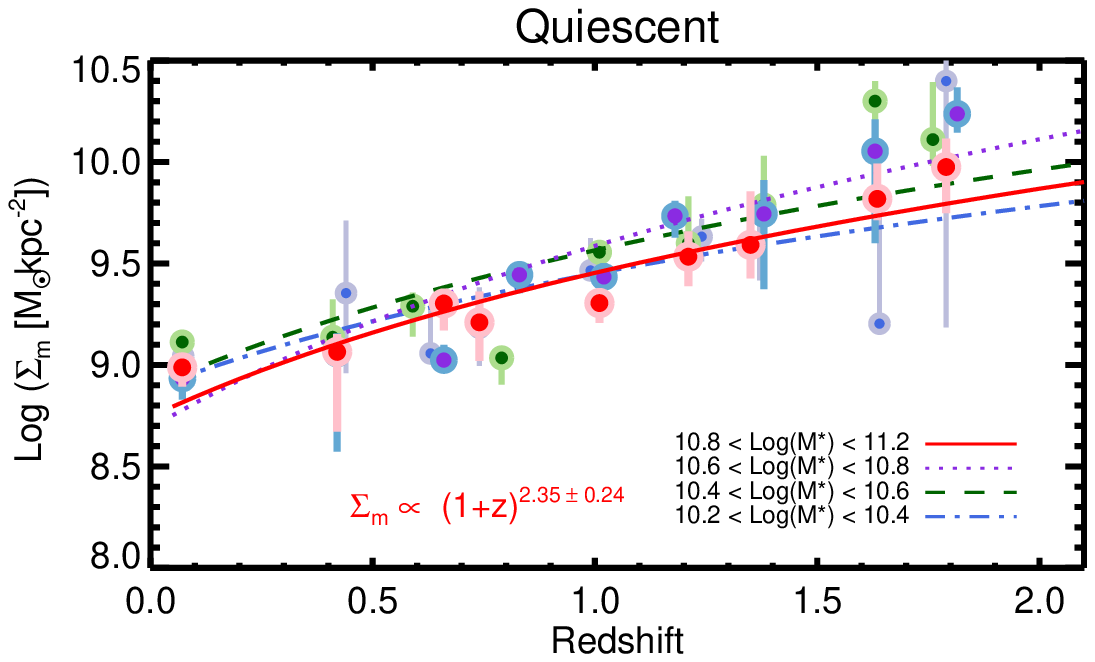}
\caption{Left Panel: The evolution of central mass density with 1 kpc, $\Sigma_{1 kpc}$  from $z\sim2$ to $z\sim0$ for quiescent galaxies, for different stellar mass bins (different colors). The central densities of quiescent galaxies decline slightly with redshift, with the rate of $(1+z)^{0.55\pm0.08}$ for most massive ones. Right Panel:  The evolution of surface density at $r_{m}$ ($\Sigma_{m}$) for quiescent galaxies at different masses with time. The $\Sigma_{m}$ declines sharply from $z\sim2$ to present due to higher surface densities at outer regions of quiescent galaxies in the lower redshifts  compare to their analogous at high-$z$.}

\label{fig11}
\end{figure*}

\begin{figure*}
\centering
\includegraphics[width=0.48\textwidth]{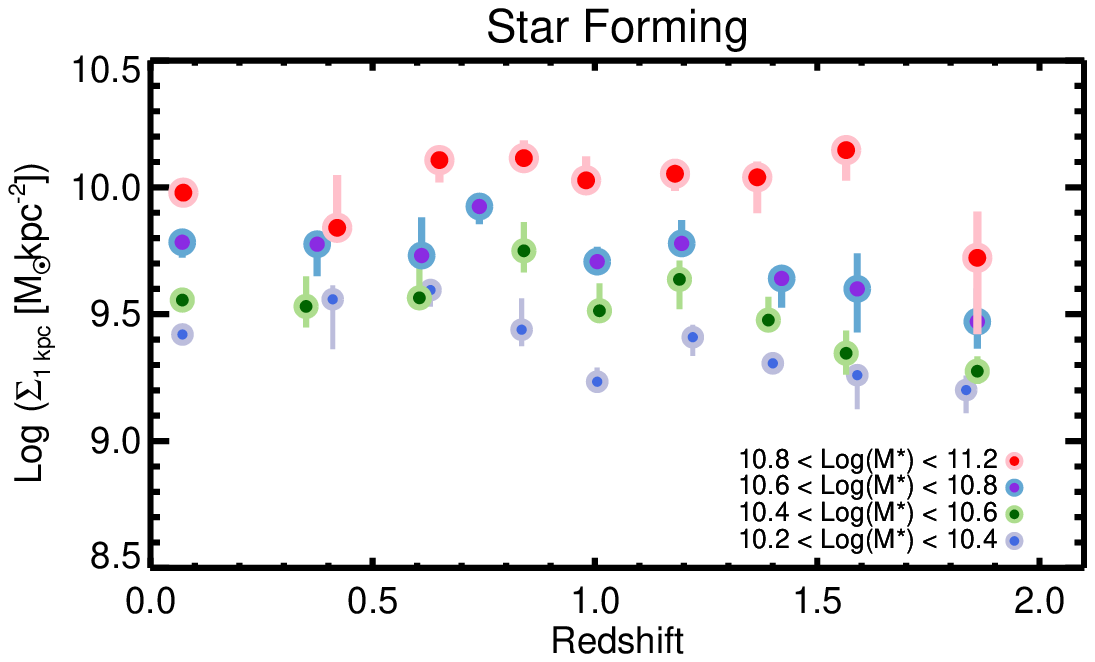}
\hspace{5. mm}
\includegraphics[width=0.48\textwidth]{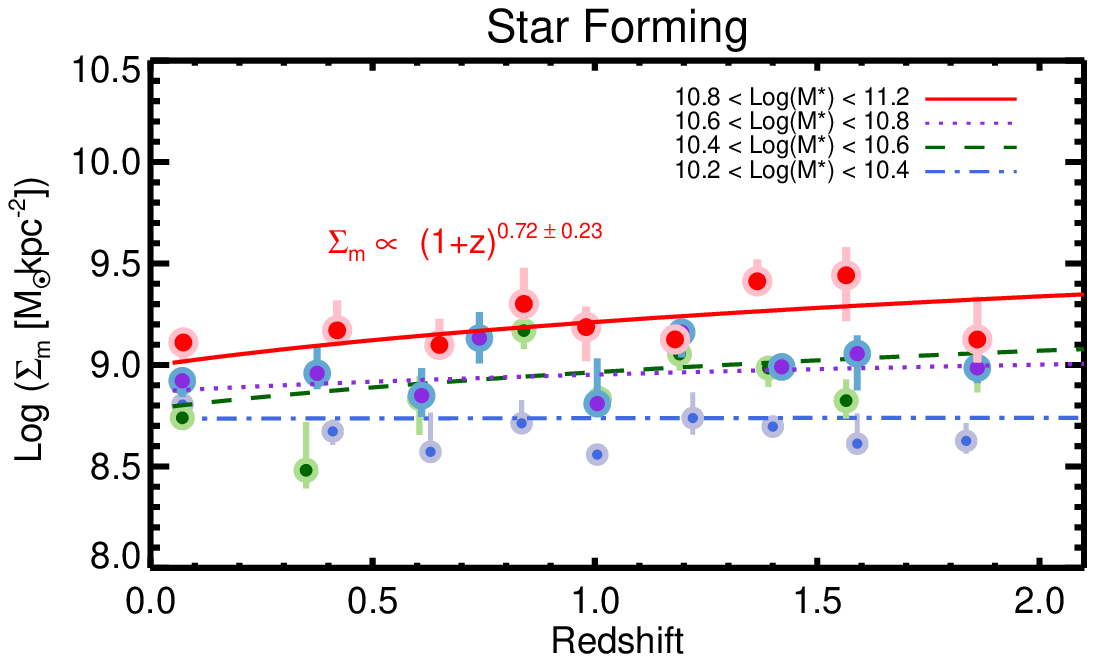}
\caption{The same as Figure \ref{fig11}, but for star-forming galaxies. The surface density at effective mass radii ($\Sigma_{m}$) declines slightly, specially for the most massive ones, however the changes are very slow for intermediate and low masses. The central densities ($\Sigma_{1}$)  increases with time for these galaxies up to $z\sim0.5$ and remain relatively constant. This shows again the changes of stellar masses in the inner and outer regions. We emphasize that the trends are very mild.}

\label{fig12}
\end{figure*}

\subsection{Central \& Effective Surface Density Evolution}

The evolution of mass profiles can also be studied by means of the evolution of surface and central mass densities of galaxies. Stellar mass densities within central 1 kpc region, $\Sigma_{1}$, is used by some authors \citep[][]{saracco2012, cheung2012, fang2013, tacchella2015} as a proxy for central densities. In addition surface density $\Sigma_{m}$ is another parameters which can representative of the mass profiles evolution. In this paper, $\Sigma_{m}$ is defined to be the mass surface density at the half-mass radii ($r_{m}$). Moreover, we also use the traditional ``effective surface density'', $<\Sigma_{m}>$ which is the effective surface density within half-mass radii ($\int_{0}^{r_{m}} 2 \pi \Sigma r dr/  \pi r_{m} ^{2} = \mstar/2\pi r_{m}^{2}$) .  

As galaxies half-mass sizes increase towards lower redshift, it is expected that the surface densities $\Sigma_{m}$ decline correspondingly. This is shown in the right panels of Figure \ref{fig11} and \ref{fig12}, for quiescent and star-forming ones, respectively. At fixed mass, quiescent galaxies have higher surface densities in the past compared to their counterparts in the local Universe. This evolution scales as $(1+z)^{2.35\pm0.24}$ for massive quiescent galaxies and is similar for lower mass bins. However, the changes of $\Sigma_{m}$ is less significant for massive star-forming galaxies with $\Sigma_{m} \propto  (1+z)^{0.72\pm0.23}$ and is even slower for less massive ones (see right panel of Figure \ref{fig12}). The significant changes in the $\Sigma_{m}$ for quiescent galaxies reflects the increases of stellar masses in the outer regions, as depicted in the Figure \ref{fig7}, implying the inside-out growth of these galaxies. For massive star-forming ones, the slow evolution of  $\Sigma_{m}$ indicates that most of the effect is due to the mass assembly in the outer regions of low-$z$ galaxies compared to their high-$z$ counterparts. However, the marginal changes of $\Sigma_{m}$ for star-forming at intermediate masses is due to the increase of surface density at all radii, specifically in their central regions. 

This can be further examined by the evolution of the central densities, i.e. $\Sigma_{1}$. In the left panel of Figure \ref{fig11}, the central mass density evolution of quiescent galaxies is shown at fixed masses. The central densities decline slightly and the rate of evolution is slow,  $\Sigma_{1} \propto  (1+z)^{0.55\pm0.08}$ for massive quiescent galaxies, from $z\sim2$ to $z\sim0$. This indicates that the central regions of the massive quiescent galaxies were slightly denser at higher redshifts compared to their counterparts at low-$z$. It is worth reminding that by stellar mass one means here the time integral of the SFR. Using stellar masses of live and remnant stars will introduce more significant changes.  We caveat that this is not reflecting the true evolution of properties of individual galaxies. Overall, the slow decline of central densities might be described by arrival of the newly quenched systems (with slightly lower $\Sigma_{1}$). \cite{Lilly2016} model (see their Fig. 12) predicts $\Sigma_{1} \propto (1+z)^{0.5}$ for the evolution of the population of quiescent $\sim~10^{11}$ galaxies, which is in agreement with our findings, within the errors. Another scenario could be the possible decline in the velocity dispersion of galaxies as a results of dynamical frictions by minor mergers \citep{bezanson2009,naab2009, sande2013}, though the contributions of mergers needs to be quantified. The $\Sigma_{1}$ is expected to change very slightly in that case of the inside-out scenario \citep{tacchella2015}. We will address this in more details in Section 6. 

For the star-forming ones, there is a weak trend of increasing $\Sigma_{1}$ with time up to $z\sim0.5-1$, specifically for intermediate and less massive ones (left panel of Figure \ref{fig12}), $\Sigma_{1}$ increases from $z\sim2$ to $z\sim0.5$. The most massive star-forming galaxies have already reached to their maximum by $z\sim 1.5$ and remain roughly constant for redshift range of $z\sim 1.5 - 0.5$. 

In general, assuming $\Sigma_{1}$ as a bulge density, star-forming galaxies at lower redshifts have more masses in their central regions (massive bulges) compare to their analogous at high-$z$, at fixed mass. Therefore, the results above show that there should be physical processes of mass growth (as well as re-distributions of stellar masses) that increase both central and outer parts of star-forming galaxies with time. \\

\begin{figure*}
\centering
\includegraphics[width=\textwidth]{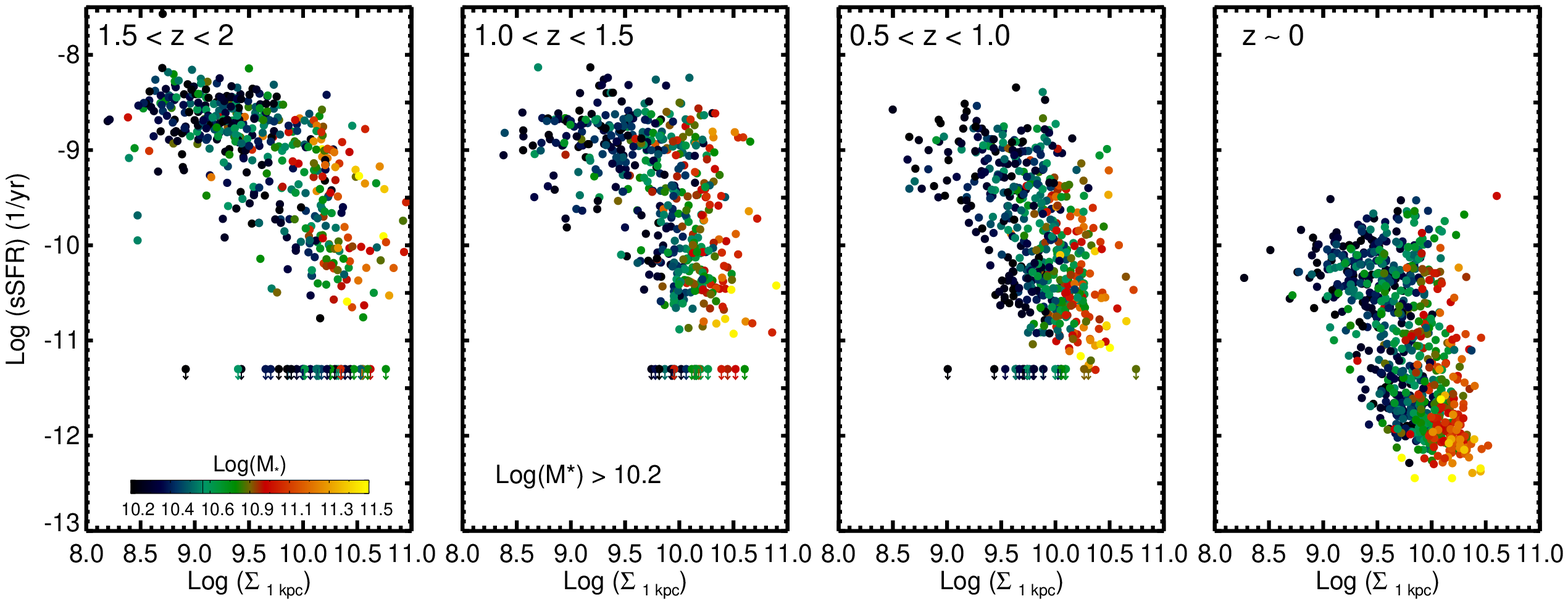}
\vspace{2. mm}
\includegraphics[width=\textwidth]{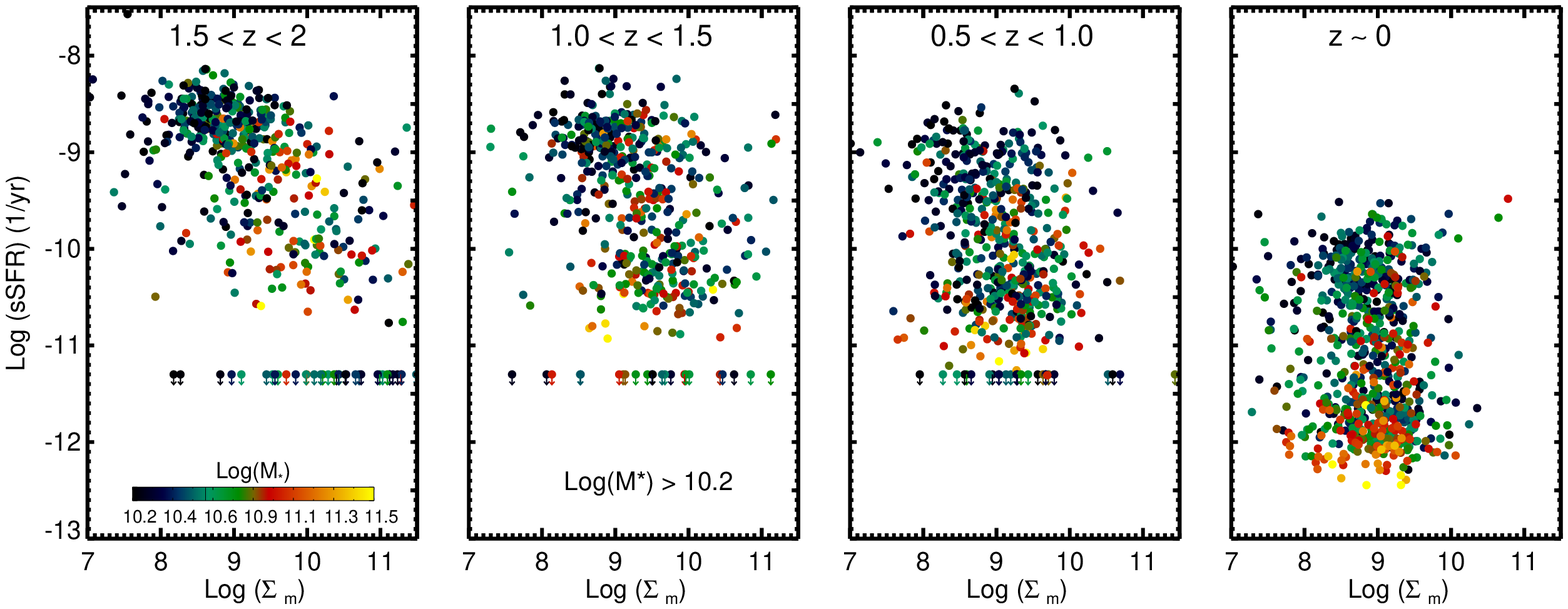}
\vspace{2. mm}
\includegraphics[width=\textwidth]{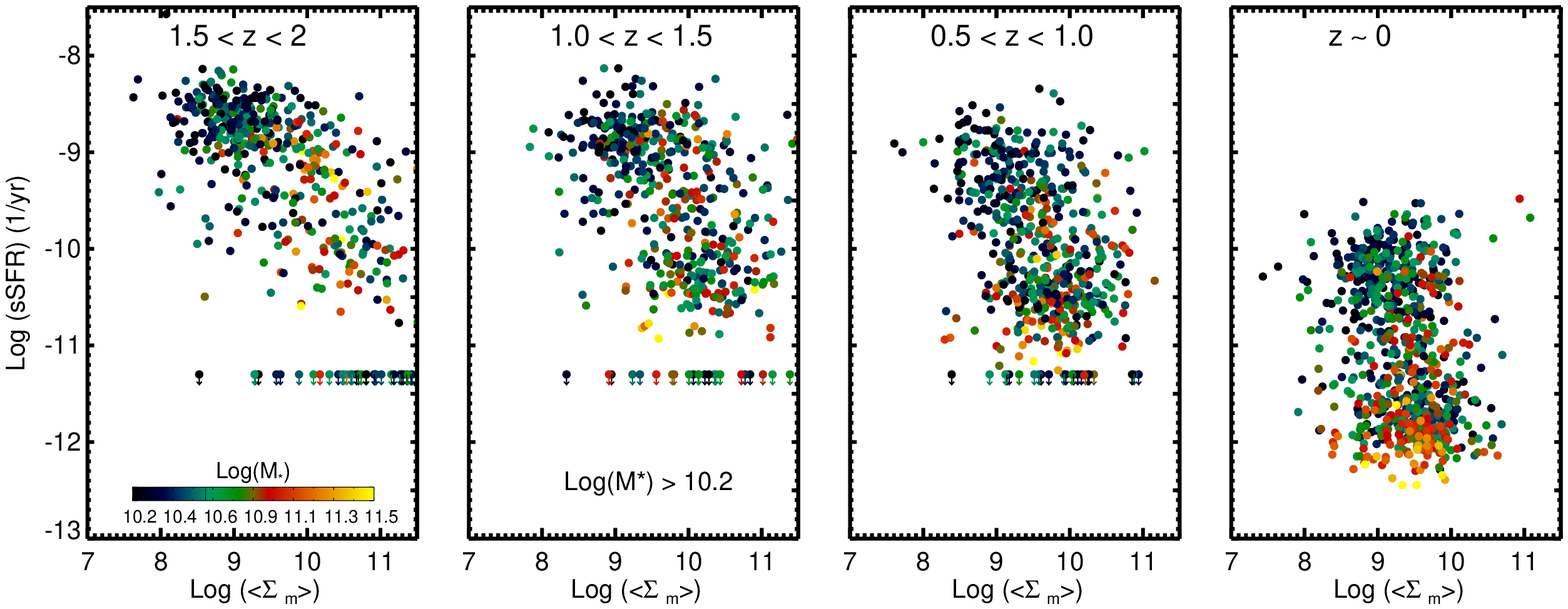}
\caption{Distributions of specific star formation rates (sSFRs) versus central stellar mass density $\Sigma_{1}$ (top panels), stellar mass density at half-mass radii $\Sigma_{m}$ (middle panels) and average mass density withing half-mass radii $<\Sigma_{m}>$ (bottom rows) are shown for galaxies in our sample at different redshift bins. The galaxies are also color-coded according to their total stellar masses. It can be seen that galaxies with higher central densities have lower sSFR at all redshifts. However, the tight (``L''-shape distributions) of objects on sSFR-$\Sigma_\mathrm{1 kpc}$ plane suggest that quenching is more correlated with central densities than $\Sigma_{m}$ and $<\Sigma_{m}>$.}

\label{fig13}
\end{figure*}

\begin{figure*}
\centering
\includegraphics[width=\textwidth ]{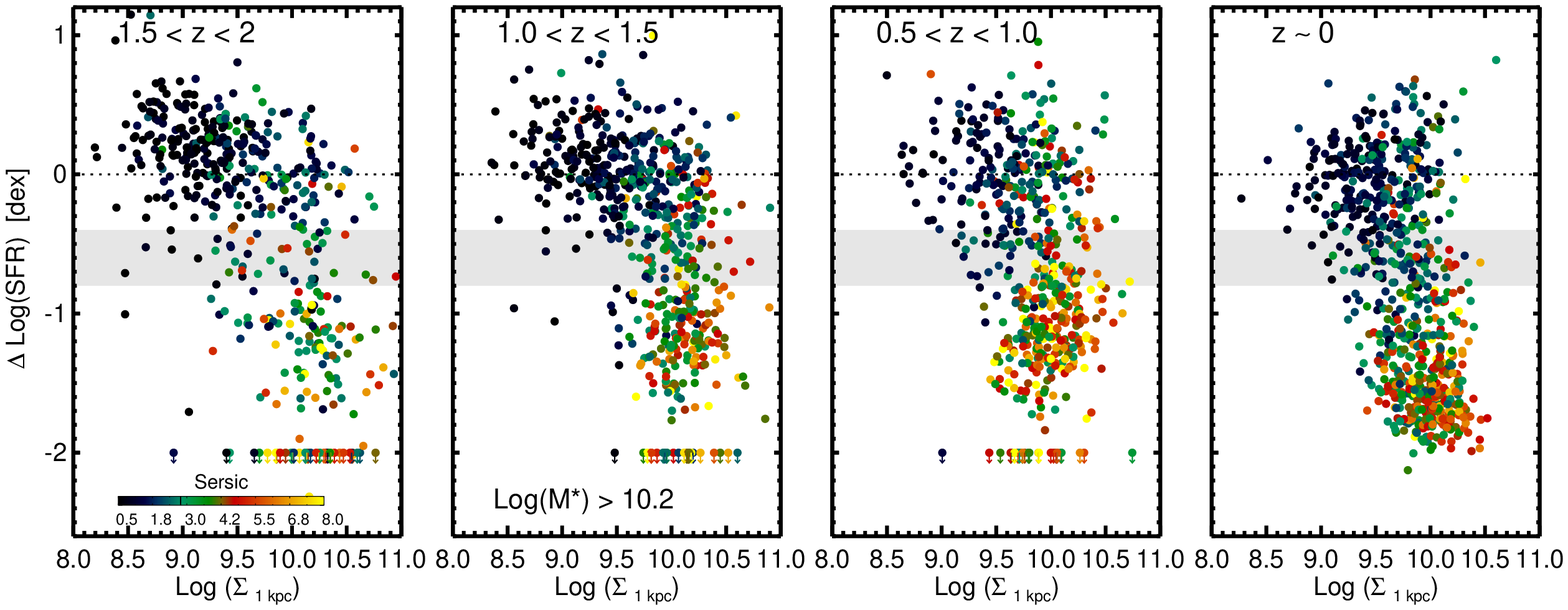}
\vspace{1. mm}
\includegraphics[width=\textwidth ]{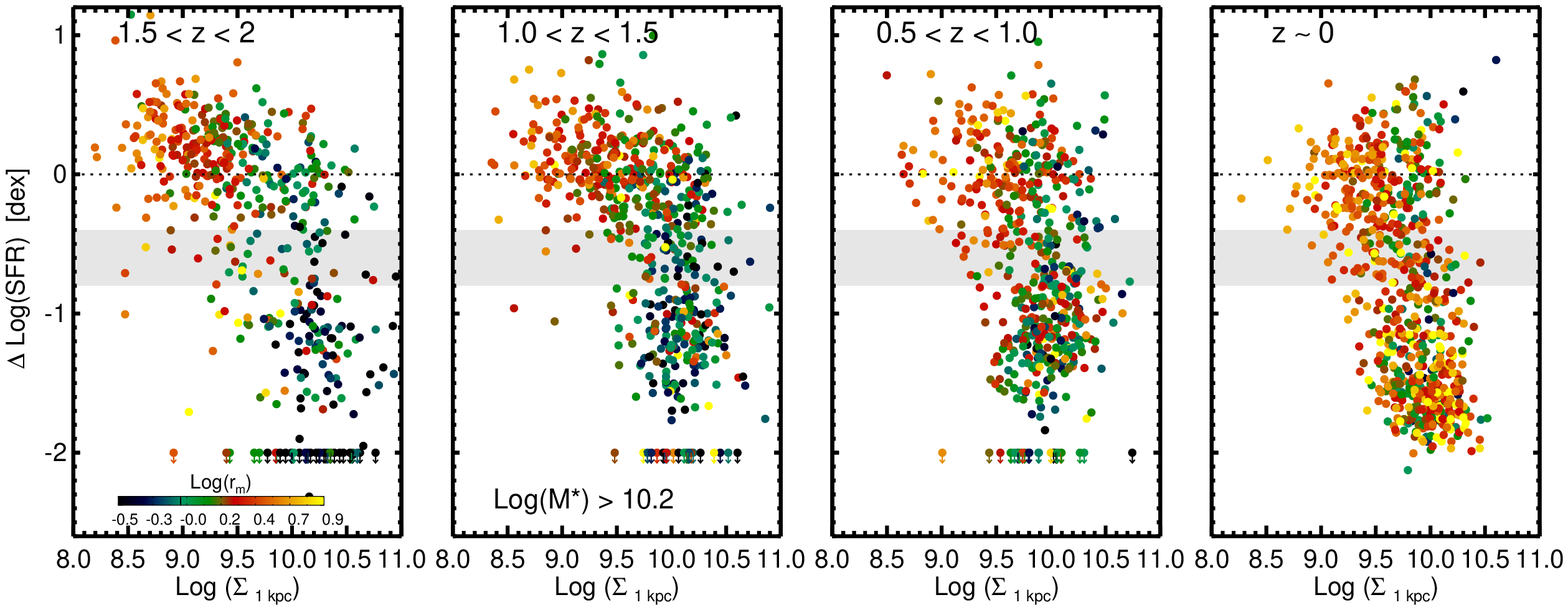}
\vspace{1. mm}
\includegraphics[width=\textwidth ]{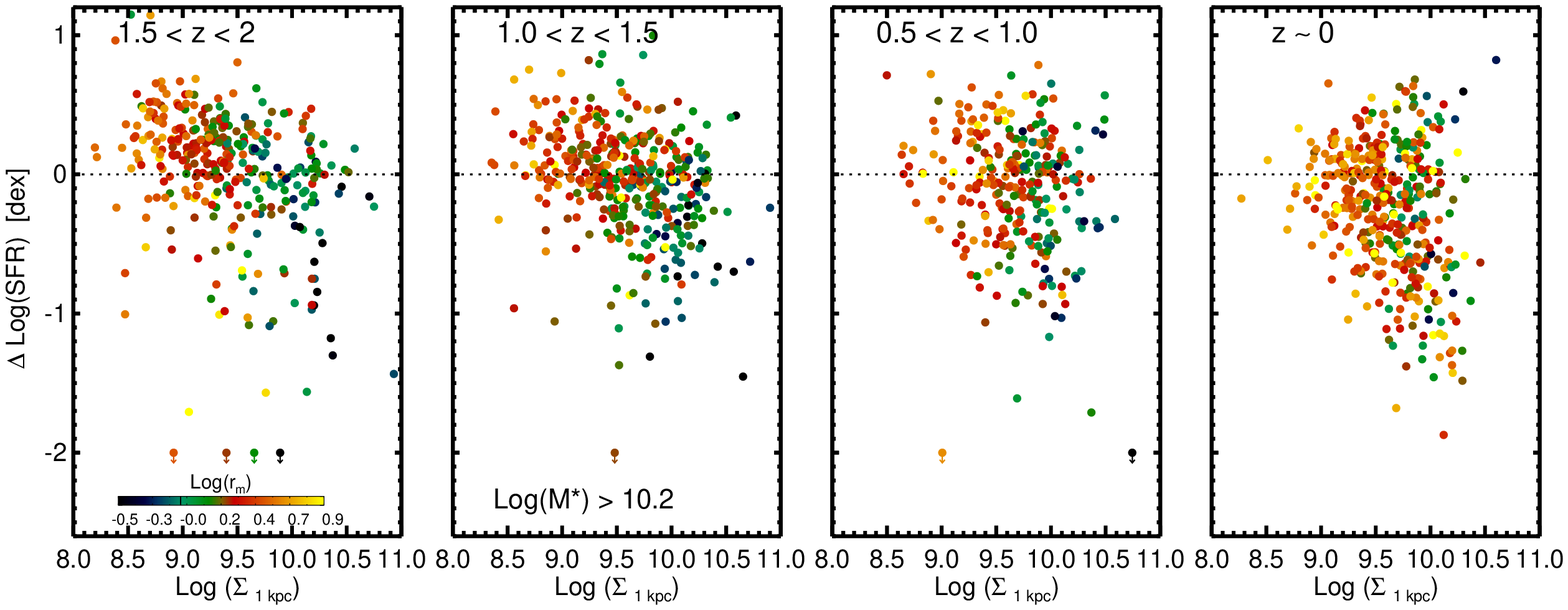}
\vspace{1. mm}
\caption{The differences between star-formation rate and the the main sequence relation (at each redshifts) versus stellar mass central densities ($\Sigma_{1}$). In top rows, galaxies are color coded according to their \ser indices and in the middle and bottom rows, they are color-coded according to their half-mass sizes. The gray strips represent green valley regions ($-0.8 < \Delta$SFR$<-0.4$). The bottom row panels, show only galaxies selected as star-forming in our samples. This figure indicates that galaxies with lower SFR have more concentrated mass distributions and smaller half-mass sizes compare to those with high SFRs at all redshifts ($z<2$).}
\label{fig14}
\end{figure*}


\begin{figure*}
\centering
\includegraphics[width=0.8\textwidth]{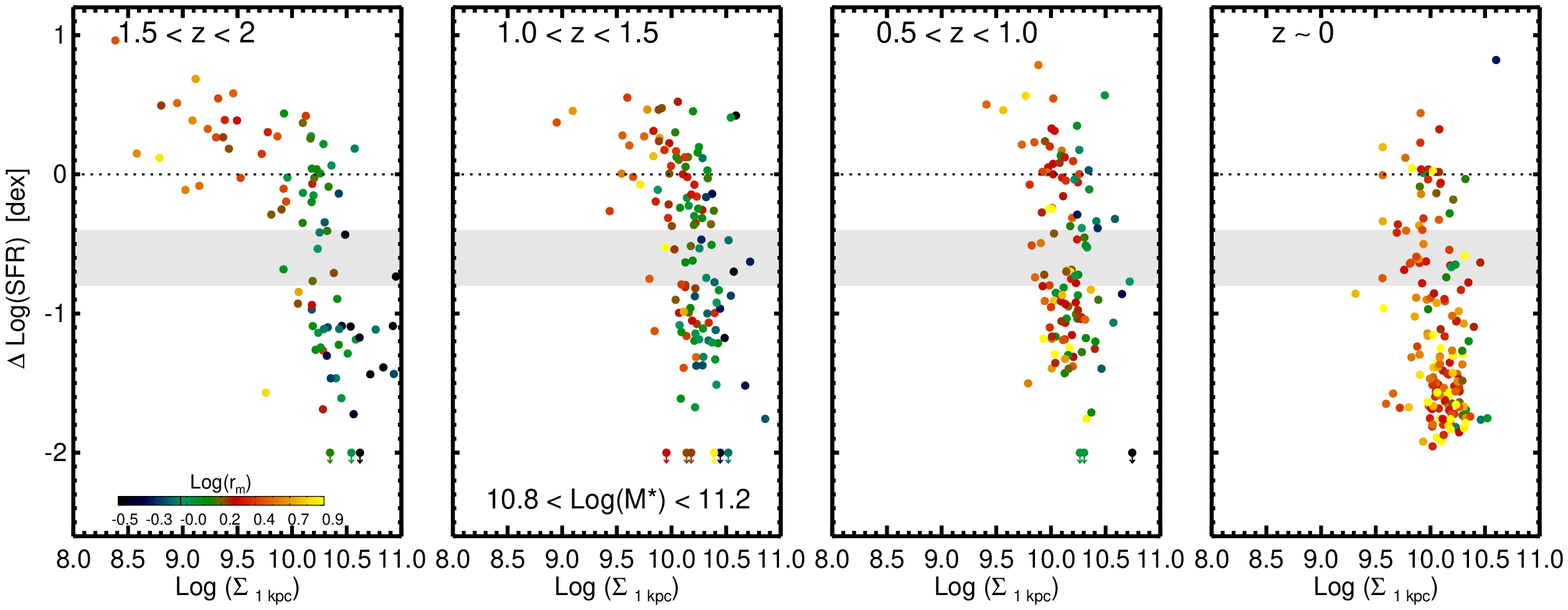}
\vspace{0. mm}
\includegraphics[width=0.8\textwidth]{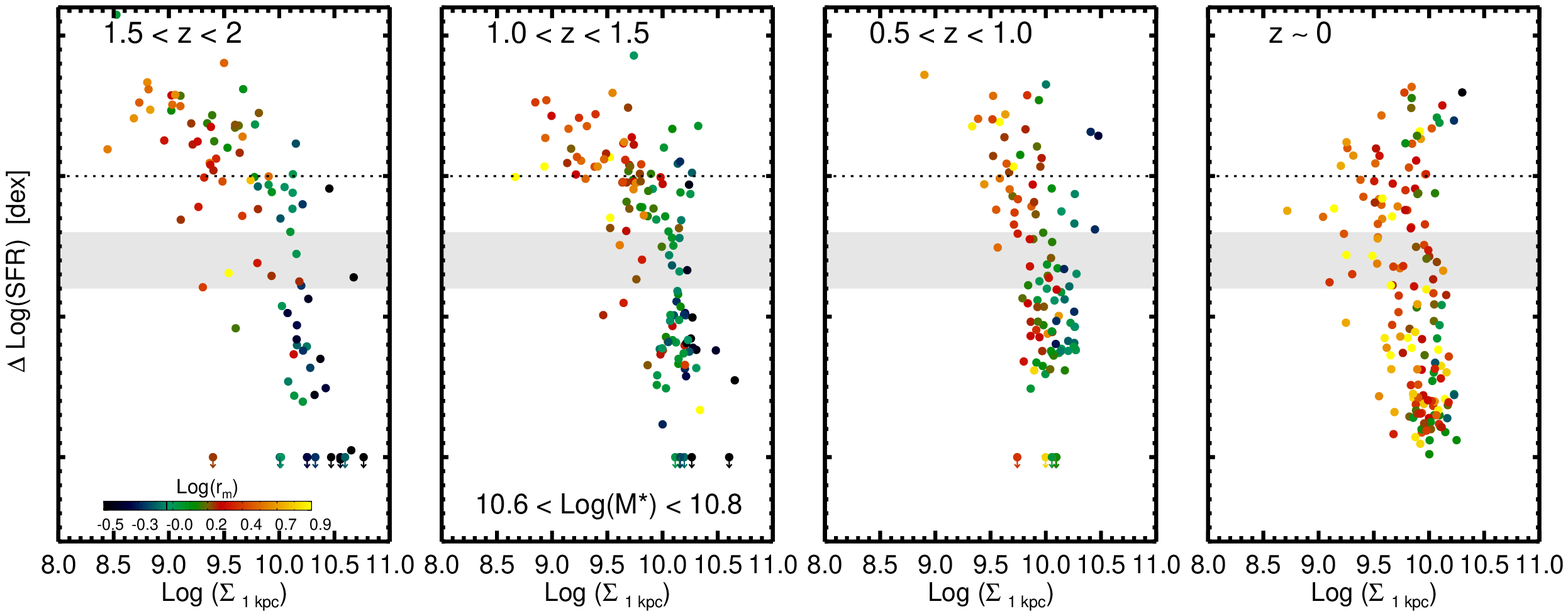}
\vspace{0. mm}
\includegraphics[width=0.8\textwidth]{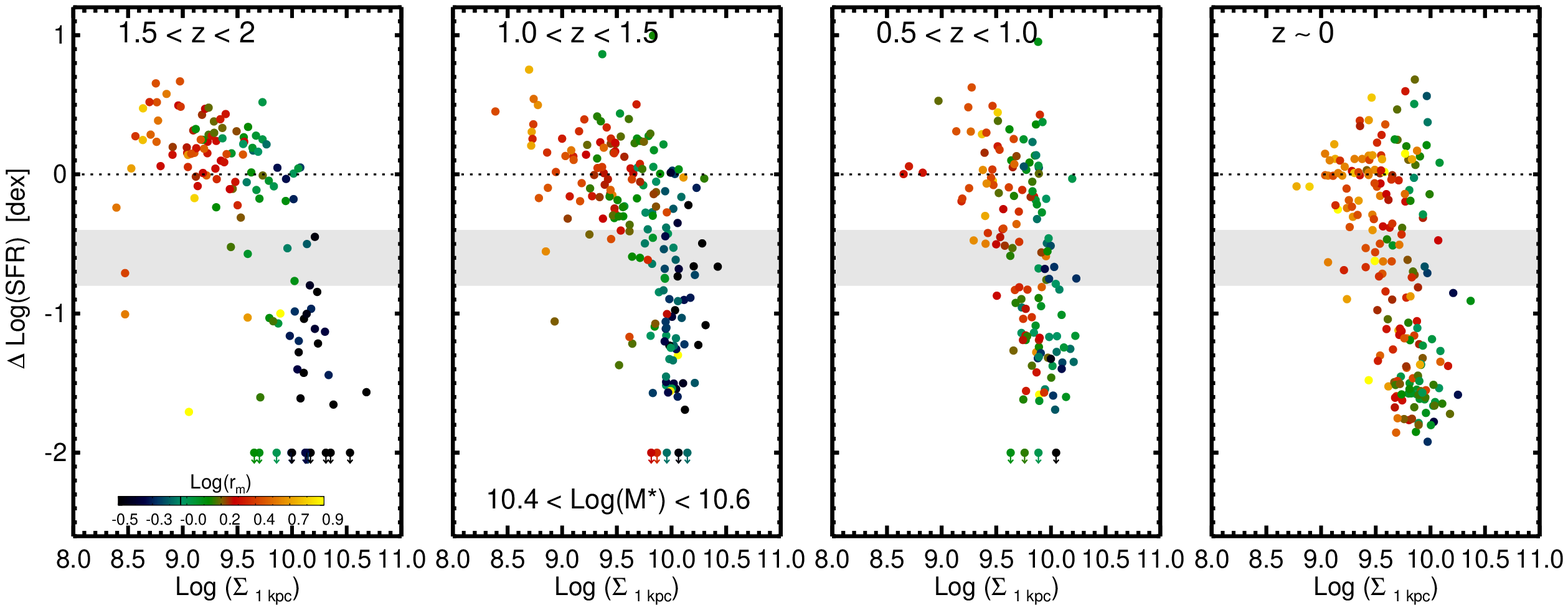}
\vspace{0. mm}
\includegraphics[width=0.8\textwidth]{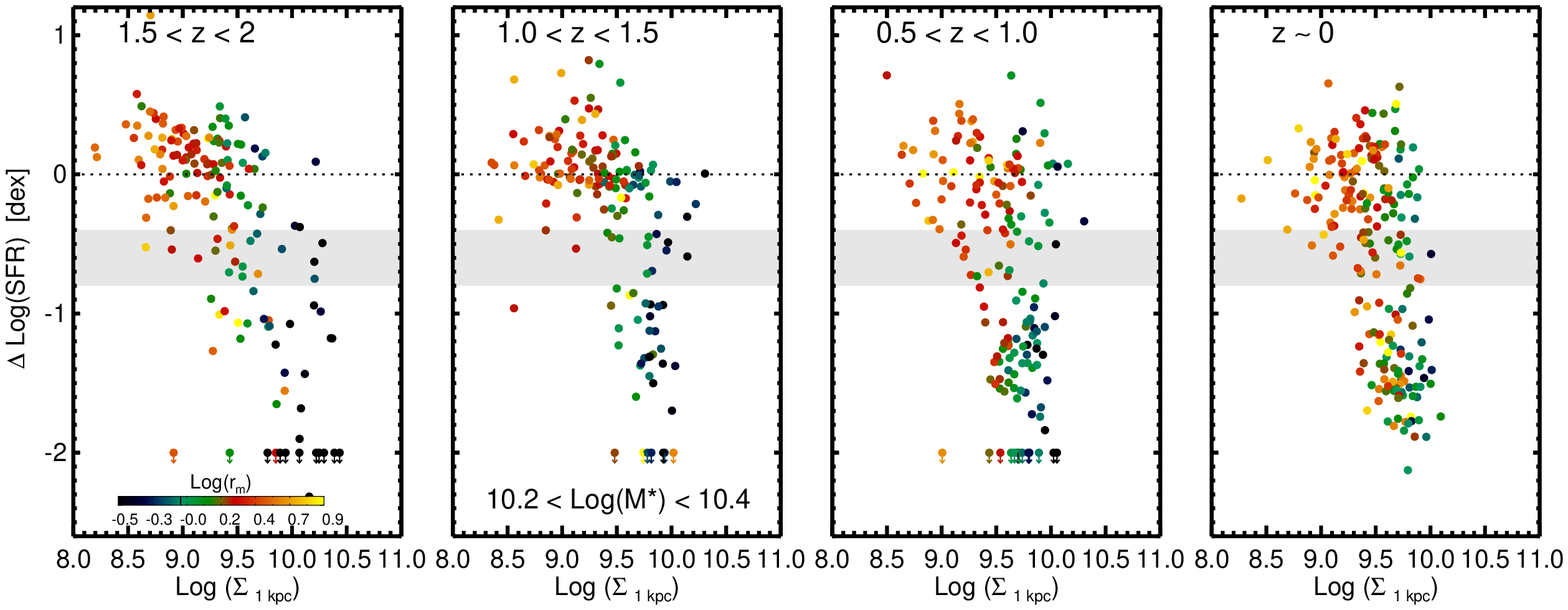}

\caption{The same as Figure \ref{fig14} but for different stellar mass bins. At fixed mass and fixes SFR, galaxies with higher $\Sigma_{1}$ have smaller half-mass sizes.}

\label{fig15}
\end{figure*}

\begin{figure*}
\centering
\includegraphics[width=\textwidth]{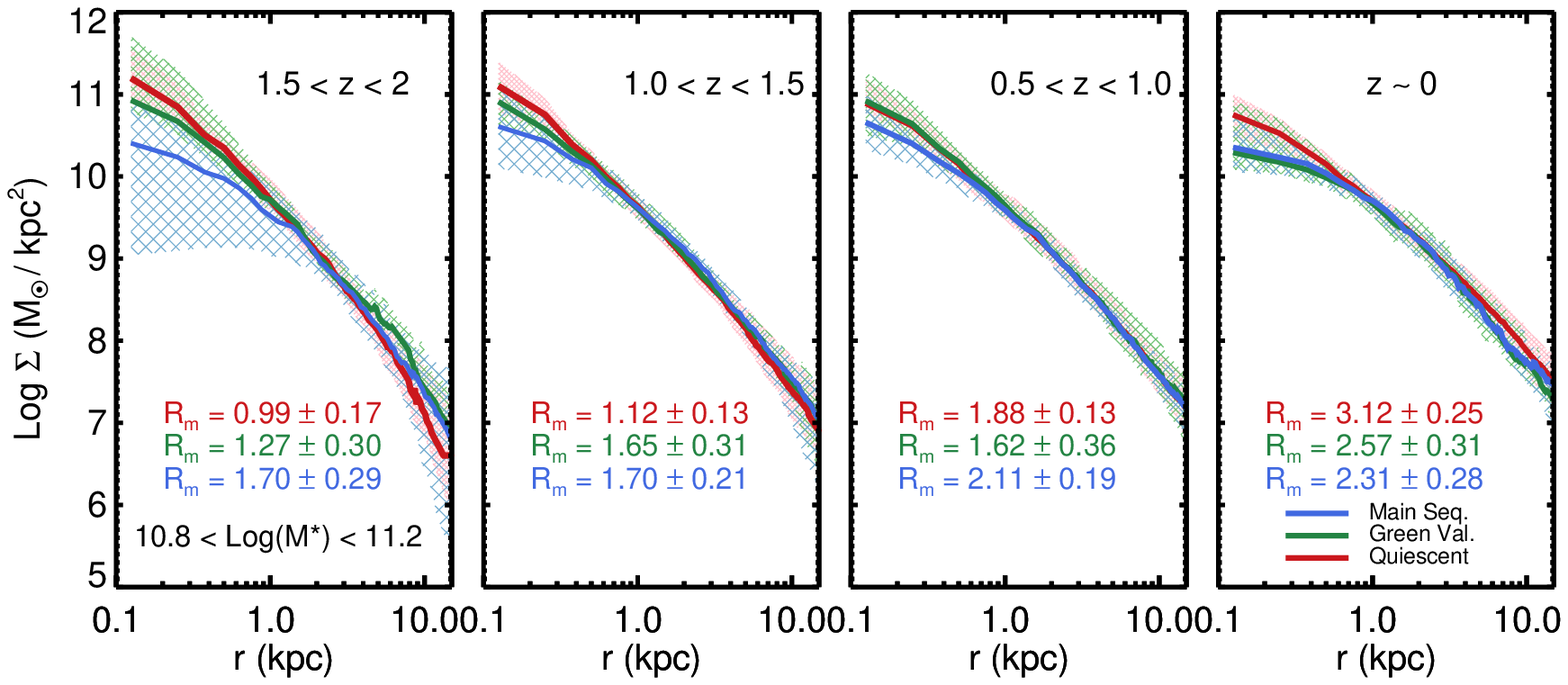}
\vspace{0. mm}
\includegraphics[width=\textwidth]{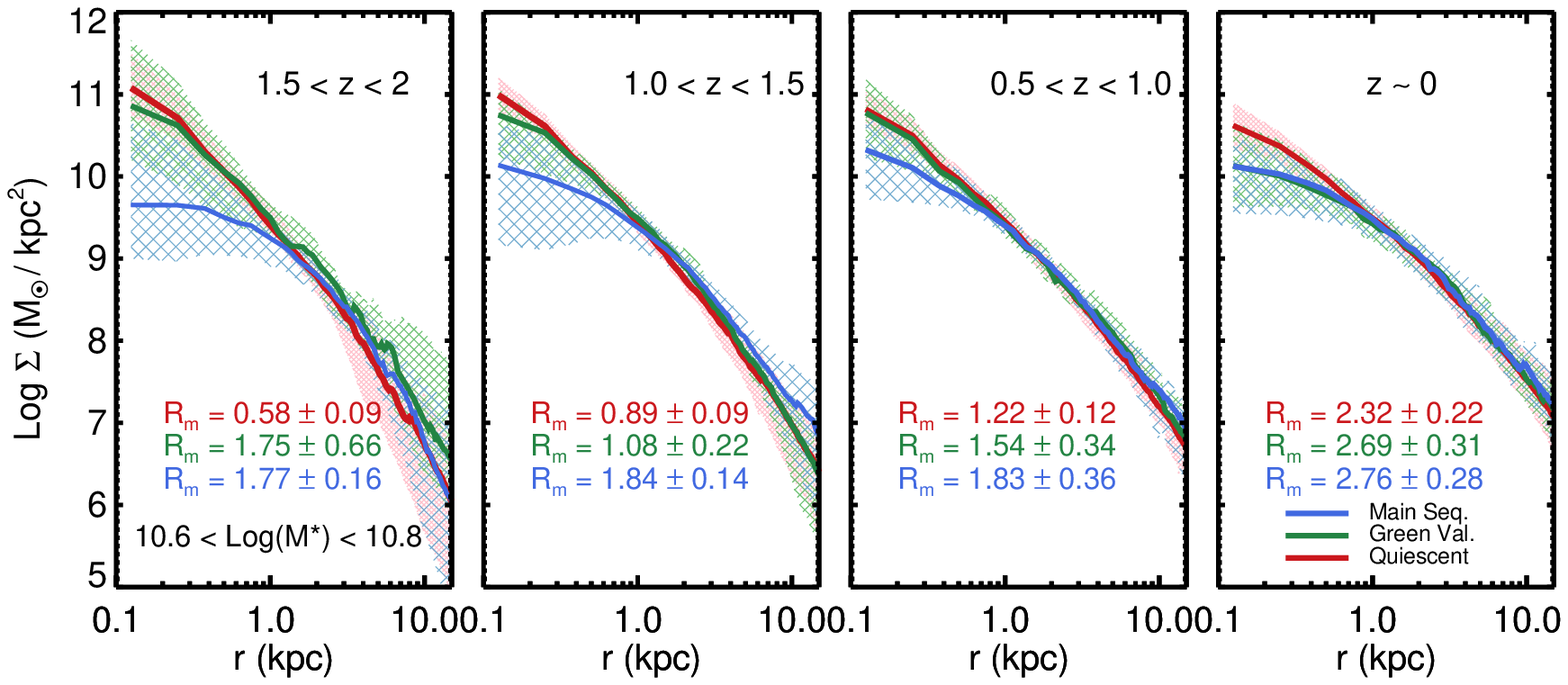}
\caption{The stellar mass profiles of galaxies at fixed mass (upper and lower rows) in the regions of main sequence, green valley and quenched defined in Figure \ref{fig14} and at different redshift intervals (from left to right). Galaxies in the green valley have comparable central densities to quenched galaxies.}

\label{fig16}
\end{figure*}

\begin{figure*}
\centering

\includegraphics[width=\textwidth]{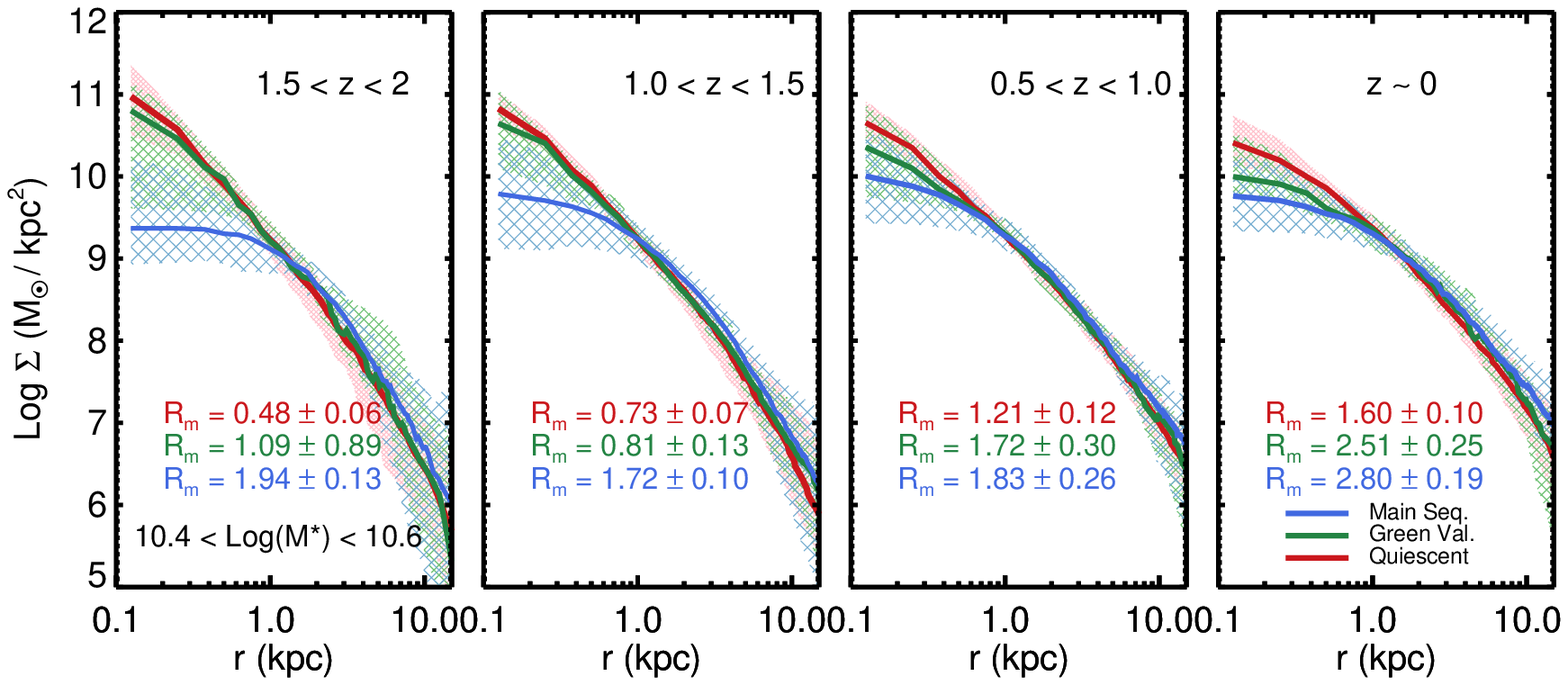}
\vspace{0. mm}
\includegraphics[width=\textwidth]{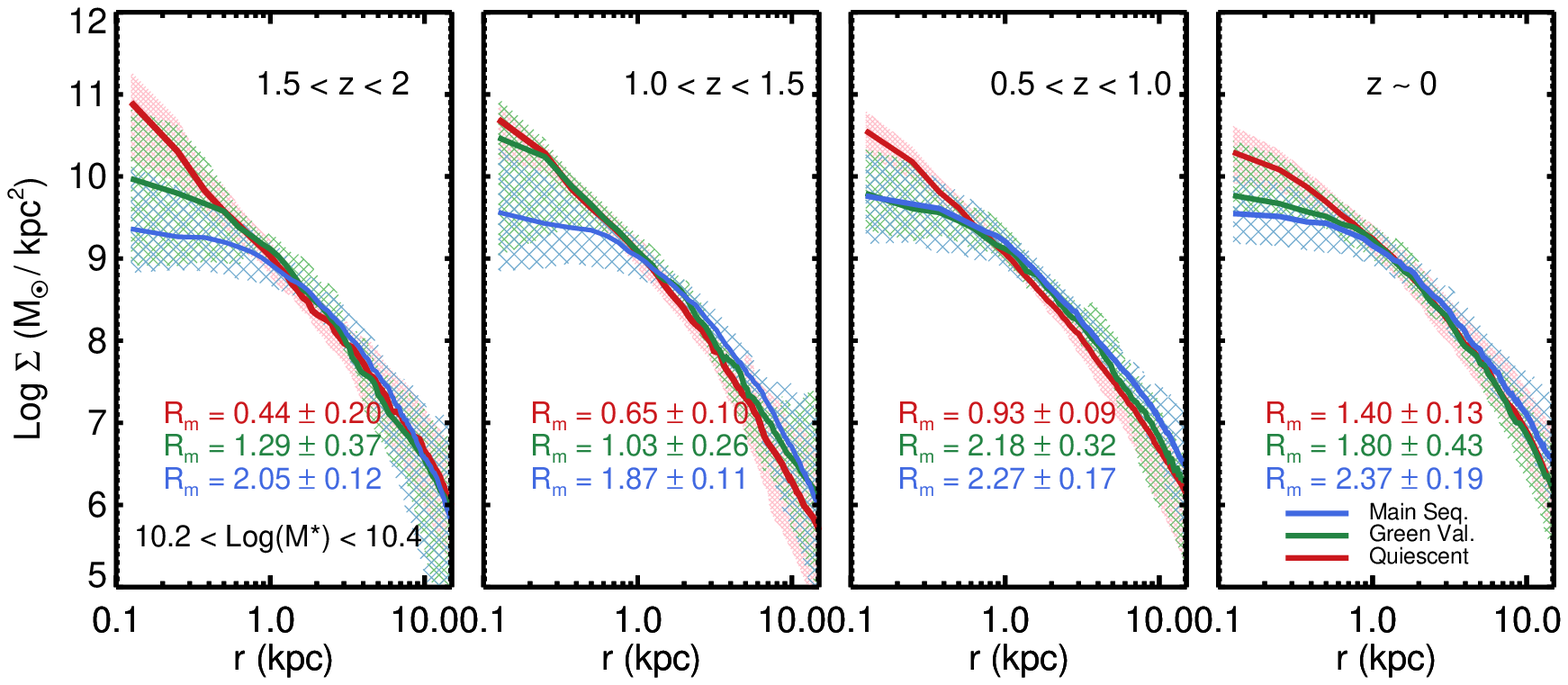}
\caption{The same as Figure \ref{fig16} but for lower mass bins.}
\label{fig17}
\end{figure*}


\section{Relation between Stellar Mass Densities and Quenching}

As mentioned early, it is important to understand any relation between the stellar mass assembly of galaxies and the star-formation quenching. Therefore, comparing mass profile parameters of star-forming and quiescent galaxies at different epochs is an effective way for investigating the possible relation and physical mechanisms. Several studies have already shown that galaxies with higher surface densities have lower specific star formation rates and the relation hold up to high redshifs \citep[e.g.,][]{kauffmann2003, franx2008} hence quenching seems to be correlated to the structural parameters \citep[see also][ and reference therein]{bell2012, lang2014, bluck2014}. These studies is followed by \citet{cheung2012} and \citet{fang2013} to show the links between quenching and bulge mass density ($\Sigma_{1}$) for low and intermediate redshifts. \citet{Barro2015b} also investigate these relations up to $z\sim3$. We revisit these correlations using PSF-corrected mass profiles of our sample at fixed masses in different redshift bins $0 < z < 2$ measured in a consistent way.  

In Figure \ref{fig13} we compare the specific star-formation rate of galaxies in our sample ($>10^{10.2} \msun$) with surface densities within 1 kpc ($\Sigma_{1}$), surface densities at half-mass radii of galaxies ($\Sigma_{m}$) and effective surface densities within $r_{m}$ (i.e., $<\Sigma_{m}>$). The samples are divided into four redshift bins, starting from highest redshift ($1.5<z<2$) in the left to the lowest redshift ($z\sim0$) in the right panel. From top to bottom row, the distributions are shown for sSFR versus  $\Sigma_{1}$, $\Sigma_{m}$ and $<\Sigma_{m}>$, respectively, and objects are color-coded according to their total stellar masses.

As can be seen in this figure, galaxies with higher surface densities have lower sSFRs and this holds at all our studied redshifts. However, the distribution is tighter for sSFR versus $\Sigma_{1}$ compared to the distributions between sSFR and $\Sigma_{m}$ or $<\Sigma_{m}>$, i.e., $\Sigma_{1}$ of galaxies has less scattered distributions below average sSFR of star-forming galaxies at each redshift compared to the other surface density parameters \citep[see also][]{whitaker2016}. The distributions of galaxies on sSFR-$\Sigma_{1}$ plane is following a knee shape \citep[or ``L''-shape as already seen in][]{Barro2015b} (top panels of Figure  \ref{fig13}). This is also similar to the color-central density trend found by \citet{fang2013}.   

The ``L''-shape  distributions of galaxies is originated by the two correlations, 1) star-formation rate - stellar mass relation (SFR-M) of star-forming galaxies known as main sequence relation \citep[e.g.,][and references therein]{Noeske2007, Daddi2007, whitaker2014, Renzini2015, schreiber2015} and 2) the correlation between $\Sigma_{1}$ and stellar masses \citep{saracco2012, fang2013}. The correlation between central densities ($\Sigma_{1}$) and stellar masses is shown to exist up to high-$z$ by several authors \citep{tacchella2015, Barro2015b} and also in simulations \citep{zolotov2015,tacchella2016a}. It is also indicated that  $\Sigma_{1}$-stellar mass correlation is tighter than the correlation between average surface density and total stellar masses \citep[see][]{Barro2015b}, specifically for quiescent ones. The tighter distribution between $\Sigma_{1}$ and stellar masses of quiescent galaxies (low sSFRs) in comparison to their $\Sigma_{m}$, originated from the fact that $\Sigma_{1}$ of these galaxies do not change significantly, in contrast to $\Sigma_{m}$ which increased significantly. 

Therefore, in the top panels of Figure \ref{fig13}, galaxies below central density thresholds, which depends on total masses and redshifts, e.g., $\Sigma_{1} \lesssim 9.5$ $\msun/\mathrm{kpc}^{2}$, for $1<z<1.5$, retain relatively constant sSFR with slight negative slopes (reflecting the main-sequence relation and the $\Sigma_{1}$-total mass relation). Galaxies with a critical central density threshold have left or about to leave the main sequence relation and the star-formation is ceased in these galaxies. The distributions above $\Sigma_{1}$ thresholds remain less scattered by the tight correlation between $\Sigma_{1}$ and stellar mass of quiescent ones. It is tempting to interpret the tighter $\Sigma_{1}$-sSFR correlations in terms of cause and effect between central density and quenching. However, as $\Sigma_1$ also correlates with the total stellar mass, it will be difficult to say whether central density or mass drive the quenching. Therefore, the causal links are difficult to establish. We shall return to this point in Section 6.

To gain more insight on the relation between being quenched and central density, Figure \ref{fig14} shows the distributions of galaxies over $\Sigma_{1}$ and the relative distances of galaxies to the main sequence, i.e., $\Delta$SFR plane, for each redshift separately. We use the best-fit main sequence relation for galaxies at high-$z$ and low-$z$ from \citet{whitaker2014} and \citet{Renzini2015}, respectively to find the relative distances. Using $\Delta$SFR cancels out the redshift dependence of the distributions on SFR. Objects in the top row panels of Figure \ref{fig14} are color-coded according to their \ser indices (form their light-profiles), emphasizing previous results that objects with large distance from the main sequence have higher \ser indices (more concentrated) \citep[e.g.,][]{wuyts2011}. 

The middle panels of Figure \ref{fig14} is the same as the top panels, but color-coded according to their half-mass sizes. This reveals that at fixed SFR (or $\Delta$SFR) galaxies with higher central densities ($\Sigma_{1}$) are more compact, in other words, their stellar masses are more concentrated in their central regions. At fixed mass, this is also valid; if we split samples into four stellar mass bins (Figure \ref{fig15}), galaxies with higher central densities at fixed $\Delta$SFR, have smaller half-mass sizes. As central densities increase the SFR declines for star-forming galaxies. This leads to nominate $\Sigma_{1}$ thresholds to start quenching mechanism at least, up to $z\sim2$. 

The bottom panels of Figure \ref{fig14} shows the same distributions but only for those classified as star-forming in our sample.  At fixed central mass density, there is no indication that star-forming galaxies below and above main sequence are more (or maybe less) compact (see also Figure \ref{fig15}) consistent with the results of recent simulations \citep{tacchella2016b}. However, these plots point out that star-forming galaxies with higher central densities have smaller mass-weighted sizes with a negative tilt between $\Delta$SFR and $\Sigma_{1}$. 

If increasing the central mass density is correlated with quenching, then at fixed mass, galaxies which leave the main sequence should have similar mass profiles, particularly in their central regions. To investigate this, we first split the plane of $\Sigma_{1}$-$\Delta$SFR into three main parallel regions, i.e., galaxies on the main sequence ($\Delta$SFR$>-0.4$), galaxies in the green valley regions ($-0.8 < \Delta$SFR$<-0.4$) and quenched galaxies ($ \Delta$SFR$<-0.8$). We compare the median  mass profiles of galaxies at fixed masses of these three regions in Figure \ref{fig16} and \ref{fig17}, for each redshift bins, separately (left to right panels, from high-$z$ to low-$z$, respectively). 

In these figures, red, green and blue lines represent the mass densities of quenched, green valley and main sequence galaxies, respectively. The median half-mass sizes for each profile is mentioned in each panel accordingly. As can be seen, at fixed masses, quenched galaxies have higher central densities and smaller half-mass sizes compare to the main-sequence ones.  The green valley objects (recently quenched or in transition) also have higher central density profiles compared to the main sequence ones, and their mass profiles are similar to the quenched ones, and this holds at all redshifts up to $z\sim2$. The half-mass size values of green valley galaxies are in between main-sequence and quenched ones. We caveat that at $z\sim0$, massive galaxies on the main-sequence and green valley (with $\log(\mstar/\msun) > 10.6$) have similar mass density profiles indicating that these galaxies are potentially subject to quenching.

These results seem to indicate a high central density as a prerequisite for quenching. However, \cite{Lilly2016} have shown that there may be no direct causality link between central density and quenching, as both may be driven by a third factor, such as the galaxy mass. Still, we emphasize that the relatively empty region at low $\Sigma_{1}$ and low $\Delta$sSFR implies that at least for our studied mass range, almost no galaxies can be quenched prior of reaching to a certain $\Sigma_{1}$ threshold, so there might always be a ``compaction'' phase prior to quenching. If star-forming galaxies quench after reaching a required bulge mass density, then their mass profiles should become similar to that of the quenched galaxies at later times. We will address this in next section, in addition to the mass and redshift dependence of $\Sigma_{1}$ thresholds.\\


\section{Discussion}
The results of previous sections are tempting to indicate the necessity of bulge formation beforehand the secession of star-formation, but the causality links are difficult to establish. It is possible that ``mass quenching'' process \citep{peng2010} alone is able to establish a density-quenching correlation, simply because the first galaxies to (mass) quench are the most dense. Still, it remains important to first understand how galaxies build up their central regions and how the mass profiles of galaxies change once they start quenching by reaching to a certain central density limit. Secondly, what mechanism(s) related to the bulge growth can play a role in preventing the gas to cool and for how long? We try to address these questions in the following. 

\subsection{Formation \& Evolution of Central Concentration}

In Section 4 we have shown that at fixed mass, central mass concentration (resembling bulges) of star-forming galaxies increase with time by a factor of about $2-4$ between $z\sim2$ and $z\sim0.5$ (depending on the total stellar mass). Star-forming galaxies can build and growth their bulges via different proposed mechanisms, including violent disk instabilities  and mergers of giant clumps at high redshifts \citep[e.g.,][]{noguchi1999, Elmegreen2008, Bournaud2007, ceverino2010}, weak disk instabilities a.k.a secular evolution \citep{kormendy2004} and inflow of gas via mergers \citep[e.g.,][]{Fisher2008, hopkins2009c,hopkins2010a}. These processes depends on the redshift, for instance due to the high gas fraction at high redshifts \citep{Daddi2010, Tacconi2010}, formation of giant $10^{8-9}\msun$ clumps by strong gravitational instabilities and hence coalescence of them to build bulges \citep[e.g.,][]{genzel2011,guo2012}. However, at low-$z$ weak instabilities of galaxies sub-components such as bars, can likely trigger formation/evolution of bulges \citep[see also recent reviews by][]{Bournaud2016, Brooks2016, Fisher2016, Falcon2016, kormendy2016} and references therein.

For simulated galaxies \cite{tacchella2016a,tacchella2016b} argued that oscillations about the main-sequence ridge are important, i.e., when galaxies are above the main sequence, they form most of their stars in the bulge, while when then are below the main-sequence, they form most of stars in the disk. The outcome is that the mass in the central and outer regions grow concurrently. As shown in Figure \ref{fig12}, there are some hints that lower mass star-forming galaxies form their bulges at later times than massive ones, concurrently with their disk. Star-forming galaxies with moderate-mass form their bulges about 2-3 Gyr later, comparing to the massive ones. Therefore, viable mechanisms might also depend on total stellar mass of galaxies and hence the \textit{``compaction''} phase \citep[e.g., ][]{Dekel2014, zolotov2015, tacchella2016a,tacchella2016b} can be different processes at different redshift and mass.  In addition, we have also seen that at fixed mass, disks of star-forming galaxies grow with time in lockstep with bulges \citep{vandokkum2013, patel2013b}, hence indicating that the bulge formation mechanism(s), such as violent gas inflow induced by counter rotating streams and/or merging supported by violent disk instabilities presumably should be able to preserve the disks. Note that in this work, we have compared similarly selected galaxies at fixed masses, therefore the true evolution of individual galaxies with different stellar masses needs to be understood. Nevertheless, bulges of star-forming galaxies grow with time up to a certain values (stellar mass dependent) and possibly retain their bulge masses roughly constant thereafter.

\begin{figure}
\centering
\includegraphics[width=0.48\textwidth]{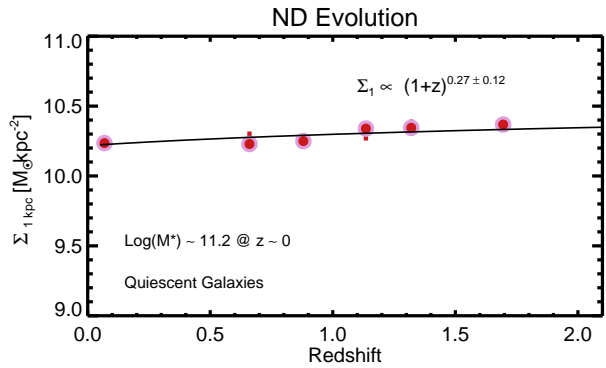}
\caption{Using abundance matching approach to select progenitors of massive ($\sim10^{11.2}\msun$) quiescent galaxies at $z\sim0$ and trace the evolution of their central densities from $z\sim2$ to $z\sim0$. The central densities ($\Sigma_{1}$) of these galaxies change as $(1+z)^{0.27\pm0.12}$ consistent with little to no evolution. However, this method has limitations as many galaxies ending up at this massive end at $z\sim0$ (see text).}

\label{fig18}
\end{figure}

\begin{figure}
\centering
\includegraphics[width=0.48\textwidth]{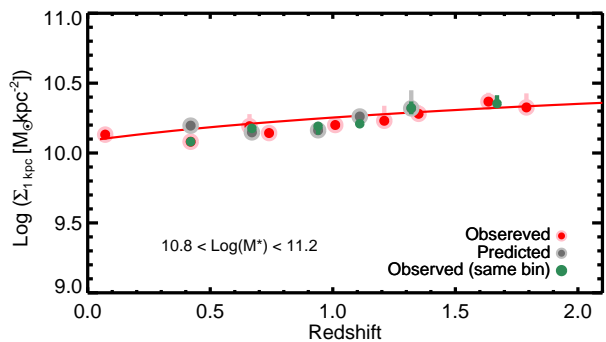}
\caption{The observed evolution of the average central density of quenched galaxies in the indicated mass bin as from Figure \ref{fig11} (red symbols), compared to the prediction from Eq. \ref{equation6} (gray circles). The green circles are again the observed surface densities, now using the same redshift binning as for the model.}

\label{fig19}
\end{figure}


On the other hand, at fixed mass, quiescent galaxies at $z\sim0$ have slightly lower $\Sigma_{1}$ values compare to their high-$z$ counterparts (see Figure \ref{fig11}). If quiescent galaxies grow inside-out, then it is expected that $\Sigma_{1}$ remain roughly constant, i.e., galaxies central densities remain almost intact after they become quenched. However, as pointed earlier the newly quenched galaxies could introduce biases and decrease the average $\Sigma_{1}$ values. To further examine this, it is needed to trace back individual galaxy out to early times which is observerationally difficult. Statistically, it is proposed to use ``constant'' cumulative number density approach \citet{dokkum2010} however, this method fails to consider different galaxy mergers and scatter in mass accretion histories \citep[see e.g.,][]{Torrey2015}. \citet{marchesini2014} takes into account these effects by using abundance matching method \citep{behroozi2013} and cumulative number density evolution of massive galaxies to match progenitors and descendants.  If we use this later approach only for massive quiescent galaxies (Figure \ref{fig18}), we see little to no evolution of the central densities for quiescent galaxies ending up at $10^{11.2}\msun$ quiescent galaxies at $z\sim0$ (i.e., $\Sigma_{1} \propto  (1+z)^{0.27\pm0.12}$). For that we have used \citet{baldry2008} ($z\sim0$) and \citet{tomczak2014} (high-$z$) mass functions to estimate the cumulative number densities at different redshifts. Similar results also found by \citet{sande2013} \citep[see also][]{bezanson2009, oser2012}.  However, this approach is still problematic as so many galaxies of different stellar masses can end up at $z\sim0$ with $>10^{11}\msun$.

We use now a simple empirical model to interpret the observed evolution of the central density of quiescent galaxies ($\Sigma_{1_Q}$). For each pair of redshifts $z_{1}$ and $z_{2}$ ($z_{1}>z_{2}$), we write:

\begin{equation}
\Sigma_{1_Q} (z_{2}) = \frac{n_{Q}(z_{1}) \cdot \Sigma_{Q} (z_{1}) + (n_{Q} (z_{2})-n_{Q}(z_{1})) \cdot \Sigma_{1_{SF}}(z_{1})}{n_{Q} (z_{2})},
\label{equation6}
\end{equation}

where $n$ is the comoving number density of galaxies in the considered mass bin. This assumes that the central density of quiescent galaxies at $z_{1}$ remain identical at $z_{2}$ and that the star-forming galaxies that quench between $z_{1}$  and $z_{2}$ maintain the central mass density they had at $z_{1}$. In this way we model the effect of the so-called progenitor bias. We have used the number density of quiescent and star-forming galaxies at different redshifts from \citet{tomczak2014} and the result for our higher mass bin are shown in Figure \ref{fig19}. The predicted central density of quiescent galaxies is shown by gray circles and compared to the observational results as in Figure \ref{fig11} (red circles). The green circles also shows the observed ones but with different redshift binning (i.e., binning similar to the ones used for the model). Any difference between the value calculated with this equation and the actual data could be ascribed to other effects, such as a structural evolution of individual galaxies. However, the predicted central densities agree quite well the observed ones, implying that the central regions of galaxies do not evolve much following quenching.\\


\begin{figure*}
\centering
\includegraphics[width=0.32\textwidth]{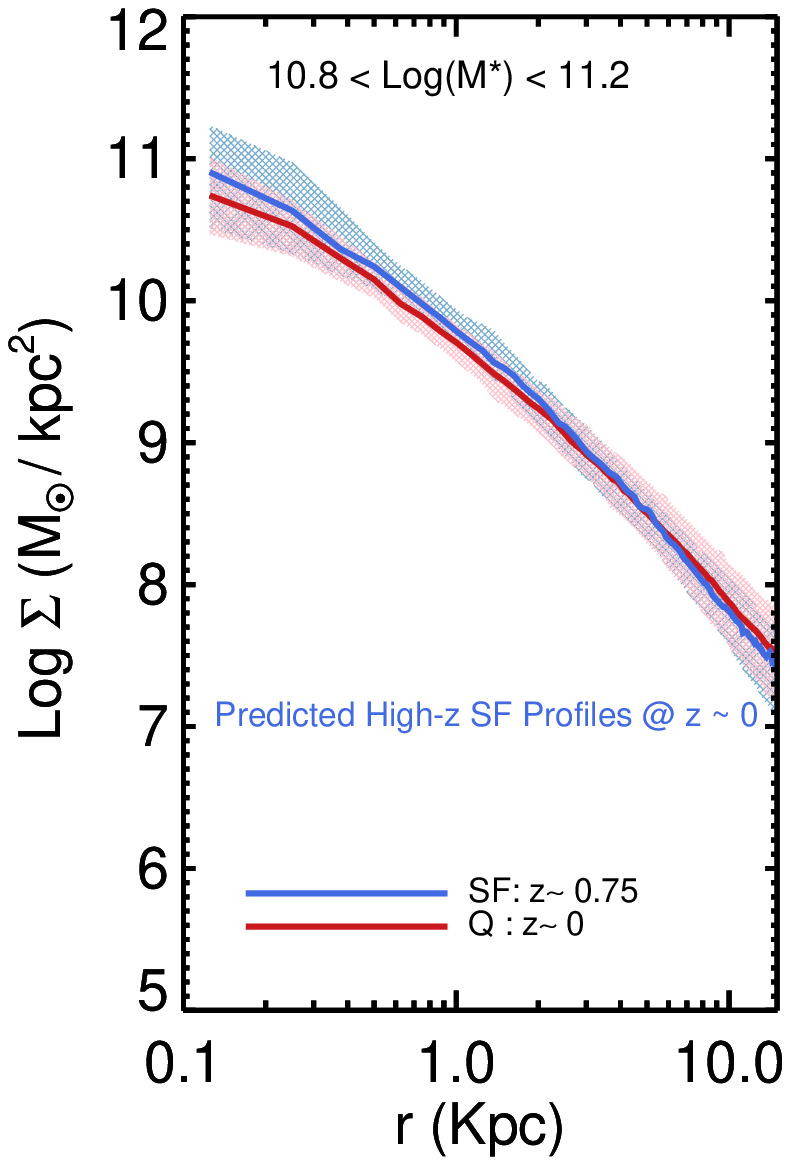}
\hspace{0. mm}
\includegraphics[width=0.32\textwidth]{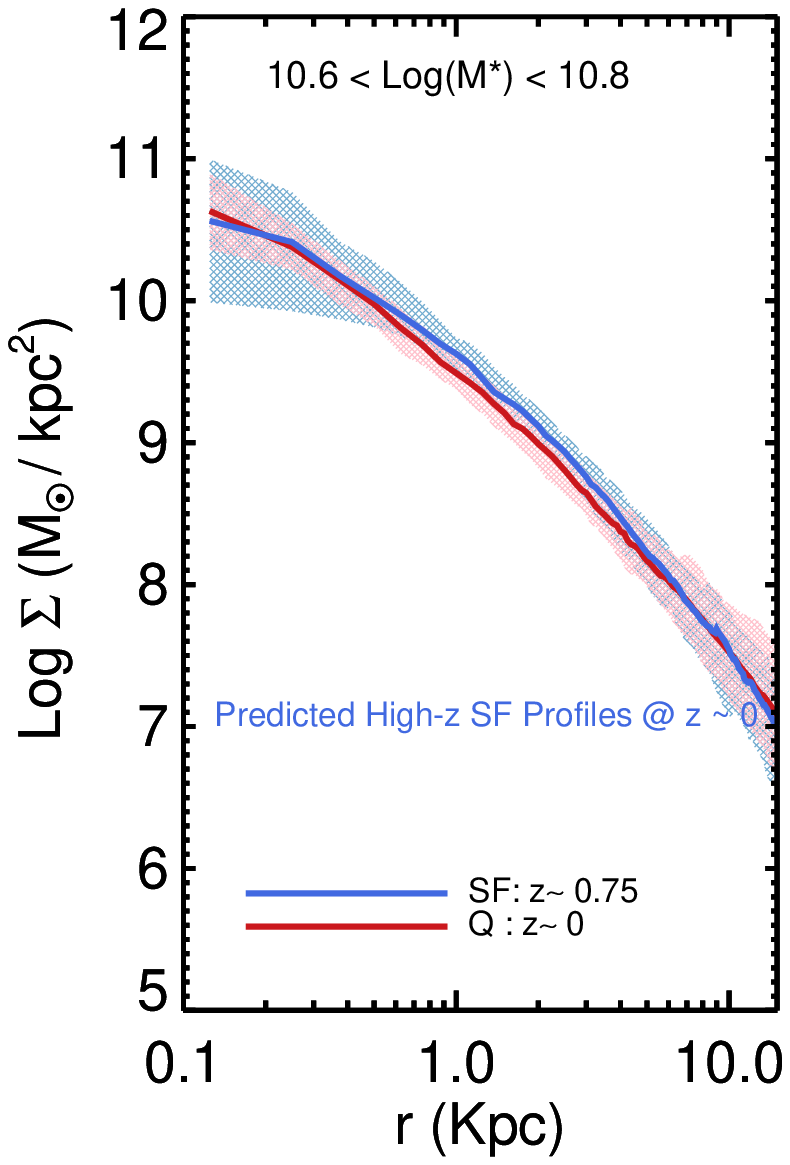}
\hspace{0. mm}
\includegraphics[width=0.32\textwidth]{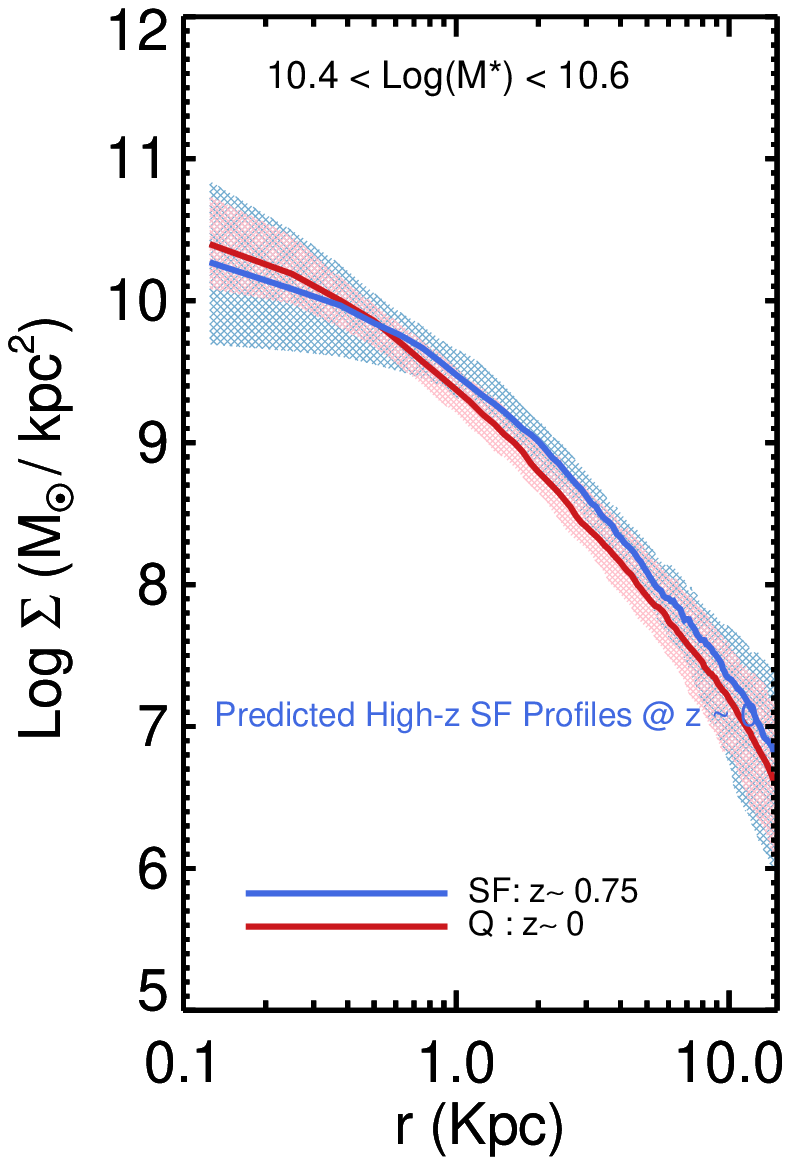}
\caption{Comparing the expected $z\sim 0$ stellar mass profiles of star-forming galaxies at $z\sim 0.75$ (blue lines) with the observed local quiescent galaxies at $z\sim0$ (red lines) at fixed masses (different panels, decreasing from left to right). The star forming galaxies at these mass ranges reached their maximum central densities by $z\sim0.75$. Hence, this simple test illustrates how the mass profiles of star-forming
galaxies at high-$z$ will look like once they become completely quenched by $z\sim0$ (ignoring the effects of major mergers). The stellar mass profiles of these galaxies are very similar specifically in the central regions, indicating the transformation of star-forming galaxies to quenched ones objects after reaching $\Sigma_{1}$ threshold is plausible. The solid lines are the median of profiles and shaded regions show the 1-$\sigma$ scatter.}

\label{fig20}
\end{figure*}

\begin{figure*}
\centering
\includegraphics[width=0.32\textwidth]{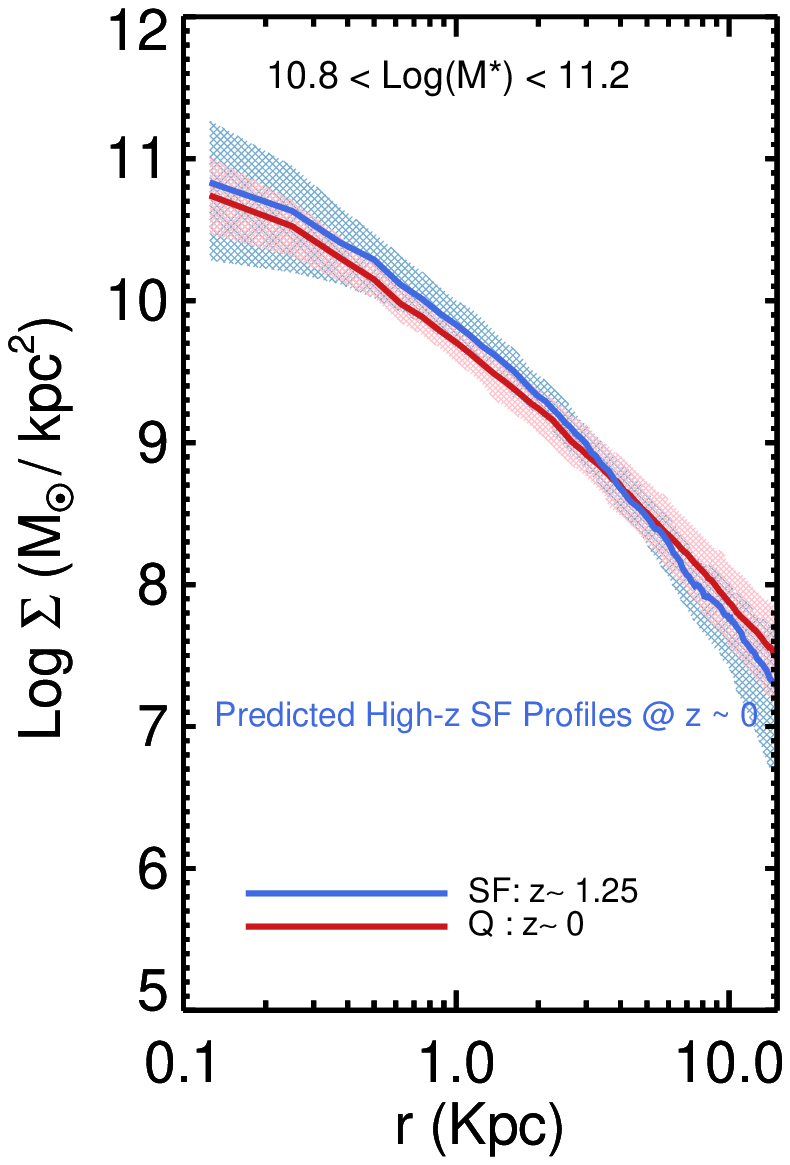}
\hspace{0. mm}
\includegraphics[width=0.32\textwidth]{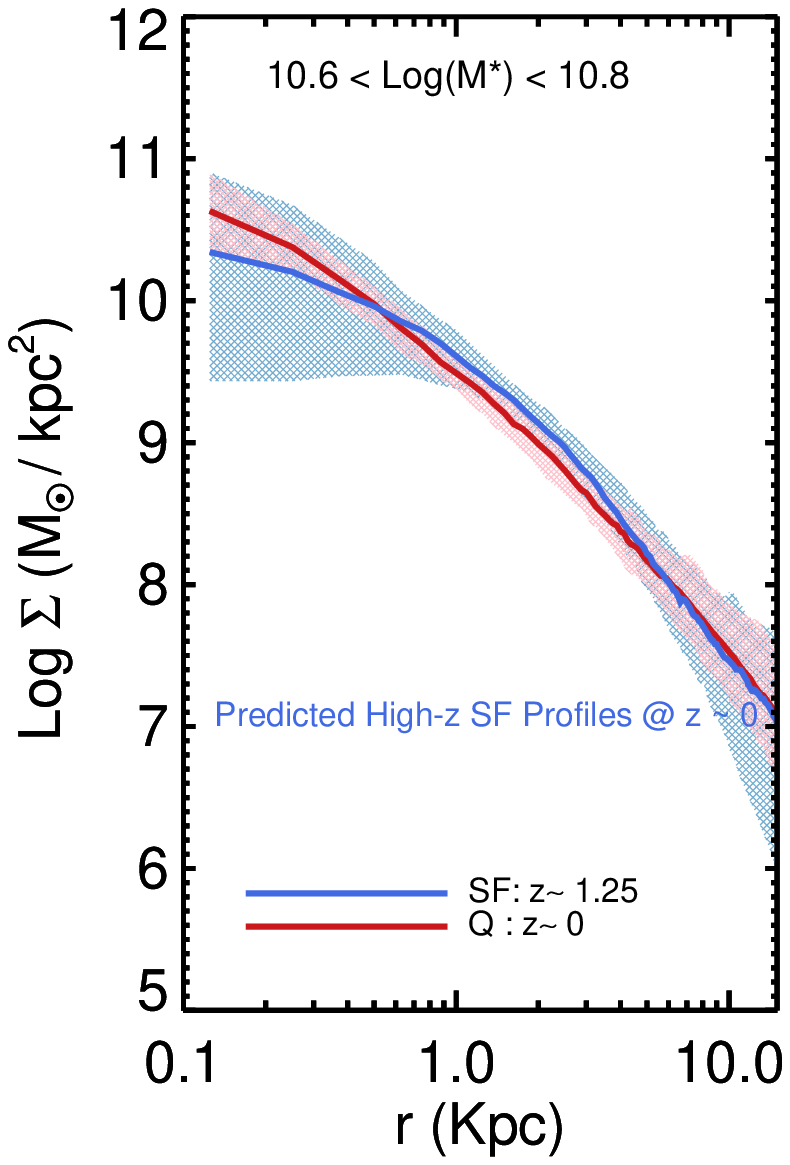}
\hspace{0. mm}
\includegraphics[width=0.32\textwidth]{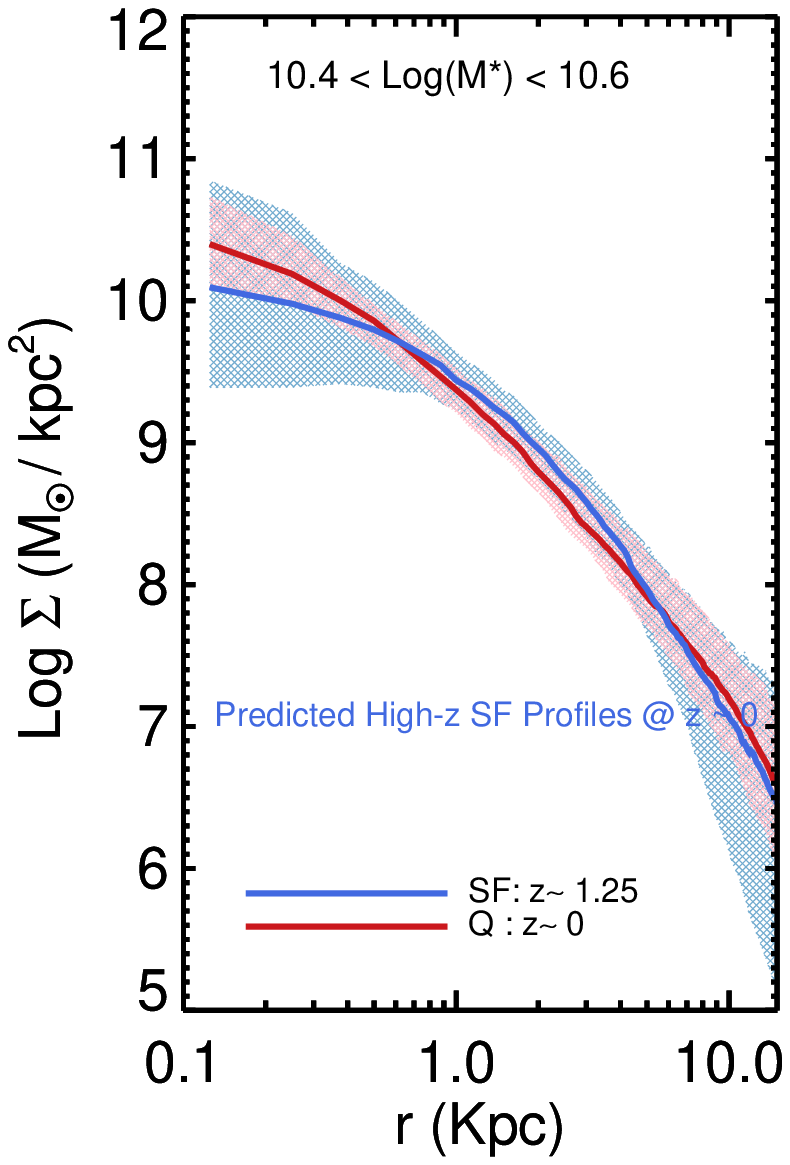}
\caption{Similar to Figure \ref{fig20} but comparing the expected $z\sim 0$ stellar mass profiles of star-forming galaxies at $z\sim 1.25$ (higher redshift bins) with the local quiescent galaxies at $z\sim0$ at fixed masses. Except the most massive bin (left panel) in which galaxies reached to their maximum central densities, intermediate star-forming galaxies have different mass profiles, specifically in their central regions, indicating the necessity for building the central bulges.}

\label{fig21}
\end{figure*}

\begin{figure}
\centering
\includegraphics[width=0.48\textwidth]{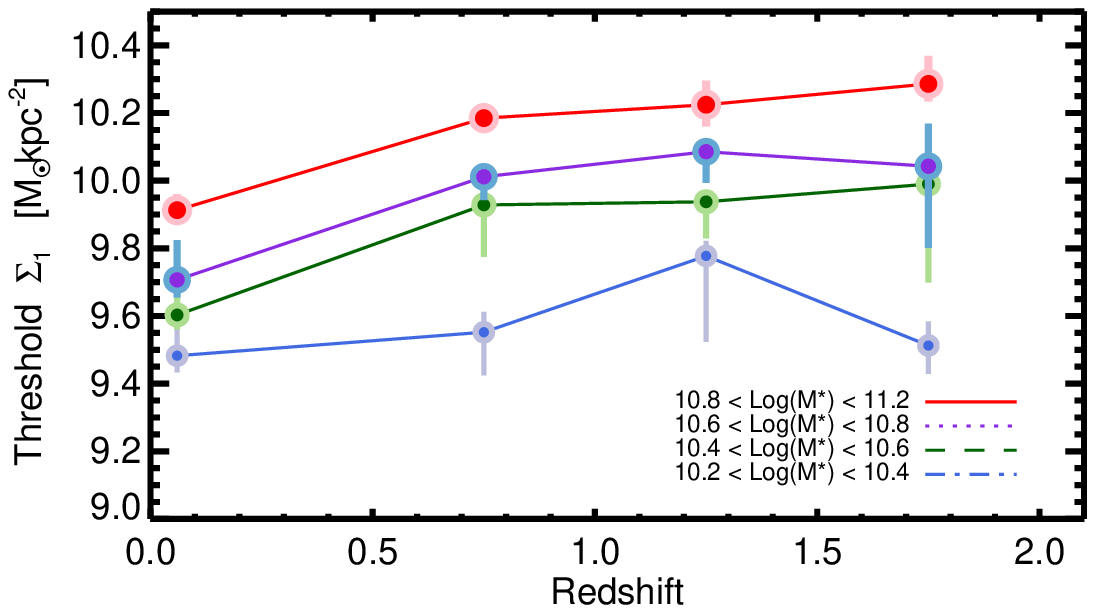}
\caption{Evolution of $\Sigma_{1}$ threshold values at fixed masses form high-$z$ to low-$z$, using the median $\Sigma_{1}$ values of galaxies in green valley (gray regions of Figure \ref{fig15}). At fixed mass, the threshold is higher at earlier times.}
\label{fig22}
\end{figure}


\subsection{Relation between $\Sigma_{1}$ \& Quenching}

As shown in Figure \ref{fig14} and \ref{fig15}, at fixed mass, the star formation of galaxies declines sharply around specific $\Sigma_{1}$ threshold. This can suggest a ``knee-shaped'' or ``L-shaped''  evolutionary path on sSFR-$\Sigma_{1}$ or color-$\Sigma_{1}$ planes at all our studied redshifts and different mass bins \citep[as also suggested by][]{Barro2015b}. The total stellar mass and central density of galaxies are correlated \citep[e.g.,][]{tacchella2015}. In addition, simulations show that while the total stellar masses of star-forming galaxies increase, their central densities also grow (via star-formation or redistribution of stellar masses) till they saturate around specific central density \citep[see e.g.,][]{tacchella2016a}. Therefore, at first sight this may suggest that the galaxies start to quench once they reach around a certain $\Sigma_{1}$ threshold. This looks to be the same across all our studied redshifts, although the threshold values depend slightly on the redshifts (higher values for high-$z$ galaxies).  This is consistent with early observational studies on galaxies mass profiles at low and intermediate redshifts \citep{cheung2012, fang2013}, at higher redshifts \citep{tacchella2015, Barro2015b} and simulations \citep{tacchella2016a} \citep[see also][]{bell2012, lang2014, bluck2014, vandokkum2014, Brennan2015, teimoorinia2016, whitaker2016}. However, as mentioned earlier the links between cause and effects are difficult to establish. We note that the advantage of this work compared to previous studies is to use PSF-corrected mass-profiles and reducing the uncertainties due to $M/L$ and color-gradients in a consistent way from $z\sim2$ to $0$.

To further understand this transition, we compare galaxy mass profiles. From Figure \ref{fig12} it can be seen that at $z\sim0.7$ star-forming galaxies (with $\log(\mstar/\msun)>10.4$) have already reached to their maximum central densities at fixed masses compared to their counterparts at high-$z$. Therefore, if these galaxies will become quenched at later times, then their expected mass profiles should be comparable to the quiescent ones at $z\sim0$ with similar masses (assuming galaxies will not experience major mergers which is less frequent at low redshifts). In other words, assuming quiescent galaxies in the local universe once have been star-forming at high redshifts and they have gradually become quenched without any significant changes in their internal structures, then it is expected that their mass distributions become comparable to the local quenched galaxies at fixed masses (inside-out quenching scenario, see also \citealt{tacchella2016a} and \citealt{Onodera2015}).

We put this in a simple test and investigate it by using stellar population synthesis models of \cite{BC2003} to predict their mass profiles at $z\sim0$ and compare them to local quiescent galaxies at fixed masses. For this purpose, we used the best-fit parameters of the star-formation rate and the age derived from SED fitting, in addition to the stellar population model created with the same parameters. Using the table of the age and stellar mass from the stellar population model, the expected stellar mass at an specific time (here $z\sim0$) are estimated. The comparison between predicted mass-profiles at $z\sim0$ of high-$z$ star-forming galaxies ($z\sim0.75$) are shown in Figure \ref{fig20}, divided into three mass bins, decreasing from left to right. For most massive bins (left and middle panels), the expected mass profiles are very similar to local quiescent ones, at all radii. For lowest mass bin (right panel), the predicted mass profile of star-forming ones is slightly extended and less compact as the quiescent galaxies at low-$z$. Perhaps, additional process(es) required to redistribute stellar masses or assemble masses in their central regions. Note that these less massive galaxies build their central densities later, hence using a lower redshift samples would help better for this test. 

In Figure \ref{fig21}, we show the same but for the star-forming galaxies at $z\sim1.25$. The $z\sim0$ expected profiles of most massive star-forming galaxies at $z\sim1.25$ is very similar to the quiescent ones in the local Universe. These massive star-forming galaxies have reached their relative maximum central densities by $z\sim1.5$. This shows indeed that stellar mass and central density correlate, hence it is difficult to disentangle whether it is mass or density (or even the halo mass) that control the quenching. However, for intermediate mass bins (middle and right panels), which have not reached to their maximum central densities by $z\sim1.25$, their expected profiles differs from the low-$z$ quiescent ones. They have a more extended and less concentrated mass profiles. Note that objects in the low mass bins might change their ``mass bin'' before they quench. This test recalls that star-forming galaxies at this high-$z$ need to grow their central densities and reach to a specific $\Sigma_{1}$ threshold value, prior to their quenching.

The specific threshold of $\Sigma_{1}$ values depends on stellar mass and redshifts.  It can be defined approximately by the median values of galaxies in green valleys (gray shaded region of Figure \ref{fig15}) for each stellar masses and is shown in Figure \ref{fig22}. The specific threshold values increase with redshift at fixed mass, implying higher central density threshold requirements for galaxies at high-$z$ to become quenched.  \citet{franx2008} showed similar evolution for the effective surface density threshold and pointed the reason for this evolution could be due to the higher specific star-formation rate at higher redshift.  

It worth noting that central density strongly correlated with central velocity dispersion ($\sigma_{1}$), as derived by \citet{fang2013} from virial theorem. It has also pointed by \citet{wake2012} that central velocity dispersion is better correlated with color of galaxies. Therefore, possibly $\sigma_{1}$-sSFR might have similar or tighter correlation than $\Sigma_{1}$. This needs to be examined observationally for a large redshift range, though it is very expensive. \\


\section{Summary \& Conclusions}

In this paper we have derived PSF-corrected stellar mass profiles of 2391 mass-complete sample of $>10^{10}\msun$ galaxies up to $z\sim2$ from 3D-HST catalog and randomly selected of $\sim1000$ mass-matched $>10^{10}\msun$ galaxies at $z\sim0$. We examined the half-mass radii of star-forming and quiescent galaxies in small stellar mass bins. We also examined the relation between stellar mass structural parameters and star-formation activity. We find that:

\begin{itemize}
\item Half-mass radii of all types of galaxies are on average smaller that their half-light radii (by $\sim30-50\%$) with a weak dependence of stellar masses.
\item At fixed mass, the average half-mass size of quiescent galaxies increase by a factor of 4 since $z\sim2$ in a similar pace to their half-light radii ($R_{m}\propto (1+z)^{-1.43\pm0.12}$ for massive quiescent galaxies). Mass-weighted sizes of star-forming galaxies change very little at fixed mass ($R_{m}\propto (1+z)^{-0.46\pm0.11}$ for massive star-forming galaxies).  
\item Star-forming galaxies build up their central densities with cosmic time concurrently with their outer regions. Quiescent galaxies build up most of the stellar mass in their outskirts.
\item Stellar mass central density within 1 kpc of galaxies ($\Sigma_{1}$) is better correlated with sSFR than effective surface density at all studied redshift. 
\item Galaxies follow a ``knee-shaped'' path on the $\Sigma_{1}$-$\Delta$sSFR plane at all masses and independent of redshift, i.e., galaxies need to build central densities before quenching. Threshold $\Sigma_{1}$ depends on redshift and stellar mass.
\item  We emphasize that the correlation between $\Sigma_{1}$ and sSFR does not imply a causal link.
\end{itemize}


\begin{acknowledgments}
We thank the anonymous referee for helpful comments and suggestions. This work is based on observations taken by the 3D-HST Treasury Program (GO 12177 and 12328) with the NASA/ESA HST, which is operated by the Association of Universities for Research in Astronomy, Inc., under NASA contract NAS5-26555.

Funding for the SDSS and SDSS-II has been provided by the Alfred P. Sloan Foundation, the Participating Institutions, the National Science Foundation, the U.S. Department of Energy, the National Aeronautics and Space Administration, the Japanese Monbukagakusho, the Max Planck Society, and the Higher Education Funding Council for England. The SDSS Web site is http://www.sdss.org/.

The SDSS is managed by the Astrophysical Research Consortium for the Participating Institutions. The Participating Institutions are the American Museum of Natural History, Astrophysical Institute Potsdam, University of Basel, University of Cambridge, Case Western Reserve University, University of Chicago, Drexel University, Fermilab, the Institute for Advanced Study, the Japan Participation Group, Johns Hopkins University, the Joint Institute for Nuclear Astrophysics, the Kavli Institute for Particle Astrophysics and Cosmology, the Korean Scientist Group, the Chinese Academy of Sciences (LAMOST), Los Alamos National Laboratory, the Max-Planck-Institute for Astronomy (MPIA), the Max-Planck-Institute for Astrophysics (MPA), New Mexico State University, Ohio State University, University of Pittsburgh, University of Portsmouth, Princeton University, the United States Naval Observatory, and the University of Washington.

\end{acknowledgments}

\appendix
\section{Different Stellar Formation History}
\begin{figure*}
\centering
\includegraphics[width=0.48\textwidth]{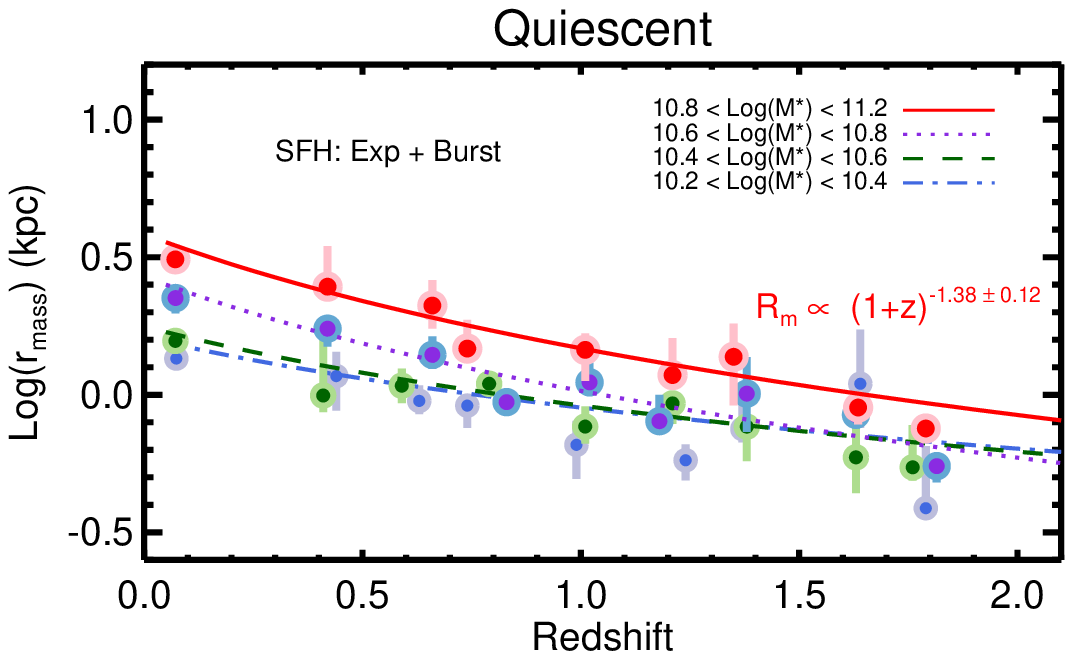}
\hspace{5. mm}
\includegraphics[width=0.48\textwidth]{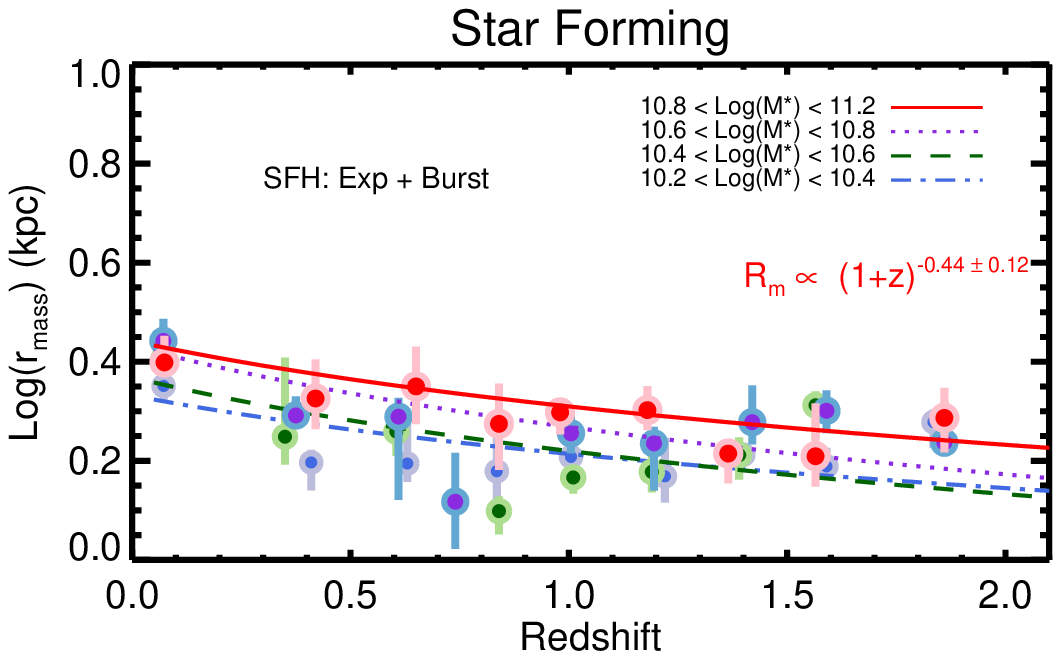}
\caption{The evolution of half-mass radii of galaxies at fixed masses, assuming different star formation history at higher redshifts, i.e,  using random burst in addition to exponential declining star formation history. The results will not alter by assuming different star-formation history. }

\label{figA1}
\end{figure*}

Using difference star-formation history (SFH) at low and high redshift could introduce systematic effects on the derived parameters. As discussed in the text, we have assumed exponentially declining SFH for high redshift sample. However, at low redshift we assumed stochastic burst in addition to the declining SFH. Assuming additional burst could introduce systematic in total stellar masses \citep{Pozzetti2007}. To examine whether our results depend on these choices, we repeat analysis at high redshift assuming similar SFH at low-$z$, using iSEDfit code. We remeasure half-mass radii using the new stellar mass profiles and show their evolution at fixes masses for quiescent and star-forming galaxies in left and right panel of Figure \ref{figA1}, respectively. It can be seen that our results do not depends on assumption of different SFH.    \\

\section{Residual-Corrected Method for Deriving Light Profiles}

The complexity of the light profiles of galaxies due to sub-structural components (e.g., inner disks, clumpy features, etc.) could have introduced deviations of galaxies surface brightness profiles from the best-fit single-\ser profiles and consequently on the derived stellar mass densities. To test whether this could have affected  the results of this work, we follow \cite{szomoru2010, szomoru2012} and apply the same technique to derive the corrected surface brightness profiles in different filters. The de-convolved profiles of galaxies are derived by fitting PSF convolved single-\ser models to the observed two dimensional surface brightness images of galaxies. We then derived the residual profile of galaxies, taking into account the geometry and position of the best-fit model and  added this to the best-fit de-convolved \ser model. Where the uncertainties of the sky background dominate, we extrapolated the profiles at the larger radii, using a \ser model profile. The stellar mass surface density was  then derived from these corrected surface brightness profiles in the same way as described in section 3.2.

\begin{figure*}
\centering
\includegraphics[width=0.48\textwidth]{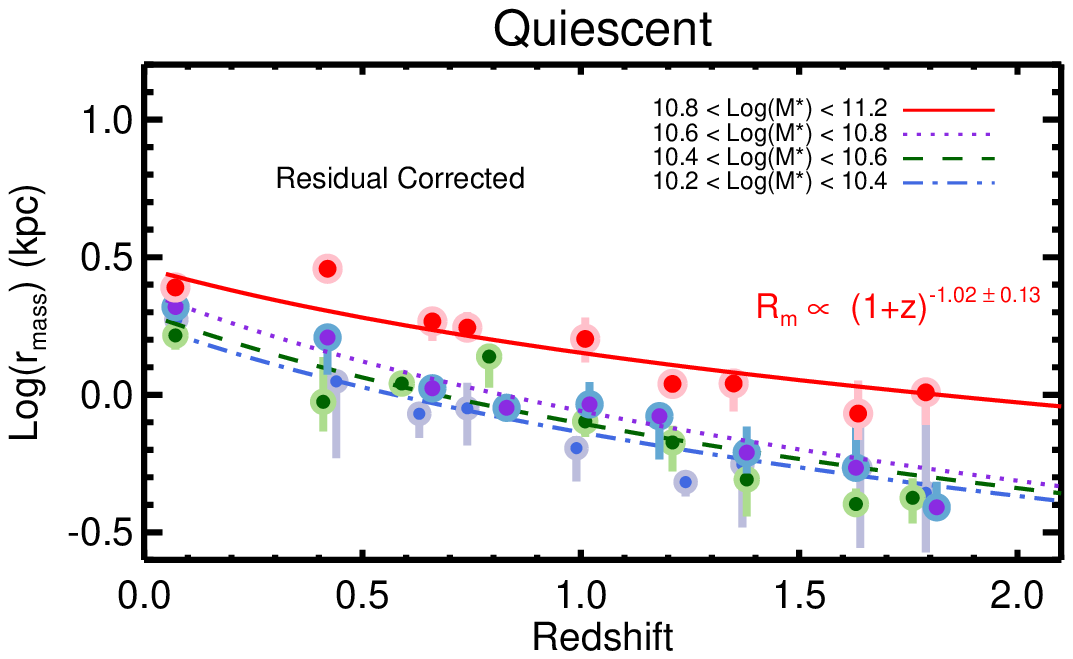}
\hspace{5. mm}
\includegraphics[width=0.48\textwidth]{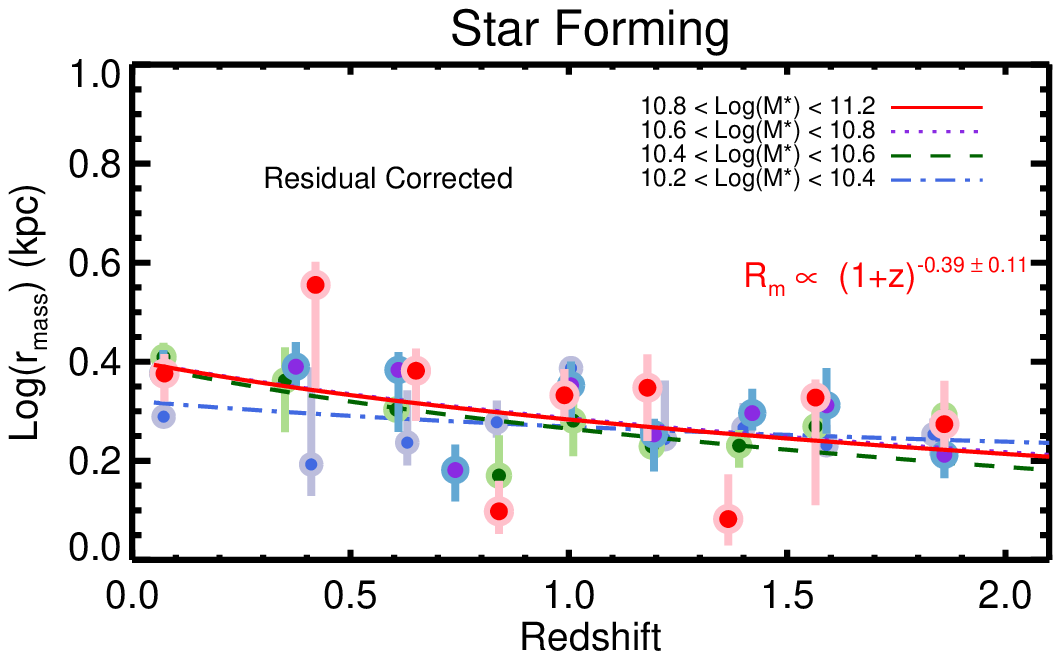}
\caption{The evolution of half-mass radii of galaxies at fixed masses, derived using residual-corrected profile method by \cite{szomoru2010}. The rate of evolution for star-forming and quiescent at different mass bins are consistent with the results in Figure \ref{fig9} and \ref{fig10} assuming single \ser models.}

\label{figA2}
\end{figure*}

The right and left panels of Figure \ref{figA2} show the evolution with redshift of half-mass radii of star-forming and quiescent galaxies, respectively. A comparison with the left panels of Figures \ref{fig9} and \ref{fig10} shows that the rate of evolution of half-mass radii of both quiescent and star-forming galaxies are consistent with the results based on the best-fit single-\ser models. This indicates that independent of the method for deriving the stellar mass profiles, the half-mass radii of quiescent galaxies at fixed masses increase with time. Similarly, for the star-forming galaxies, the rate of evolution of the half-mass at fixed masses is slower than for quiescent galaxies. 

\begin{figure*}
\centering
\includegraphics[width=0.48\textwidth]{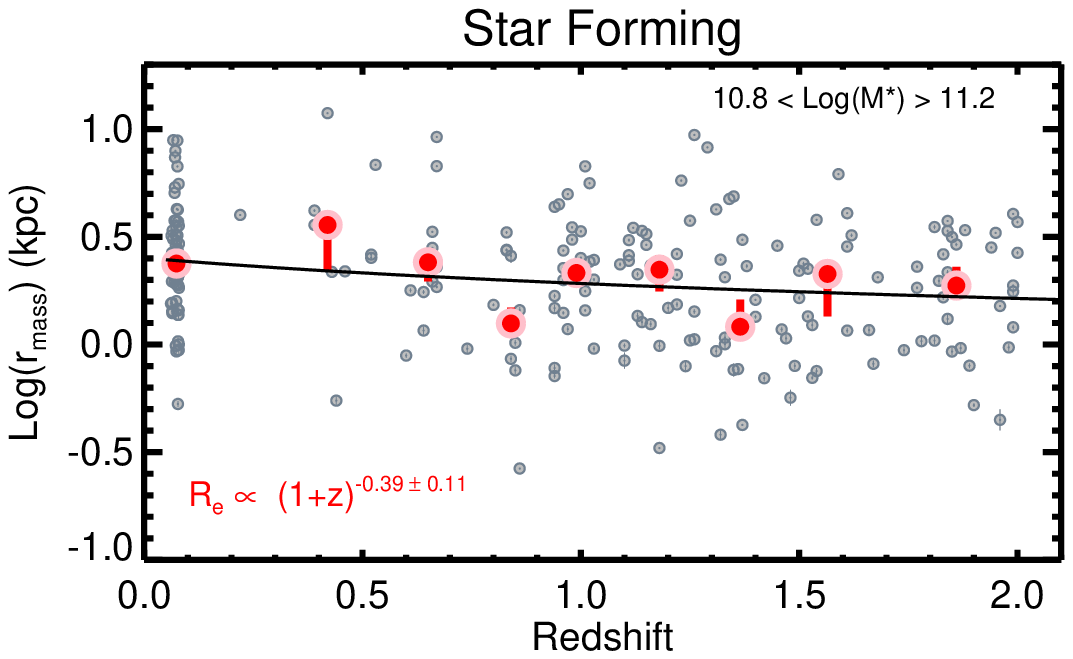}
\hspace{5. mm}
\includegraphics[width=0.48\textwidth]{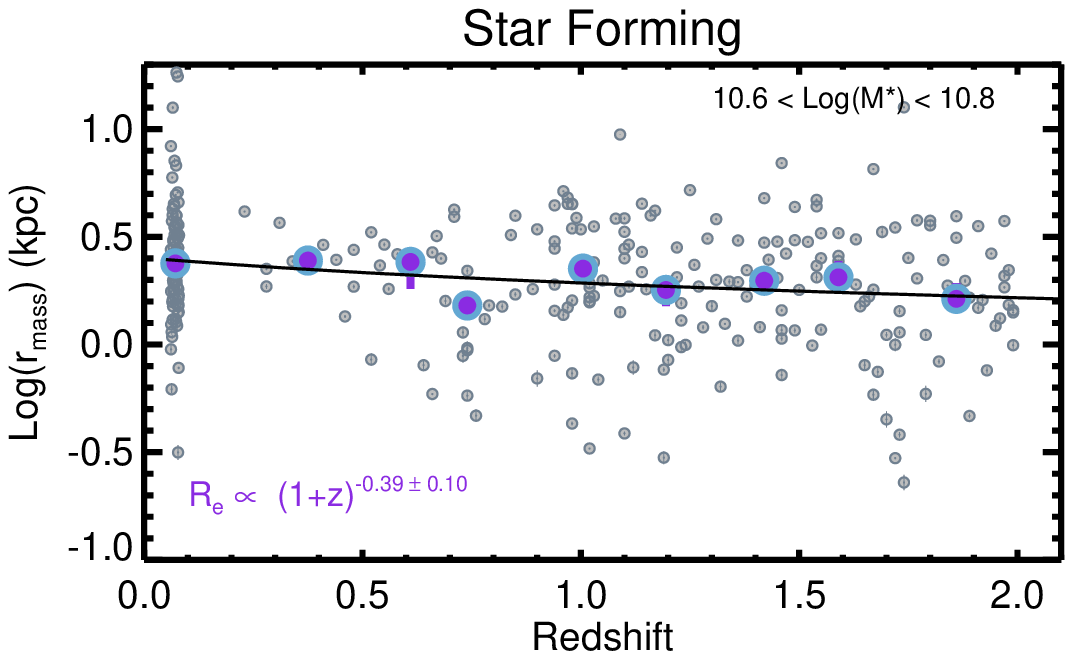}
\hspace{5. mm}
\includegraphics[width=0.48\textwidth]{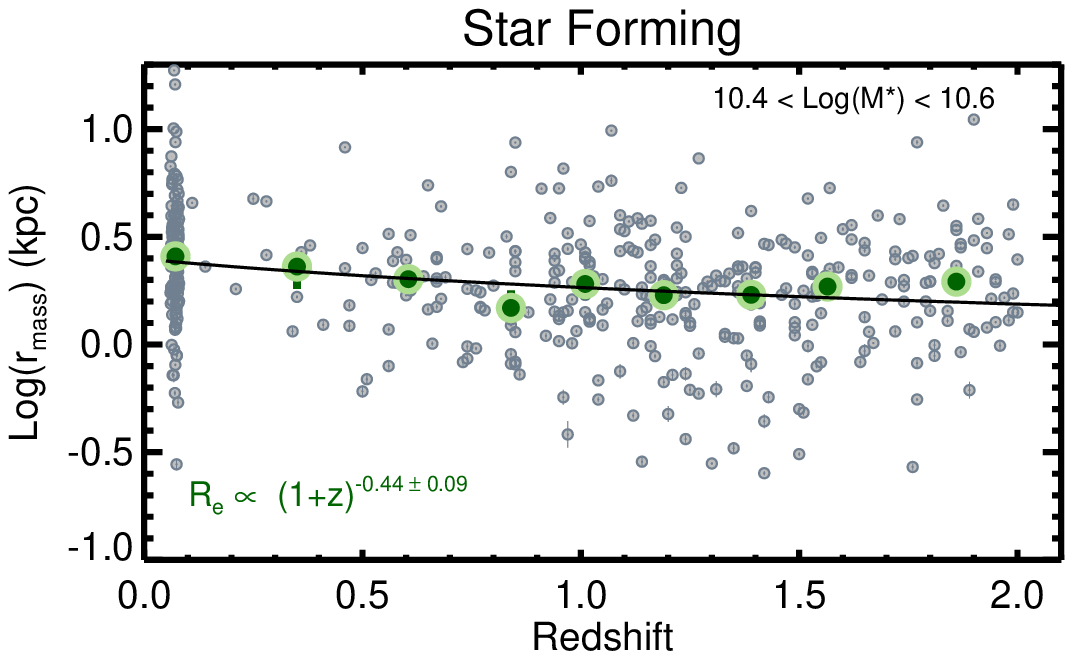}
\hspace{5. mm}
\includegraphics[width=0.48\textwidth]{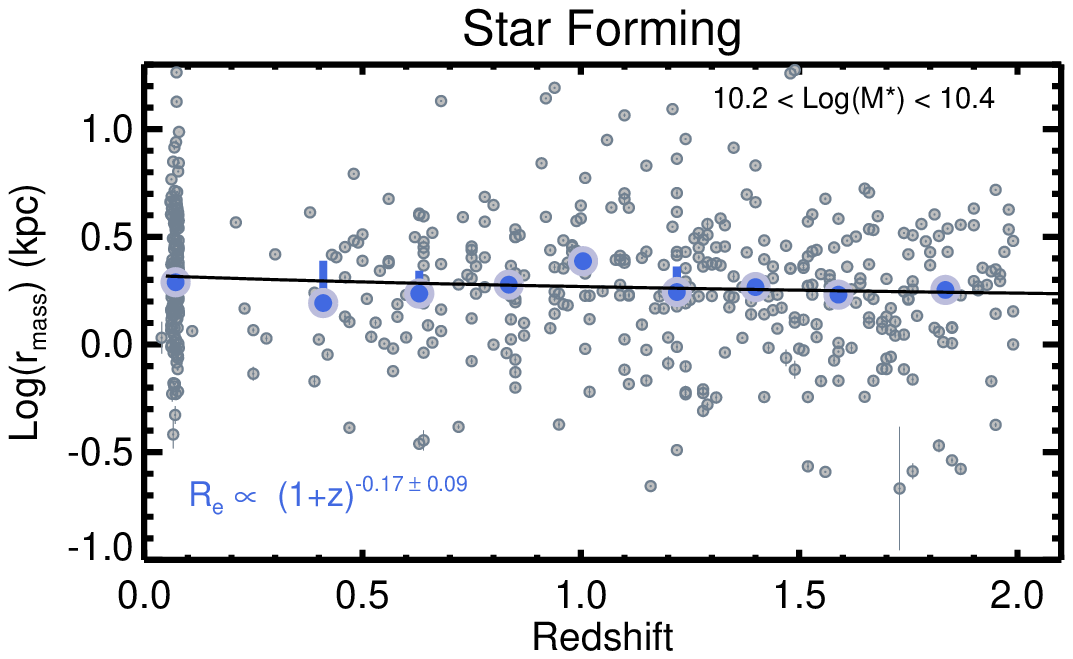}

\caption{The evolution of mas-weighted sizes of star-forming galaxies at different stellar mass bins (different panels). The gray points represents the half-mass sizes of individual galaxies and the color circles are their medians at different redshifts. The solid lines are the best-fit size evolution to the data points. The star-forming galaxies show little size evolution at fixed masses with redshift.} 

\label{figA3}
\end{figure*}

In Figure \ref{figA3}, we show in detail the evolution of half-mass radii of star-forming galaxies using residual-corrected method. The grey points show the mass-weighted sizes for individual galaxies and the color symbols show the medians in each redshift bin.

\begin{figure*}
\centering
\includegraphics[width=0.8\textwidth]{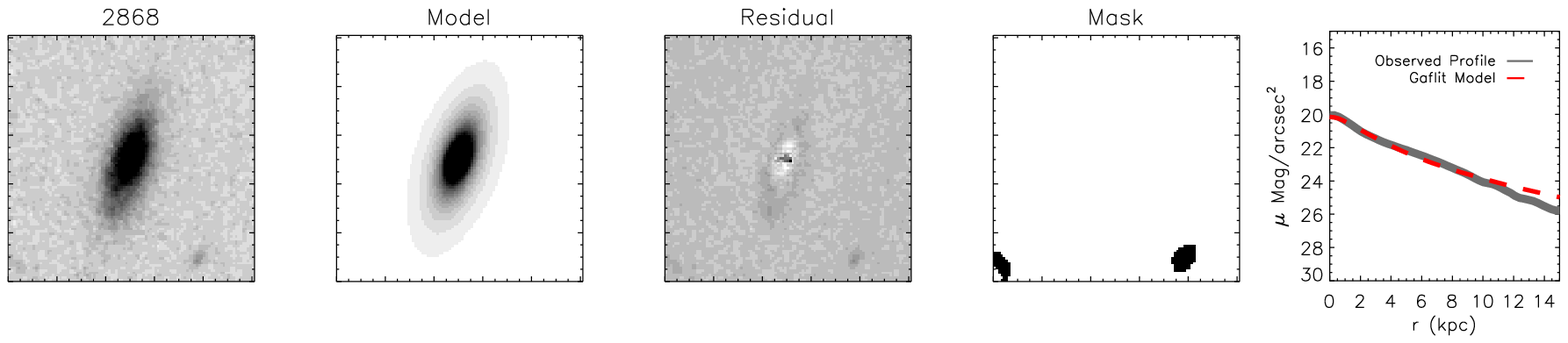}
\hspace{5. mm}
\includegraphics[width=0.8\textwidth]{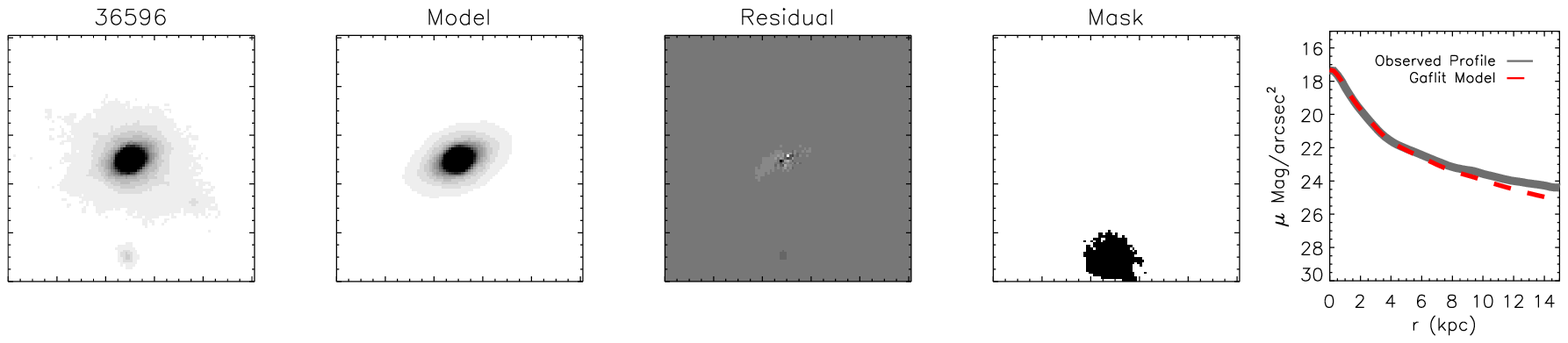}
\hspace{5. mm}
\includegraphics[width=0.8\textwidth]{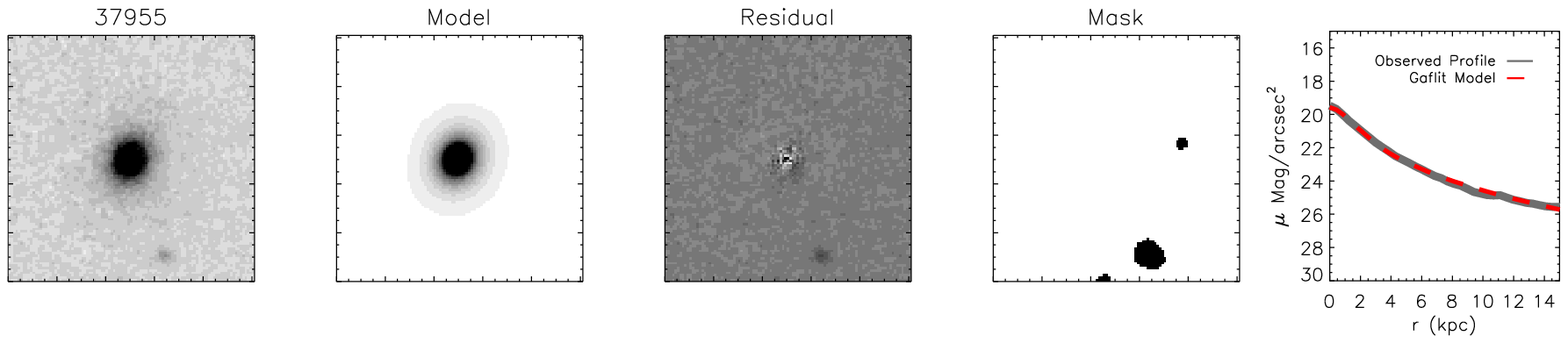}
\hspace{5. mm}
\includegraphics[width=0.8\textwidth]{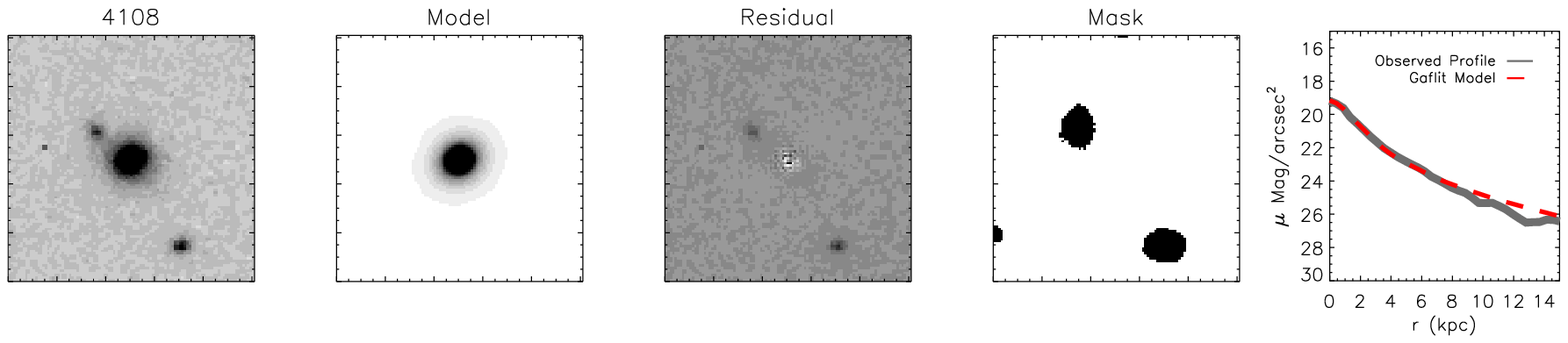}
\hspace{5. mm}
\includegraphics[width=0.8\textwidth]{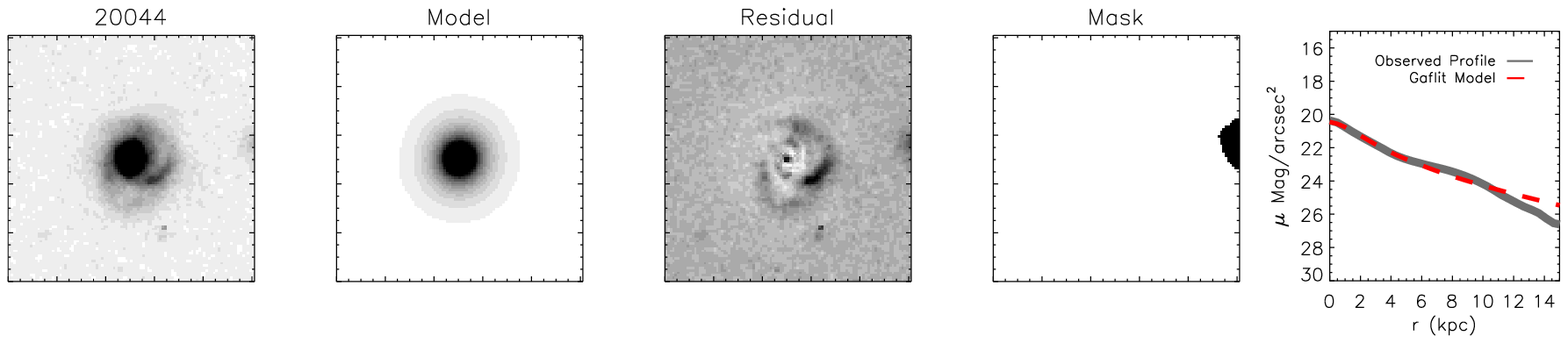}
\caption{From left to right, we show the $H$-band postage stamps of few galaxies in our sample, their best-fit single \ser models, the residuals, mask maps and the comparison of 1D observed light profiles (gray lines) with their best-fit models (dashed red lines). Using single \ser models for galaxies in this study can recover the true properties of galaxy light profiles.} 

\label{figA4}
\end{figure*}

The best-fit single \ser models are compared in 1D and 2D to the observed light profiles of few galaxies in Figure \ref{figA4}. We conclude that, with the current resolution of HST images, the results of this paper are robust when using the stellar mass profiles derived from single-\ser models.\\

\end{document}